\documentclass[12pt]{style/spieman}  
\usepackage{amssymb,amsbsy,amsmath,amsfonts,amscd,mathrsfs,upgreek}
\usepackage{graphicx}
\usepackage{setspace}
\usepackage{tocloft}
\usepackage{lineno}
\usepackage{placeins}
\usepackage{color}
\usepackage{xcolor}
\usepackage{soul}
\usepackage{booktabs}
\usepackage{tabularx}
\usepackage{makecell}
\usepackage{multirow}
\usepackage{orcidlink}
\usepackage[ruled,linesnumbered]{algorithm2e}
\usepackage{threeparttable}
\colorlet{soulred}{red!40}

\definecolor{darkred}{rgb}{0.75, 0, 0}

\title{Physics-Informed Fusion-MIONet with IFENN: A Generalized Framework for Coupled Poroelasticity Problems}

\author[a,b,c]{Fouad M. Amin\,\orcidlink{0000-0001-6903-4629}}
\author[d]{Rigoberto Moncada\,\orcidlink{0000-0001-7655-5406}}
\author[b]{Panos Pantidis\,\orcidlink{0000-0001-8722-2535}}
\author[b,e]{Diab W. Abueidda\,\orcidlink{0000-0003-3594-2455}}
\author[d]{Fan Fei\,\orcidlink{0000-0001-7273-4458}}
\author[a,b]{Mostafa E. Mobasher\textsuperscript{*}\,\orcidlink{0000-0002-5108-4713}}

\affil[a]{Civil and Urban Engineering Department, Tandon School of Engineering, New York University, Brooklyn, NY11201, US}
\affil[b]{Civil and Urban Engineering Department, New York University Abu Dhabi, Abu Dhabi, P.O. Box 129188, UAE}
\affil[c]{Structural Engineering Department, Faculty of Engineering, Mansoura University, Mansoura 35516, Egypt}
\affil[d]{Lawrence Livermore National Laboratory, 7000 East Avenue, 94550, Livermore, CA, USA}
\affil[e]{National Center for Supercomputing Applications, University of Illinois at Urbana-Champaign, Urbana, IL, USA}

\begin{document}

\maketitle

\begin{abstract}
Complex, high-dimensional, large-scale poroelasticity simulations using traditional numerical solvers can be prohibitively expensive in terms of computational resources. We introduce an innovative framework for poroelasticity problems that integrates physics-informed machine learning (PI-ML) models with high-performance FEM solvers. This integration is achieved through the Integrated Finite Element Neural Network (IFENN) setup, which uses a staggered scheme between an FEM solver for mechanical equilibrium and a PI-ML model for fluid-flow predictions. We introduce a novel Physics-Informed Fusion-MIONet (PIF-MIONet) architecture, in which a fusion network encodes the outputs of the MIONet branches. The MIONet encodes different input functions, and the fusion network enables early interaction between the inputs, supporting generalization across multiple scenarios while adhering to the underlying physics of the poroelasticity problem. We study different examples, covering 2D and 3D domains and considering variable material properties and time discretizations. We test the models on unseen loading histories, demonstrating accuracy and generalizability across scenarios. For complex geometries and boundary conditions, we show that with both physics and data considered in the loss function, a PI-ML model can be trained using only 0.1\% of the training increments used for pure data-driven training. In addition, we expand the PIF-MIONet to incorporate a domain decomposition approach, which was found to be particularly useful when modeling flow over more complex geometries with more localized phenomena. For the first time, IFENN is implemented using the open-source multiphysics simulator GEOS for the FEM part, providing a path for the framework's application to industrial-scale simulations and complex multiphysics problems.     
\end{abstract}


\begin{flushleft}
\footnotesize
\textsuperscript{*}\textbf{Corresponding author:}
Mostafa E. Mobasher,
\linkable{mostafa.mobasher@nyu.edu}
\end{flushleft}

\section{Introduction}
\label{section:introduction}

\subsection{Background and literature review}
\label{section:introduction:literature_review}

Poromechanics is a multiphysics domain that studies interactions between mechanics and fluid flow, covering essential problems in geomechanics, such as slope and borehole stability analysis~\cite{moradi_comparing_2024, maneepong_rock_2025, ding_borehole_2019}, seepage~\cite{guo_uncertainty_2021}, and consolidation analysis~\cite{potts_finite_1999}. These problems involve complex interactions that hinder comprehensive experimental investigations. Numerical simulations offer ways to investigate these problems through approaches such as the Finite Difference Method (FDM)~\cite{shwetank_coupled_2023, liu_generalized_2023}, Finite Volume Method (FVM)~\cite{terekhov_finite_2022, boon_solving_2026}, and Finite Element Method (FEM)~\cite{he_new_2025, ge_two-level_2022}. That said, these conventional approaches have their own challenges, including multiscale phenomena in time and space, which can make some simulations prohibitively expensive~\cite{moradinia_navigating_2024, meinecke_data-driven_2024, lyu_multiscale_2025, du_addressing_2026}. Researchers have introduced several approaches to address these numerical challenges, such as adaptive remeshing, parallel computing, domain decomposition, and staggered solvers~\cite{tang_review_2021,permann_moose_2020, castellazzi_staggered_2021, stershic_adaptively_2025}. However, improving these approaches and offering better alternatives remains an open research area with many possible avenues. Machine Learning (ML) is one avenue currently being extensively explored to address the challenges in coupled multiphysics simulations. In this paper, we introduce a novel formulation of the Integrated Finite Element Neural Networks (IFENN) framework to expedite poroelasticity simulations, utilizing an innovative machine learning model. 

Machine learning approaches can be classified as \textit{data-driven}, using only labeled data in the loss function, or \textit{physics-informed}, where residuals of the governing physics laws contribute to the loss function. Data-driven approaches may generalize poorly, while physics-informed machine learning (PI-ML) yields physically consistent predictions and improved generalization~\cite{karniadakis_physics-informed_2021}. PI-ML can integrate both labeled data and governing physical laws, improving data efficiency and stabilizing predictions~\cite{karniadakis_physics-informed_2021, sharma_review_2023}. In addition, PI-ML has shown promising results in several applications in porous media~\cite{delpisheh_leveraging_2024}, including fluid flow~\cite{fuks_limitations_2020, fraces_physics_2021, shan_physics-informed_2023} and poroelasticity~\cite{bekele_physics-informed_2020, bekele_physics-informed_2021, haghighat_physics-informed_2022, millevoi_physics-informed_2024, roy_physics-informed_2025} simulations. On the other hand, PI-ML is associated with numerous challenges, such as slow convergence, hyperparameter sensitivity, complex loss landscapes, difficulty scaling to multiphysics problems, enforcing boundary conditions, and neural network architectural design~\cite{cuomo_scientific_2022, luo_physics-informed_2025}. To address these challenges, different techniques were introduced in the literature, including adaptive weights~\cite{bischof_multi-objective_2025, chen_gradnorm_2017, heydari_softadapt_2019, nath_amore_2026, wang_understanding_2021}, dimensionless forms~\cite{haghighat_physics-informed_2022, amini_physics-informed_2022, amini_inverse_2023}, shape function derivatives~\cite{he_use_2023, pantidis_i-fenn_2024}, and domain decomposition~\cite{wu_physicsinformed_2024, sharma_ensemble_2024}.

Focusing on poroelasticity, several data-driven and physics-informed approaches were introduced to solve inverse and forward problems. Among data-driven approaches, Vasilyeva and Tyrylgin~\cite{vasilyeva_machine_2021} used a convolutional neural network (CNN) to solve an inverse problem, estimating macroscale properties from local microscale stochastic fields. Kadeethum et al.~\cite{kadeethum_framework_2021} adopted conditional generative adversarial networks (cGAN) to address both forward and inverse problems for steady-state response. Choi et al.~\cite{choi_operator_2026} investigated different architectural designs, using data-driven training, for a one-dimensional (1D) consolidation problem.  Physics-informed neural networks (PINNs) have been utilized in several forward problems. For example, Bekele~\cite{bekele_physics-informed_2020, bekele_physics-informed_2021} used fully connected deep neural networks to predict displacements and fluid pressure for the 2D Barry-Mercer injection-production problem and a 1D consolidation problem. Haghighat et al.~\cite{haghighat_physics-informed_2022} introduced a sequential training technique based on the stress-split approach in poromechanics, resulting in improved training convergence, and implemented it for consolidation, injection-production, and two-phase drainage problems. Millevoi et al.~\cite{millevoi_physics-informed_2024} explored incorporating data into the loss function, resulting in faster training and more accurate results. To address heterogeneous media, Roy et al.~\cite{roy_physics-informed_2025} introduced a composite neural network (CoNN) in which different networks are used for each homogeneous subdomain.  More recently, Ren et al.~\cite{ren_poro_2026} addressed an inverse problem by identifying material parameters through processing the observed system response.

Despite the valuable contributions of the aforementioned studies, data-driven approaches limit the generalization capabilities of the trained models~\cite{karniadakis_physics-informed_2021} and require large datasets~\cite{amin_i-fenn_2026}. On the other hand, studies utilizing PINNs ~\cite{bekele_physics-informed_2020, bekele_physics-informed_2021, haghighat_physics-informed_2022, millevoi_physics-informed_2024, roy_physics-informed_2025}, have two main limitations. First, they do not explore 3D spatial domains or complex boundary conditions for poroelastic modeling of geomaterials. Second, all the introduced PINNs are trained for fixed loading scenarios. The inputs are only spatial and temporal coordinates, without the ability to encode other input fields, such as a strain field, a variable material property, or a random boundary condition. These limitations highlight the need for advanced ML architectures and training techniques that can handle complex geometries, multiple inputs, and random loading histories while adhering to the underlying physics of poroelasticity problems. 

As ML has evolved, operator-based architectures have emerged, offering greater generalization and expressivity~\cite{raonic_convolutional_2023, sarkar_spatio-spectral_2025, li_fourier_2020, lu_learning_2021}. The deep operator network (DeepONet)~\cite{lu_learning_2021} is one such architecture, with promising capabilities for learning different operators. A classic (vanilla) DeepONet consists of two networks: a \textit{branch} for encoding input functions, and a \textit{trunk} for encoding the coordinates for the output points. The branch and trunk outputs are combined via a dot product operation to produce the DeepONet output. The DeepONet architecture is a high-level modular design, with many possible designs for the branch or the trunk, including multilayer perceptrons (MLPs), convolutional neural networks (CNNs), recurrent neural networks (RNNs), and Kolmogorov-Arnold networks (KANs)~\cite{lu_learning_2021, abueidda_deepokan_2025}. Many variants of DeepONet have been introduced, such as the separable physics-informed DeepONet (Sep-PI-DeepONet) for high-dimensional problems~\cite{mandl_separable_2025}. Another variant is the multiple-input deep neural operator (MIONet)~\cite{jin_mionet_2022}, which handles multiple input functions through multiple branches. The MIONet architecture has been utilized in several applications, including structural dynamic response~\cite{ahmed_physics-informed_2025}, phase-field hydraulic fracturing~\cite{wang_spi-mionet_2024}, multiphase flow modeling~\cite{jiang_fourier-mionet_2024}, and poroelasticity~\cite{choi_operator_2026}. 

In this paper, we introduce a Physics-Informed Fusion-MIONet (PIF-MIONet) that integrates fusion concepts~\cite{zhang_multimodal_2020, kiziltepe_integration_2024,liu_image_2025, peyvan_fusion-deeponet_2026, choi_operator_2026} into the MIONet. The PIF-MIONet inherits MIONet's ability to encode multiple input functions, including variable material properties and random injection-production flux histories. The fusion component enables early interaction between the input variables. In addition, the trunk and branch outputs are aggregated using the same approach as in the standard DeepONet~\cite{lu_learning_2021}, making PIF-MIONet a suitable extension of many existing models. The PIF-MIONet has two main aspects, namely, the Fusion-MIONet architecture and the physics-informed training process. To improve training stability and efficiency, we investigate and integrate adaptive weights, dimensionless forms, shape function derivatives, and domain decomposition within the PIF-MIONet. 

Despite the computational efficiency of pure ML-based methods, they face several limitations, including concerns regarding reliability~\cite{pantidis_error_2023, daw_rethinking_2022}, generalizability~\cite{fuhg_physics-informed_2022}, limited interpretability~\cite{sun_data-driven_2022, manfren_data-driven_2022}, sensitivity to hyperparameters~\cite{pantidis_error_2023, markidis_old_2021}, training stability~\cite{krishnapriyan_characterizing_2021, wang_understanding_2021}, and network convergence~\cite{shin_convergence_2020, mishra_estimates_2022}. To bridge this gap, a promising research trend has emerged: hybrid approaches that integrate ML and classical methods in different ways for different problem types. For single-physics applications, several approaches have been introduced, including FEMIN~\cite{thel_introducing_2024} and PI-DeepONet-FEM~\cite{wang_time-marching_2025} for dynamic loading and crash simulations, as well as HINTS~\cite{zhang_blending_2024}, ANN-FEA~\cite{quan_integrating_2025}, and NN-PGD~\cite{daby-seesaram_finite_2026}. For coupled multiphysics applications, the hybrid approaches include NN-Nested-FEM~\cite{he_machine_2025}, NeuroSEM~\cite{shukla_neurosem_2025}, PDNNs~\cite{zhang_pdnns_2023}, and the Integrated Finite Element Neural Network (IFENN)~\cite{pantidis_integrated_2023}. 

IFENN is a general-purpose framework providing efficient and accurate simulations for coupled multiphysics problems. The core idea of IFENN is to use both FEM and a pre-trained NN to solve the governing PDEs. Following a hybrid staggered scheme, the FEM solver typically handles the mechanical field equilibrium, while an NN approximates the solution of the coupled field. The two solvers exchange information at every load or time increment, creating tight coupling throughout the analysis. This design choice (i.e., FEM for the mechanical field and an NN for the coupled field) is made because the mechanical response is more complex and has multiple components, while the other coupled field is usually scalar. Thus, given the limited reliability of ML models, a strategic choice is to use an NN for scalar field predictions and an FEM solver for the complex mechanical response. 

The IFENN framework has been successfully applied across several applications, including damage~\cite{pantidis_integrated_2023, pantidis_i-fenn_2024} and phase-field fracture modeling~\cite{pantidis_integrated_2026, pantidis_hybrid_2026}, thermoelasticity~\cite{abueidda_i-fenn_2024, abueidda_variational_2024, amin_i-fenn_2026}, and poroelasticity~\cite{amin_i-fenn_2026}. Each of these studies has its own limitations, usually related to the training challenges or generalization capabilities. For poroelasticity, IFENN has exclusively addressed 2D problems utilizing large datasets for data-driven DeepONet training. Despite the promising capabilities of PI-ML, and due to the aforementioned challenges across this review, research gaps remain in its application to poroelastic geomechanical problems. In particular, the feasibility of generalized PI models for poroelasticity and their implementation in 2D and 3D spatial domains, while accounting for temporal and material variations, remains insufficiently explored. In this paper, we introduce the PIF-MIONet, address its training challenges, and integrate it within the IFENN for 2D and 3D domains, exploring improved generalizability and relevance to real-world problems.

\subsection{Scope and outline}
\label{section:introduction:scope_and_outline}
In this study, we introduce the Physics-Informed Fusion-MIONet (PIF-MIONet) as a surrogate solver for the fluid mass balance equation. We integrate PIF-MIONet within IFENN for coupled poroelasticity problems. This novel integration is introduced for the first time with the following contributions: 

\begin{itemize}

\item The PIF-MIONet training incorporates adaptive weights, dimensionless forms, and shape function derivatives.

\item We incorporate spatial domain decomposition into PIF-MIONet, improving accuracy and training stability for complex geometries while maintaining a comparable number of trainable parameters. 

\item The framework can simultaneously consider multiple time-increment sizes, material properties, flux histories, and mesh resolutions. 

\item The IFENN integration leverages the open-source FEM simulator GEOS~\cite{settgast_geos_2024}, enhancing scalability and portability for hybrid FEM-ML approaches and bridging academic research with industrial-scale simulations.

\end{itemize}

The rest of the document is structured as follows:
Section~\ref{section:problem_statement} outlines the mathematical formulation for poroelasticity problems, followed by Section~\ref{section:methodology}, which details the IFENN framework implementation for poroelasticity problems.
Section~\ref{section:pif_mionet_design} and Section~\ref{section:pif_mionet_training} provide details of the proposed PIF-MIONet design and training, respectively.
Section~\ref{section:numerical_examples} explores the training performance and the IFENN framework capabilities through numerical examples. Section~\ref{section:discussion_conclusion} discusses and concludes the key takeaways from this work. 

\FloatBarrier
\section{Mathematical formulation for poroelasticity problems}  
\label{section:problem_statement}

\subsection{Strong form}
\label{section:poroelasticity:strong}
In poroelasticity problems, the mechanical deformations and fluid flow through a porous medium interact and affect each other. For a domain composed of solid grains and pores filled with a fluid, the governing equations, coupling two variables: the displacements ($\boldsymbol{u}$) and fluid pressure ($p$), can be written as~\cite{potts_finite_1999, coussy_poromechanics_2004,cheng_poroelasticity_2016, yi_consistent_2020, chen_thermodynamically_2022}:  
\begin{align}
	\label{eq:poro_st_blm}
	\boldsymbol{\nabla} \cdot \boldsymbol{\sigma}_t +\boldsymbol{b} &=\boldsymbol{0}
    \\
    \label{eq:poro_st_cont}
	\dot{\zeta}+\boldsymbol{\nabla} \cdot \boldsymbol{v}_{f} &= Q_f
\end{align}

\noindent
where 
$\boldsymbol{\nabla}$ is the gradient operator, 
$\boldsymbol{\nabla}\cdot$ is the divergence operator,
$\boldsymbol{\sigma}_t$ is the total Cauchy stress due to deformation and fluid pressure, 
$\boldsymbol{b}$ is the mechanical body force vector, 
$\zeta$ is the increment of fluid content, 
$\boldsymbol{v}_{f}$ is the fluid velocity vector, and $Q_f$ represents any fluid sink or source. 

\begin{figure}[hbt!]
\centering
\includegraphics[width=.40\textwidth]{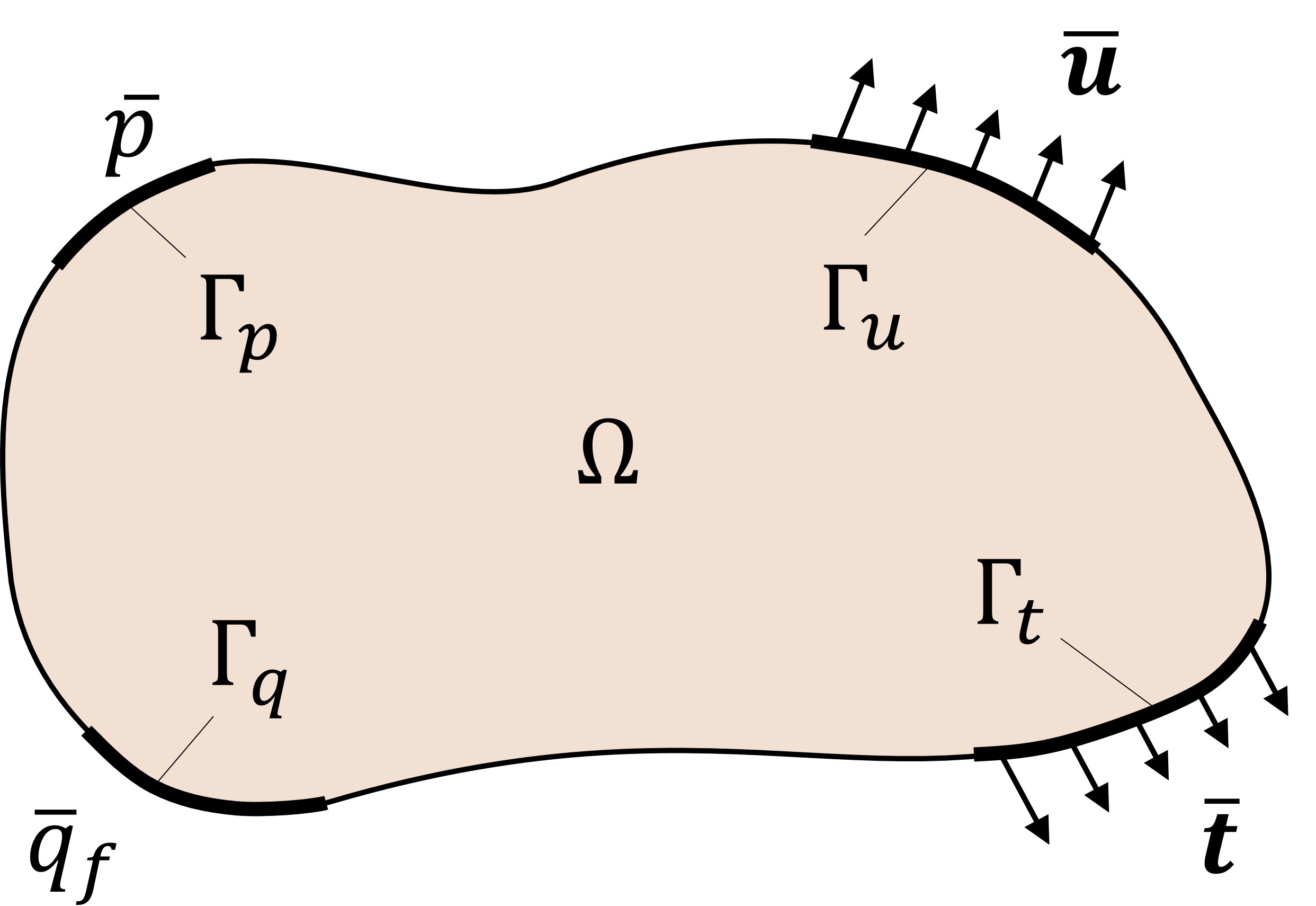}
\caption{Schematic of a poroelasticity problem domain with prescribed boundary conditions}
\label{fig:methods_poro}
\end{figure}

Boundary conditions are defined over the boundaries $\Gamma=\Gamma_u \cup \Gamma_t=\Gamma_p \cup \Gamma_q$, shown in Fig.~\ref{fig:methods_poro}, as follows:
\begin{equation}\label{eq:thermo-strong-bc}
    \begin{aligned}
        \boldsymbol{u} &= \boldsymbol{\overline{u}}, \quad x \in \Gamma_u, 
        \\
        \boldsymbol{t} &= \boldsymbol{\overline{t}}, \quad x \in \Gamma_t,
        \\
        p &= \overline{p}, \quad x \in \Gamma_p, 
        \\
        q &= \overline{q}, \quad x \in \Gamma_q 
    \end{aligned}
\end{equation}

\noindent
where 
$\boldsymbol{\overline{u}}$ and $\boldsymbol{\overline{t}}$ are the displacement and traction boundary conditions defined over boundaries $ \Gamma_u $ and  $ \Gamma_t $, respectively.
$\overline{p}$ is the fluid pressure over boundary $ \Gamma_p $.
$\overline{q} $ is the fluid flux over boundary $\Gamma_q$, defined as $ \overline{q} = \boldsymbol{v}_{f} \cdot \boldsymbol{n}$, where $\boldsymbol{n}$ is the normal unit vector at the boundary $\Gamma_q$.

Considering Darcy's law for fluid flow, the fluid velocity ($\boldsymbol{v}_{f}$) and fluid content increment ($\zeta$) are defined as~\cite{potts_finite_1999, cheng_poroelasticity_2016, yi_consistent_2020, chen_thermodynamically_2022}:
\begin{align}
    \label{eq:fluid_velocity}
    \boldsymbol{v}_{f} &= -K_H \boldsymbol{\nabla} H_f
    \\
    \label{eq:fluid_increment}
	\zeta&= \dfrac{p}{M} + \alpha \text{tr}(\boldsymbol{\varepsilon})
\end{align}
\noindent
where $K_H$ is the isotropic hydraulic conductivity, defined as $K_H = \gamma_f K_p / \mu_f  $, and $H_f$ is fluid total head, defined as $ H_f = {p} / {\gamma_f} + \boldsymbol{x} \cdot \boldsymbol{i}_g$. 
$K_p$ is the intrinsic permeability, $\gamma_f$ is the fluid unit weight, and $\mu_f$ is the dynamic fluid viscosity, $\boldsymbol{x}$ is the coordinate vector, and $\boldsymbol{i}_g$ is a unit vector parallel to gravity and opposite to its direction. 
Notably, given the definitions in Eq.~\eqref{eq:thermo-strong-bc} and Eq.~\eqref{eq:fluid_velocity}, an injection flux $\overline{q}$ would be negative. The Biot's coefficient ($\alpha$) is defined as $\alpha = 1 - {K_b}/{K_s}$, where $K_s$ is the solid grain bulk modulus, and 
$K_b$ is the bulk modulus of the solid skeleton. The Biot's modulus ($M$) is defined as ${1}/{M} = {\phi_b}/{K_f} + {(\alpha-\phi_b)}/{K_s}$, where $\phi_b$ is the porosity of the medium, and $K_f$ is the fluid bulk modulus. The bulk modulus ($K_b$) can be defined as $K_b = \lambda + {2\mu}/{3}$.

The total stress $\boldsymbol{\sigma}_t$ can be expressed as a function of the solid effective stress $\boldsymbol{\sigma}^\prime$ and fluid pressure $\boldsymbol{\sigma}_f$:
 \begin{align}
    \label{poro_sigma}
    \boldsymbol{\sigma}_t &= \boldsymbol{\sigma}^\prime + \boldsymbol{\sigma}_f \notag
    \\
    &=\boldsymbol{C} : \boldsymbol{\varepsilon} - \alpha p \boldsymbol{\mathit{I}}
\end{align}

\noindent
where
$\boldsymbol{C}$ is the elasticity tensor defined as a function of Lamé constants ($\lambda$ and $\mu$), as $ \hat{C}_{ijkl} = \lambda \delta_{ij} \delta_{kl} + \mu \left(\delta_{ik} \delta_{jl} + \delta_{il} \delta_{jk}\right)$, for isotropic linear materials. $\boldsymbol{\varepsilon}$ is the strain defined as a function of displacement ($\boldsymbol{u}$) as $\boldsymbol{\varepsilon} = \boldsymbol{\nabla}^s \boldsymbol{u} =\frac{1}{2}(\boldsymbol{\nabla} \boldsymbol{u}+\boldsymbol{\nabla} \boldsymbol{u}^T)$ for small deformations. Substituting Eq.~\eqref{poro_sigma} into Eq.~\eqref{eq:poro_st_blm}, and incorporating definitions in Eq.~\eqref{eq:fluid_velocity} and Eq.~\eqref{eq:fluid_increment} into Eq.~\eqref{eq:poro_st_cont}, the time-discretized linear momentum equilibrium equation and the fluid flow continuity equation can be written as:

\begin{align}
    \label{eq:poro_st_blm_3}
	\boldsymbol{\nabla} \cdot (\boldsymbol{C} : \boldsymbol{\varepsilon}_{n+1} - \alpha p_{n+1} \boldsymbol{\mathit{I}})+\boldsymbol{b}_{n+1} &= \boldsymbol{0}
    \\
	\label{eq:poro_st_cont_3}
	\dfrac{p_{n+1}-p_{n}}{ \Delta t M} + \alpha \text{tr}\left(\dfrac{\boldsymbol{\varepsilon}_{n+1}-\boldsymbol{\varepsilon}_{n}}{\Delta t} \right)  - \boldsymbol{\nabla} \cdot \left( \dfrac{ K_p}{\mu_f} \cdot (\boldsymbol{\nabla} p_{n+1} + \gamma_f \boldsymbol{i}_g) \right) &= Q_f
\end{align}
\noindent
where the subscripts $n$ and $n+1$ represent two consecutive time steps with a time increment of $\Delta t$. This time discretization follows the implicit Euler scheme, where $\dot{\boldsymbol{\varepsilon}}=({\boldsymbol{\varepsilon}_{n+1}-\boldsymbol{\varepsilon}_{n}})/{\Delta t}$, and $\dot{p}=({p_{n+1}-p_{n}})/{\Delta t}$. 

\subsection{Weak form and the FEM solvers}
\label{section:poroelasticity:weak}

Given the final strong forms of the linear momentum equilibrium equation Eq.~\eqref{eq:poro_st_blm_3} and the fluid flow continuity equation Eq.~\eqref{eq:poro_st_cont_3}, the corresponding time-discretized weak forms can be written as follows, respectively:
\begin{align} 
    \label{eq:poro_w_blm_time}
    \mathcal{D}_{n+1} &= \int_{\Omega} \boldsymbol{\nabla}^{s} \boldsymbol{\widehat{w}} : \boldsymbol{\hat{C}} : \boldsymbol{\varepsilon}_{n+1} \, d\Omega  \;- \int_{\Omega} \boldsymbol{\nabla}^{s} \boldsymbol{\widehat{w}} : (\alpha p_{n+1} \boldsymbol{\mathit{I}}) \, d\Omega \nonumber 
    \\
    &- \int_{\Gamma_t} \boldsymbol{\widehat{w}} \cdot \boldsymbol{\overline{t}}_{n+1} \, d\Gamma 
    \;- \int_{\Omega} \boldsymbol{\widehat{w}} \cdot \boldsymbol{b}_{n+1} \, d\Omega = \boldsymbol{0},
    \quad \forall \, \boldsymbol{\widehat{w}} \in \mathcal{W}_u
\end{align}

\begin{align}
    \label{eq:poro_w_cont_time}
    \mathcal{H}_{n+1} &=\int_{\Omega} \widehat{p}\; \dfrac{p_{n+1}-p_{n}}{ \Delta t M} \, d\Omega 
    \;+ \int_{\Omega} \widehat{p} \alpha \text{tr}\left(\dfrac{\boldsymbol{\varepsilon}_{n+1}-\boldsymbol{\varepsilon}_{n}}{\Delta t} \right)  \, d\Omega
    \;+ \int_{\Omega} \boldsymbol{\nabla} \widehat{p} \cdot \left(\dfrac{ K_p \boldsymbol{\mathit{I}}}{\mu_f} \cdot \boldsymbol{\nabla} p_{n+1} \right) \, d\Omega 
     \nonumber 
    \\
    &+\int_{\Omega} \boldsymbol{\nabla} \widehat{p} \cdot \left(\dfrac{ K_p \boldsymbol{\mathit{I}}}{\mu_f} \cdot \gamma_f \boldsymbol{i}_g \right) \, d\Omega + \int_{\Gamma_q} \widehat{p} \;\overline{q}_{n+1} \, d\Gamma \;- \int_{\Omega} \widehat{p} \; {Q_f}_{n+1} \, d\Omega = 0, 
    \quad \forall \, \widehat{p} \in \mathcal{W}_p
\end{align}

\noindent
where $\widehat{\boldsymbol{w}}$ and $\widehat{p}$ are the displacement and pressure test functions, $\mathcal{W}_u$ and $\mathcal{W}_p$ are the displacement and pressure function spaces, respectively.

\begin{figure}[hbt!]
\centering
\includegraphics[width=.5\textwidth]{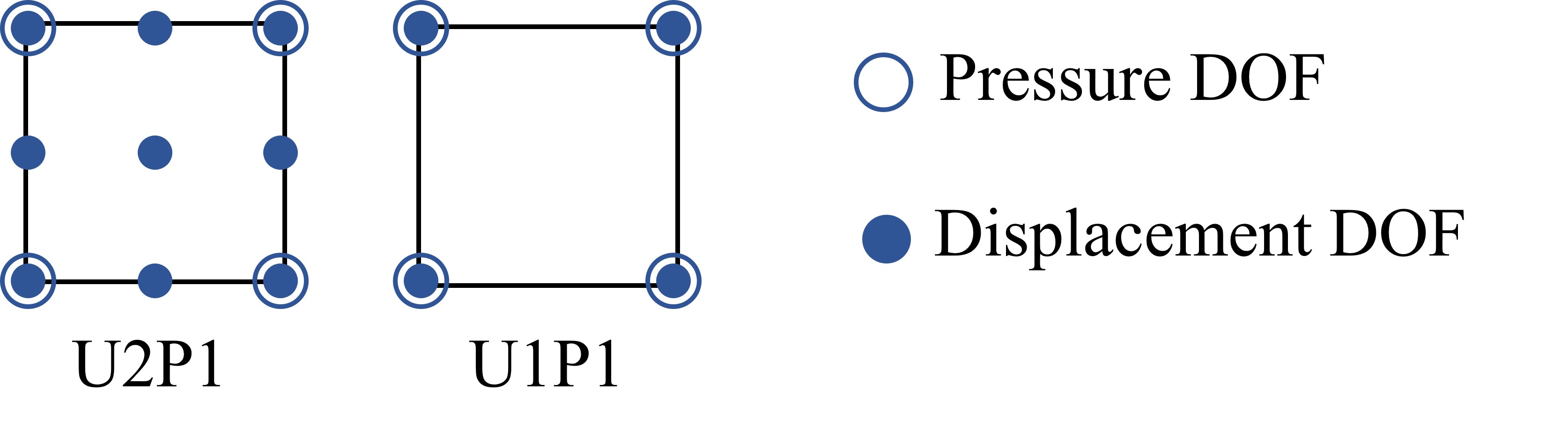}
\caption{Schematic of mixed quadrilateral elements, biquadratic-displacement/bilinear-pressure (U2P1) and bilinear-displacement/bilinear-pressure (U1P1) }
\label{fig:methods_u2p1_u1p1}
\end{figure}

Poroelasticity simulations may encounter pressure oscillations~\cite{white_stabilized_2008}, which can be mitigated by using a computationally expensive biquadratic-displacement/bilinear-pressure quadrilateral (see the U2P1 element in Fig.~\ref{fig:methods_u2p1_u1p1}). Another cost-efficient approach is to use a bilinear-displacement/bilinear-pressure quadrilateral (see the U1P1 element in Fig.~\ref{fig:methods_u2p1_u1p1}) with a stabilized weak form, introduced by White and Borja~\cite{white_stabilized_2008}, as follows:
\begin{align}
    \label{eq:poro_w_cont_time_stab}
    \mathcal{H}_{n+1} &+ \mathcal{H}_{n+1}^{stab} =0
    \\
    \label{eq:h_stab_time}
    \mathcal{H}_{n+1}^{stab} &=\int_{\Omega } \frac{\tau }{2\mu\Delta t}\left(\widehat{p}-\Pi \widehat{p}\right)\left(\left(p_{n+1}-p_{n}\right)-\Pi \left(p_{n+1}-p_{n}\right)\right)\ d\Omega
\end{align}
\noindent
where $\mathcal{H}_{n+1}^{stab}$ is the time-discretized stabilization term added to the fluid flow continuity equation, and $\tau$ is a constant multiplier to tune the stabilization level. The projection operator $\Pi$ computes the element-wise average of $\mathrm{\rho }^{h}$ over the element domain ${\Omega }^{e}$, as $\Pi {\mathrm{\rho }}^{h}|_{{\Omega }^{e}} =\frac{1}{V^{e}}\int_{{\Omega }^{e}} {\mathrm{\rho }}^{h}\,d{\Omega }^{e}$. 

For the poroelasticity problem described in Eq.~\eqref{eq:poro_w_blm_time} and Eq.~\eqref{eq:poro_w_cont_time}, a fully coupled FEM monolithic solver is built using U2P1 elements and utilizing the deal.II\cite{africa_dealii_2024} FEM library. This solver is utilized to generate data to serve as ground-truth reference values for training and error evaluation during testing. Two staggered solvers adopting U2P1 and U1P1 are developed for efficiency studies in Section~\ref{section:numerical_examples:inj_flux_2d:efficiency}. Notably, no spurious pressure oscillations were detected in all the numerical examples in this paper.

\FloatBarrier
\section{Methodology}  
\label{section:methodology}

\subsection{IFENN framework for poroelasticity problems}
\label{section:methodology_ifenn}
\begin{figure}[hbt!]
\centering
\includegraphics[width=1.0\textwidth]{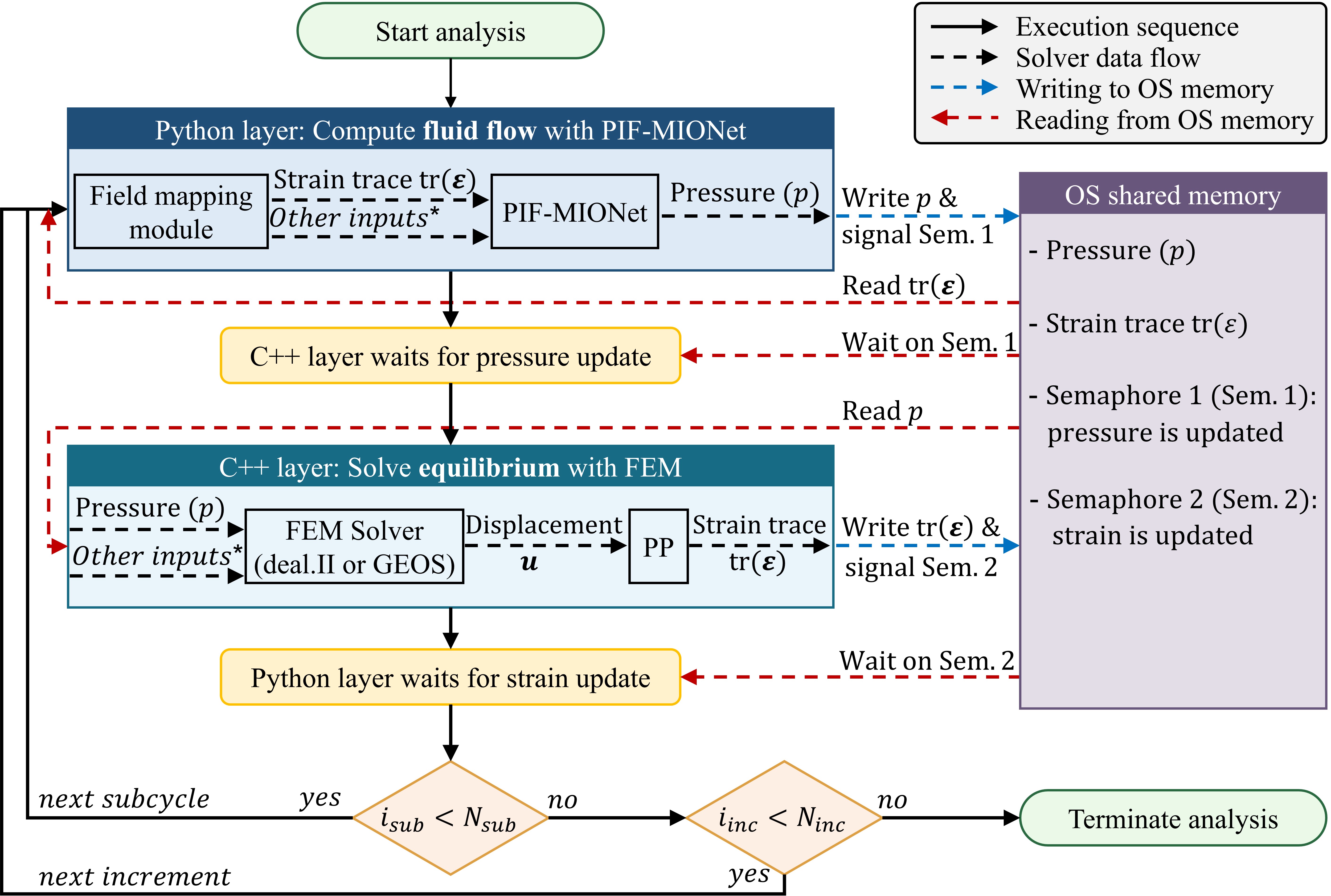}
\caption{IFENN implementation for poroelasticity problems. The Python layer and C++ layer exchange pressure $p$ and strain trace tr$(\boldsymbol{\varepsilon})$ through the operating system (OS) shared memory. Strain is computed through post-processing (PP) of the displacement. Synchronization between the two layers is achieved using semaphores that signal the availability of updated pressure and strain data. The index $i_{sub}$ is the counter for subcycles of a total count of $N_{sub}$, and $i_{inc}$ is the counter for time increments of a total count of $N_{inc}$. \textsuperscript{*}\textit{Other inputs} include the remaining information needed by the solvers, such as material properties, time increments, and loads. The field mapping module is conditionally employed based on the input discretization.}
\label{fig:methods_ifenn_implementation}
\end{figure}

The IFENN framework has been introduced and discussed for many applications in the literature~\cite{pantidis_integrated_2023, pantidis_i-fenn_2024, abueidda_i-fenn_2024, abueidda_variational_2024, pantidis_integrated_2026, pantidis_hybrid_2026}, including poroelasticity~\cite{amin_i-fenn_2026}. In brief, IFENN is conceptually similar to a staggered hybrid scheme, where the mechanical field is solved using a conventional FEM solver and the other coupled field is predicted using a pre-trained NN. Fig.~\ref{fig:methods_ifenn_implementation} illustrates the IFENN framework implementation adopted for the poroelasticity problems in this study, which uses a pre-trained PIF-MIONet. At every time increment, the solvers go through a group of subcycles. Within each subcycle, the PIF-MIONet and the FEM solver are called consecutively to compute the unknown state variables. More details on the subcycling scheme are provided in Section~\ref{section:methodology_subcycling}.

As shown in Eq.~\eqref{eq:poro_st_blm_3} and Eq.~\eqref{eq:poro_st_cont_3}, the displacement $\boldsymbol{u}$ is coupled with pressure $p$ through the strain trace rate $\text{tr}(\dot{\boldsymbol{\varepsilon}})$. Hence, the predicted pressure is shared with the FEM solver, and the strain trace (as a post-processed variable of the computed displacement) is shared with the PIF-MIONet. The FEM code runs using a C++ script, and the PIF-MIONet runs using a Python script. Communication between the Python and C++ layers is conducted through operating-system (OS) shared memory, as shown in Fig.~\ref{fig:methods_ifenn_implementation}. Synchronization between layers is done using semaphores, which act as coordination mechanisms that can block one process until the other signals that it has completed its operations.

\subsection{FEM solvers integration within IFENN}
\label{section:methodology_ifenn_fem}
In this study, we showcase the applicability of IFENN with two different platforms acting at the FEM-level: a deal.II-based solver~\cite{africa_dealii_2024} and GEOS~\cite{settgast_geos_2024}, as shown in Fig.~\ref{fig:methods_ifenn_implementation}. Both solvers are HPC-deployable and are suitable for large-scale problems. The deal.II-based solver provides more flexibility for development and testing across a wide range of target applications. On the other hand, GEOS is a high-performance simulator specializing in subsurface flows and geomechanics. In addition, it provides a set of pre-built physics solvers, enabling easy implementation of complex industrial-scale problems. Demonstrating IFENN's applicability with the GEOS solver is a key development, as it advances IFENN from a purely research-oriented package toward deployment in production-scale geomechanics simulations. This advancement facilitates future expansions and integration of other ML models. In the numerical examples presented in Section~\ref{section:numerical_examples}, both solvers are used, depending on their suitability for the problem or for research convenience. 

\begin{figure}[hbt!]
\centering
\includegraphics[width=1.0\textwidth]{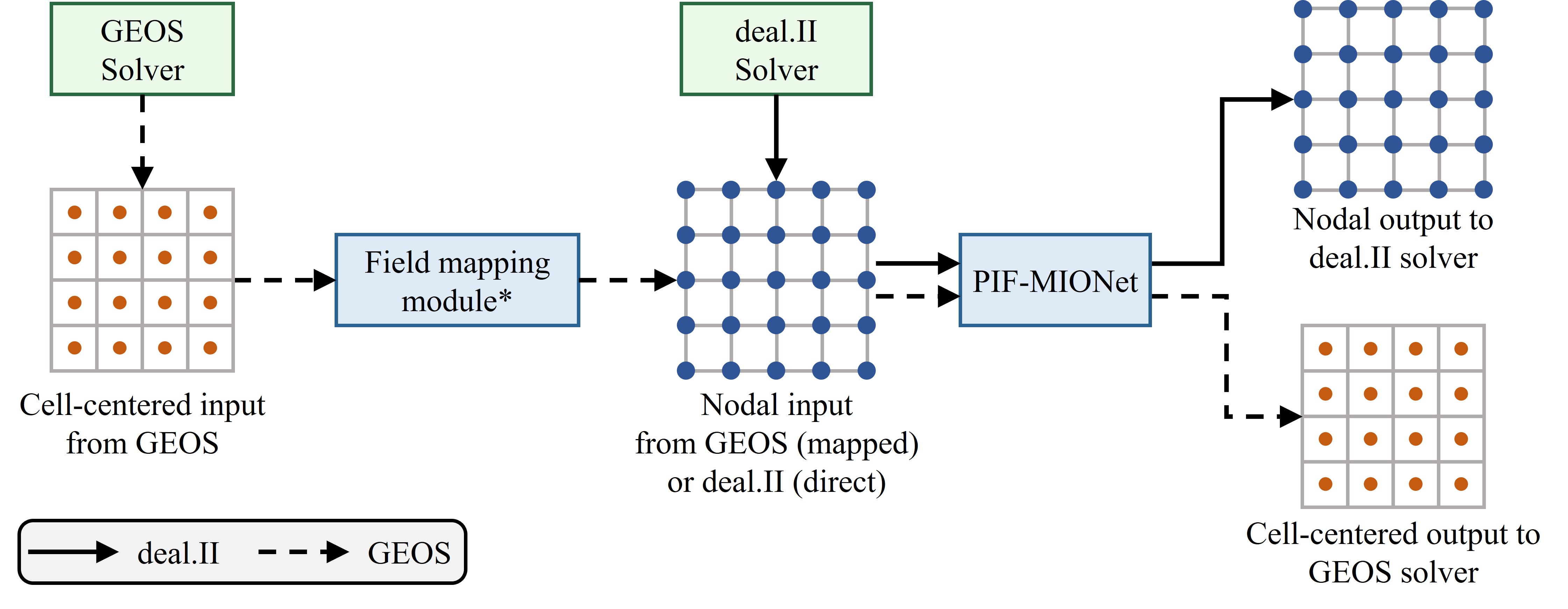}
\caption{A schematic illustration of the spatial discretization of input and output fields of the PIF-MIONet, within the Python layer of the IFENN framework. The FEM solvers deal.II and GEOS exchange data in nodal form and cell-centered form, respectively. \textsuperscript{*}The \textit{field mapping module} is required only for integration with GEOS, where input fields are transferred from cell centers to cell nodes. The PIF-MIONet can directly provide outputs at either cell nodes or cell centers.}
\label{fig:methods_ifenn_mapping}
\end{figure}

For deal.II integration, the pressure field is predicted by the PIF-MIONet at cell vertices, as shown in Fig.~\ref{fig:methods_ifenn_mapping}. The pressure values are written to the shared memory, after which a semaphore signals the FEM solver to read the nodal pressure values and interpolate them at the Gauss integration points using bilinear shape functions. The FEM solver then computes the displacement response and post-processes it to compute strain at cell nodes. The computed strain values are written to the shared memory, then the PIF-MIONet is signaled through a semaphore to read the shared strain and start a new subcycle. This approach benefits from the common vertices between adjacent cells, reducing the number of pressure evaluations that must be shared with the FEM solver. Alternatively, the pressure can be evaluated directly at the Gauss points through the PIF-MIONet operator. However, this will require sharing more data between the solvers, increasing the memory requirements for IFENN framework deployment.

The PIF-MIONet is trained using only nodal inputs and outputs. However, the existing GEOS mechanical solver is built to share and receive cell-centered values with the native FVM-based flow solver in GEOS. For IFENN integration, the FVM-based flow solver is replaced by a new custom-built solver, working as an interface that communicates with the Python layer to exchange pressure and strain-trace fields. Because PIF-MIONet and GEOS use different spatial discretizations, a field-mapping module is added between the two layers. This module is incorporated into the Python layer, where it maps the GEOS-computed strain trace to the nodal locations required by the PIF-MIONet, as shown in  Fig.~\ref{fig:methods_ifenn_mapping}. For interior nodes, the values are computed using interpolating neighboring cell-centered values, while for the boundary nodes, the values are assigned directly from the nearest cell center. For compatibility with GEOS, the pressure is evaluated directly at cell centers and shared with the GEOS solver. Notably, the PIF-MIONet is used to predict pressure at cell nodes and cell centers for the deal.II and GEOS integrations, respectively. This implementation therefore provides a practical test of IFENN portability and the flexibility of the PIF-MIONet.

\subsection{Predictor-Multi-Corrector subcycling scheme}
\label{section:methodology_subcycling}

Poroelasticity is an inherently coupled problem, and in this work, we adopt a predictor-multi-corrector scheme as detailed in Algorithm~\ref{alg:predictor_multi_corrector}. At each new increment, denoted as $n+1$, both $p_{n+1}$ and $\boldsymbol{\varepsilon}_{n+1}$ are unknowns, which creates a circular dependency between the two variables (i.e., $p_{n+1}$ and $\boldsymbol{\varepsilon}_{n+1}$ in Eq.~\eqref{eq:poro_st_blm_3} and Eq.~\eqref{eq:poro_st_cont_3}). Therefore, initial values for one of the variables should be guessed to break the loop, allowing the staggered scheme to proceed. In our implementation, we begin by setting a reasonable guess for the strain rate (predictor stage). Then, the PIF-MIONet is invoked first to compute pressure $p_{n+1}$, and the FEM solver follows to compute $\boldsymbol{\varepsilon}_{n+1}$. Further subcycles (corrector stage) are then performed to correct these predictions, for a number of subcycles. For each simulation, a fixed count of subcycles $N_{sub}$ is pre-defined as a termination criterion for the loops per each time increment. Alternatively, $N_{sub}$ can be identified automatically during the simulation given a suitable convergence criterion.

\begin{algorithm}[hbt!]
    \DontPrintSemicolon
    Start a new time increment (advancing from time step $n$ to $n+1$)\;
    Set the initial value for the subcycle index $r=0$\; 
    \While{$r < N_{sub}$}{
        \eIf{$r=0$}{
            Predictor subcycle: set the initial guess for strain rate.
            
            \eIf{$n=0$}{$\dot{\boldsymbol{\varepsilon}}=0$}{
            $\dot{\boldsymbol{\varepsilon}}=(\boldsymbol{\varepsilon}_{n}-\boldsymbol{\varepsilon}_{n-1})/{\Delta t}$\;
            }
        }{
            Corrector subcycle: set strain rate $\dot{\boldsymbol{\varepsilon}}=(\boldsymbol{\varepsilon}_{n+1}^{r-1}-\boldsymbol{\varepsilon}_{n})/{\Delta t}$\;
        }
        Predict pressure $p_{n+1}^r$, using the PIF-MIONet (trained to adhere to the strong form in Eq.~\eqref{eq:poro_st_cont_3})\;
        Solve Eq.~\eqref{eq:poro_w_blm_time} for displacement, using an FEM solver, then compute $\boldsymbol{\varepsilon}_{n+1}^r$\; 
        $r = r + 1$\;
    }
    \caption{Predictor-multi-corrector subcycling scheme for a total of $N_{sub}$ subcycles}
    \label{alg:predictor_multi_corrector}
\end{algorithm}

\FloatBarrier
\section{Design of the Physics-Informed Fusion-MIONet (PIF-MIONet)}
\label{section:pif_mionet_design}
The PIF-MIONet, with a set of parameters $\phi$, is trained with the objective of minimizing a loss function $\mathcal{L}(\phi)$, as follows:
\begin{align}
    \phi^* = \arg \min_{\phi} \mathcal{L}(\phi)
\end{align}
\noindent
where $\phi^*$ denotes the set of optimized model parameters that minimize a predefined loss function $\mathcal{L}(\phi)$. In this section, we provide in detail the adopted loss function and the PIF-MIONet architecture. 

\subsection{The loss function}
\label{section:pif_mionet_design:loss} 
The loss function can be written in a generic form as:
\begin{align}
    \label{eq:loss_function_generic}
     \mathcal{L}(\phi) = \sum_{i=1}^{N_{terms}}  \lambda_{i}  \mathcal{L}_i
\end{align}
\noindent
where $N_{terms}$ is the total number of considered loss terms, $\lambda_i$ is an adaptive weight given to each loss term ($\mathcal{L}_i$) to balance the contribution of individual loss terms and dynamically guide the optimizer's focus during the training process. The unweighted loss term $\mathcal{L}_i$ can be computed as:
\begin{align}
    \label{eq:loss_term_i_generic}
    \mathcal{L}_i = \mathcal{N}(S_i \boldsymbol{R}_i),\qquad i=1,\dots,N_{terms}.
\end{align}
\noindent
where
$\boldsymbol{R}_i$ is the loss term residual vector computed at a predefined set of evaluation/collocation points,  
$S_i$ is a scalar scaling coefficient utilized for normalizing the magnitude of the residuals, 
and $\mathcal{N}(\cdot)$ is the norm computation function, such as the second norm or the mean squared value.  
A detailed version of the loss function, in Eq.~\eqref{eq:loss_function_generic}, can be defined by replacing the subscript $i$ with $P$, $D$, $N$, and $d$ for PDE, Dirichlet boundary conditions, Neumann boundary conditions, and data loss terms, respectively, as: 
\begin{align}
    \label{eq:loss_function_detailed}
     \mathcal{L}(\phi) = \lambda_{P} \mathcal{N}(S_P \boldsymbol{R}_P)+ \sum_{j=1}^{N_{\Gamma_p}}  \lambda_{D}^j  \mathcal{N}(S_D \boldsymbol{R}_D^j)+ \sum_{j=1}^{N_{\Gamma_q}} \lambda_{N}^j \mathcal{N}(S_N \boldsymbol{R}_N^j) + \lambda_{d} \mathcal{N}(S_d \boldsymbol{R}_d)
\end{align}
\noindent
where $N_{\Gamma_p}$ and $N_{\Gamma_q}$ are the number of Dirichlet and Neumann boundary conditions, respectively. 

Notably, in this research, all of the examples are purely physics-informed, with no data contribution ($\lambda_{d}=0$), except for one example in which a data term is incorporated to regularize and enhance the training. We test two adaptive weighting algorithms, GradNorm~\cite{chen_gradnorm_2017} and ReLoBRaLo~\cite{bischof_multi-objective_2025}, for training the PIF-MIONet, with implementation details provided in Appendix~\ref{section:appendix:pif_mionet_training:adaptive}. More details on proposed approaches for computing the residuals and the scaling coefficients are provided in Section~\ref{section:pif_mionet_design:residuals} and Section~\ref{section:pif_mionet_design:dim}, respectively.

\subsection{Residuals for PDE and boundary conditions enforcement}
\label{section:pif_mionet_design:residuals}
The different residual terms are defined as:
\begin{align}
    \label{eq:residual_pde}
	\boldsymbol{R}_P &= \dfrac{\widetilde{p}_{n+1}-p_{n}}{ \Delta t M} + \alpha \text{tr}\left(\dfrac{\boldsymbol{\varepsilon}_{n+1}-\boldsymbol{\varepsilon}_{n}}{\Delta t} \right)  - \boldsymbol{\nabla} \cdot \left( \dfrac{ K_p}{\mu_f} \cdot (\boldsymbol{\nabla} \widetilde{p}_{n+1} + \gamma_f \boldsymbol{i}_g) \right) - Q_f, \quad & \boldsymbol{x} &\in \Omega   
    \\
    \label{eq:residual_dirich}
	\boldsymbol{R}_D^j &= \widetilde{p}_{n+1} - \overline{p}_{n+1}, &\boldsymbol{x} &\in \Gamma_{p}^j
    \\
    \label{eq:residual_neumann}
	\boldsymbol{R}_N^j &= \bigl(-  \dfrac{K_p}{\mu_f}\boldsymbol{\nabla} ({\widetilde{p}_{n+1}} + \gamma_f \boldsymbol{x} \cdot \boldsymbol{i}_g) \bigr) \cdot \boldsymbol{n} - \overline{q}_{n+1}, &\boldsymbol{x} &\in \Gamma_{q}^j 
    \\
    \label{eq:residual_data}
	\boldsymbol{R}_d &= \widetilde{p}_{n+1} -  \check{p}_{n+1}, &\boldsymbol{x} &\in \Omega
    \end{align}
\noindent
where $\boldsymbol{R}_P$ is the PDE residual vector computed over the domain collocation points ($\boldsymbol{x}$), $\widetilde{p}_{n+1}$ is the model prediction for the coordinates of interest. 
$\boldsymbol{R}_D^j$ and $\boldsymbol{R}_N^j$ are the residuals for the soft penalty terms for Dirichlet and Neumann boundary conditions over boundaries $\Gamma_p^j$ and $\Gamma_q^j$, respectively. $\boldsymbol{R}_d$ is the labeled-data error, where reference data from predefined nodes are incorporated to improve model accuracy. $\check{p}_{n+1}$ is the true reference response computed using a coupled FEM solver. In this approach, $\boldsymbol{R}_P$ is formulated using the strong form of the continuity equation Eq.~\eqref{eq:poro_st_cont_3}. Alternative approaches in the literature are based on variational and energy methods, or on a combination of strong and weak forms~\cite{abueidda_variational_2024, harandi_mixed_2024}.

\subsection{Dimensionless scaling}
\label{section:pif_mionet_design:dim}
Dimensionless forms have been adopted in many ML applications, including fluid flow formulas~\cite{haghighat_physics-informed_2022, amini_physics-informed_2022, amini_inverse_2023}, for two main purposes: (1) normalizing the model inputs and outputs, (2) scaling the residual equation to control the error magnitudes. If the normalization process is done before training without being tracked by the training software, new formulations of the residual forms should be introduced to account for the normalization effects. In this paper, the training software tracks all normalization operations and accounts for their impact on gradient computations. Therefore, no further modifications are needed to the residual forms, presented in Section~\ref{section:pif_mionet_design:residuals}. Hence, we focus only on the second part of the process: scaling the residuals. 

To this end, and following the processes discussed in the literature~\cite{haghighat_physics-informed_2022, amini_physics-informed_2022, amini_inverse_2023}, we identify the scaling coefficients as:
\begin{align}
    \label{eq:scale_terms}
	S_P &= \dfrac{\mu_f^* {x^*}^2}{K_p^*p^*},
    \nonumber
    \\
	S_D &= S_d = \dfrac{1}{p^*},
    \nonumber
    \\
	S_N &= \dfrac{\mu_f^* x^*}{K_p^*p^*}
    \end{align}
\noindent
where $S_P$, $S_D$, $S_d$, and $S_N$ are the scaling factors for the residuals $\boldsymbol{R}_P$, $\boldsymbol{R}_D$, $\boldsymbol{R}_d$ and $\boldsymbol{R}_N$, respectively. $\mu_f^*$, $x^*$, $K_p^*$, and $p^*$ are the normalization coefficients for the fluid viscosity, domain coordinates, permeability, and pressure, respectively. These coefficients roughly normalize their corresponding variables to unity. For example, $x^*$ is set to the largest coordinate in the domain and $p^*$ is set to the largest pressure magnitude among the training increments.

\subsection{Fusion-MIONet architecture}
\label{section:pif_mionet_design:gen_arch}
\begin{figure}[hbt!]
\centering
\includegraphics[width=1.0\textwidth]{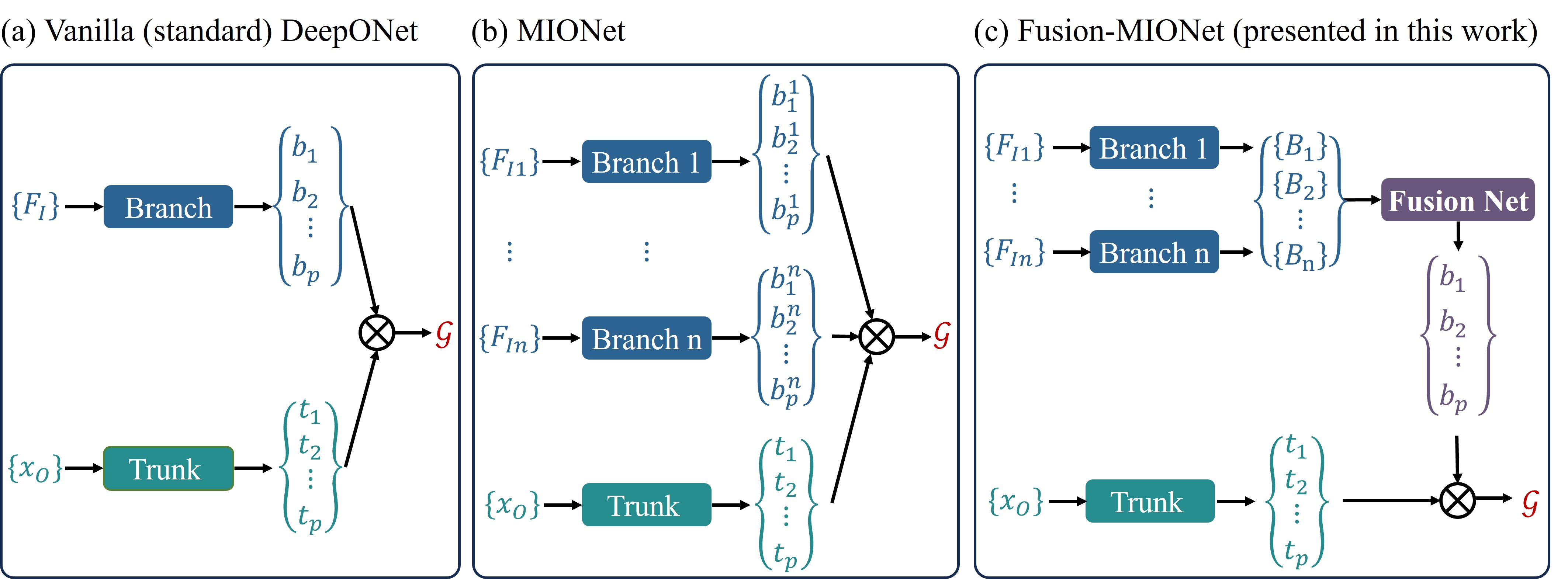}
\caption{Different variations of DeepONet: (a) Vanilla DeepONet, (b) MIONet, (c) our proposed architecture (Fusion-MIONet)}
\label{fig:deeponet_diff_arch}
\end{figure}

We start with a brief explanation of the combined constituents of the proposed architecture. A DeepONet in its simplest form (see Fig.~\ref{fig:deeponet_diff_arch}a) consists of two networks: (1) a branch network that encodes the input function signals $F_I$, and (2) a trunk network that encodes the targeted locations $x_O$ where the output function is evaluated~\cite{lu_comprehensive_2022, lu_learning_2021}. The DeepONet maps the discretized input function $(F_I)$ to the output function $(F_O)$ through an operator $\mathcal{G}$, utilizing a dot product process as follows:
\begin{align}
    \label{eq:deeponet_output}
    \mathcal{G}(F_I)(x_O) = \sum_{k=1}^{O_{to}} b_k(F_I) t_k(x_O) + b_0
\end{align}
\noindent
where $b(F_I)$ and $t(x_O)$ are the branch and trunk output vectors, respectively, of size $O_{to}$, and $b_0  \in \mathbb{R}$ is a bias. Notably, in this setup, $F_I$ is defined as a vector of discretized values at fixed locations over the domain. The output query coordinates $(x_O)$ can be arbitrarily selected, providing strong generalization when using a trained DeepONet. 

An extension of the DeepONet setup is the multiple-input deep neural operator (MIONet)~\cite{jin_mionet_2022}, which uses multiple branches to encode multiple input functions, as depicted in Fig.~\ref{fig:deeponet_diff_arch}b. Handling multiple inputs in PI-ML is particularly challenging, especially when each input has its own magnitude range, dimensionality, and spatio-temporal distribution. Using a MIONet can offer a viable solution; however, its original design limits early data exchange/interaction between input sources. This limitation can be addressed through \textit{fusion}, in which data from multiple sources are integrated into a single representation\cite{zhang_multimodal_2020, kiziltepe_integration_2024}.
Recent work has incorporated fusion concepts into DeepONet-based numerical simulations~\cite{liu_image_2025, peyvan_fusion-deeponet_2026}, using different techniques to fuse data between the branch and trunk networks. 

In this study, we propose a distinct fusion approach within the MIONet setup to enable early interaction between the input fields, as shown in Fig.~\ref{fig:deeponet_diff_arch}c. First, each branch encodes its input function $F_{Ii}$ into an output vector $B_i$. Then, all branch output vectors are concatenated into one vector $U_I=\{B_1; B_2; ..., B_n\}$ for $n$ input functions. Finally, $U_I$ is passed, as an input, to the fusion network, which encodes all the branches' information into a single vector $($similar to $b(F_I)$ in the vanilla DeepONet$)$. Therefore, an additional advantage of the Fusion-MIONet is that the final output can be aggregated using the standard DeepONet approach provided in Eq.~\eqref{eq:deeponet_output}. 

Multilayer perceptrons (MLPs) are utilized for all branch, fusion, and trunk networks. For an MLP layer with input vector $\mathbf{z}_i \in \mathbb{R}^{1 \times N_{i}}$, the output vector $\mathbf{z}_{i+1} \in \mathbb{R}^{1 \times N_{i+1}}$ can be computed as~\cite{xiang_reachable_2021} $\mathbf{z}_{i+1} = \sigma( \mathbf{z}_i \mathbf{W}_{i}  + \mathbf{b}_i)$, where $\sigma$ is an activation function, $\mathbf{W}_{i} \in \mathbb{R}^{N_{i} \times  N_{i+1}}$ is the weight parameter, $\mathbf{b}_i \in \mathbb{R}^{1 \times N_{i+1}}$ is the bias vector. Notably, a recent study by Choi et al.~\cite{choi_operator_2026} has investigated the efficacy of an architecture similar to the Fusion-MIONet in a data-driven approach, exclusively for consolidation problems.
   
\begin{figure}[hbt!]
\centering
\includegraphics[width=0.9\textwidth]{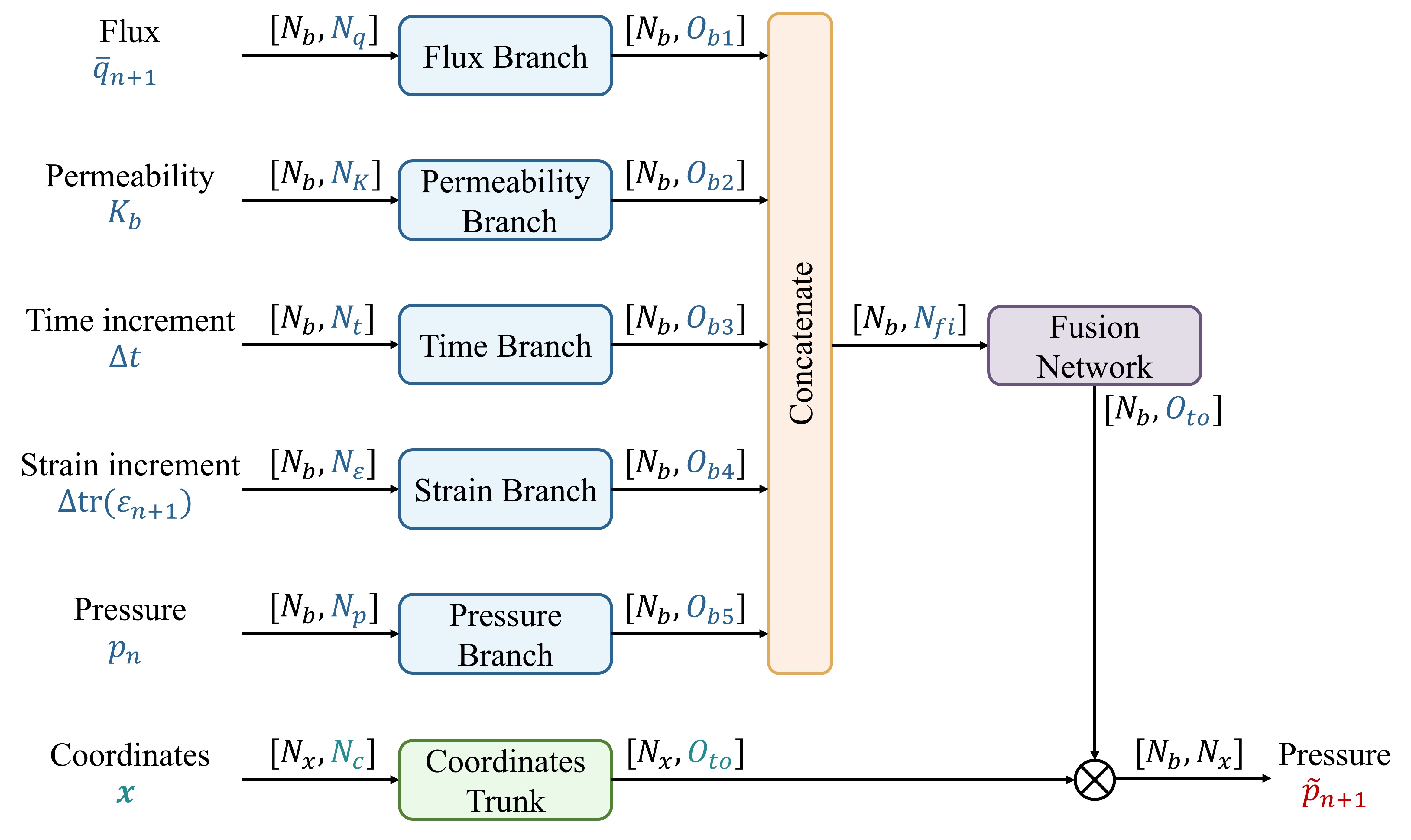}
\caption{Generalized PIF-MIONet considering variable permeability $K_p$ and time increment $\Delta t$. $N_b$ is the batch size and $N_x$ is the number of output points. The fusion network input size is equal to $N_{fi} = O_{b1} + O_{b2} + O_{b3}+ O_{b4}+ O_{b5}$. The trunk and the fusion networks have the same output size $O_{to}$.}
\label{fig:methods_pifmionet_details}
\end{figure}

We introduce the Physics-Informed Fusion-MIONet (PIF-MIONet) as a fluid mass-balance solver that accounts for the parameters governing the physical and numerical behavior of the underlying PDE. Fig.~\ref{fig:methods_pifmionet_details} illustrates our proposed PIF-MIONet in its generalized form. It focuses on the discretized continuity equation (Eq.~\eqref{eq:poro_st_cont_3}), where the output is the pressure $(p_{n+1})$ and the inputs are the flux $(\overline{q}_{n+1})$, the permeability $(K_p)$, the time increment $(\Delta t)$, the change in strain trace $\text{tr}(\Delta \boldsymbol{\varepsilon}_{n+1})=\text{tr}(\boldsymbol{\varepsilon}_{n+1}-\boldsymbol{\varepsilon}_{n})$, the pressure at previous time step $(p_{n})$, and the output query coordinates $(\boldsymbol{x})$. 

As shown in Fig.~\ref{fig:methods_pifmionet_details}, the trunk input has two dimensions, $[N_x, N_c]$, where $N_x$ is the number of output query points, and $N_c$ is the number of spatial coordinate components ($N_c=2$ or $3$). The trunk input is encoded into an output tensor of size $[N_x, O_{to}]$. All branch inputs have two dimensions: the first is $N_b$, the batch size of the sampled time steps, and the second is the number of spatially discretized input points. For example, $N_p$ is the number of spatially discretized input points of the pressure field. The branches' outputs are concatenated into one tensor of size $[N_b, N_{fi}]$, which is encoded through the MLP fusion network to end with a tensor of size $[N_b, O_{to}]$, with the last dimension equal to that of the trunk output (i.e., $O_{to}$). A common last dimension is crucial for the dot product operation in Eq.~\eqref{eq:deeponet_output}, which produces the final output pressure tensor of size $[N_b, N_x]$. 

\begin{figure}[hbt!]
\centering
\includegraphics[width=0.9\textwidth]{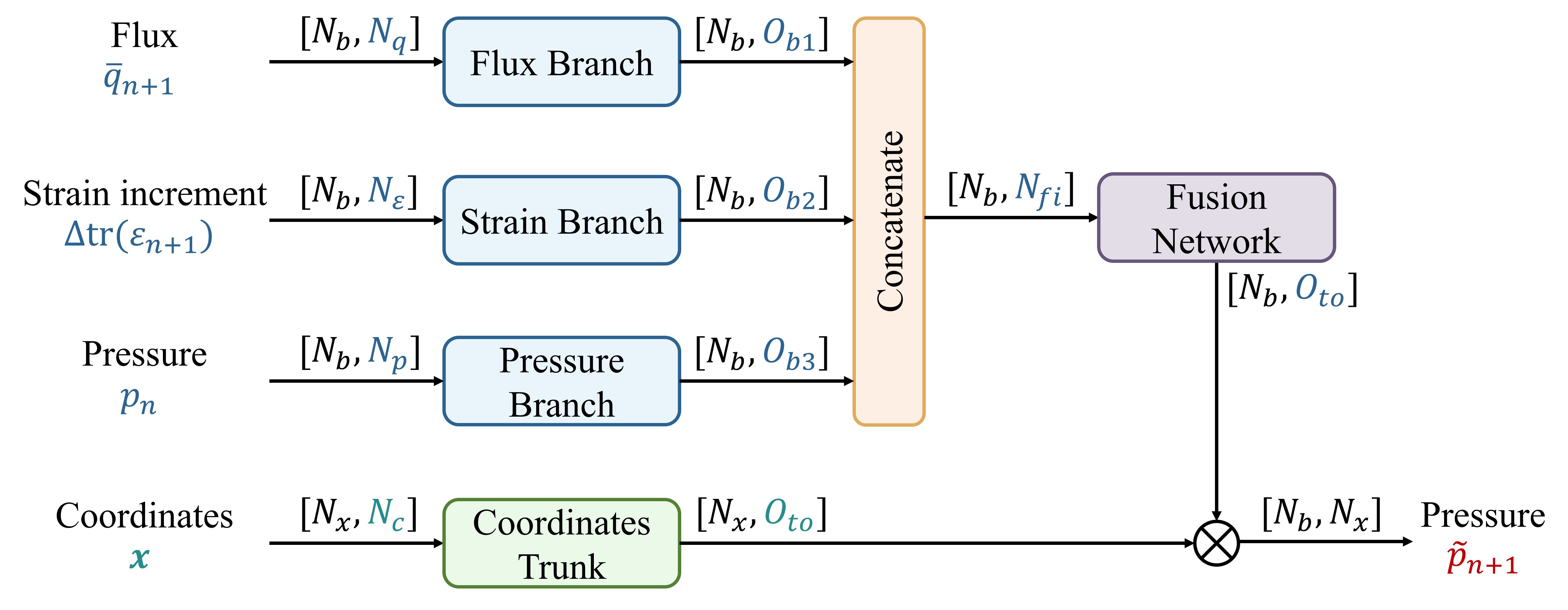}
\caption{Simplified PIF-MIONet considering constant permeability $K_p$ and time increment $\Delta t$. $N_b$ is the batch size and $N_x$ is the number of output points. The fusion network input size is equal to $N_{fi} = O_{b1} + O_{b2} + O_{b3}$. The trunk and the fusion networks have the same output size $O_{to}$.}
\label{fig:methods_pifmionet_simplified}
\end{figure}

Quantities that remain constant across all increments in this study, such as the dynamic fluid viscosity $\mu_f$, can be omitted from the network inputs and incorporated directly into the PDE residual, in Eq.~\eqref{eq:residual_pde}. Quantities that vary between increments but are spatially uniform, such as flux, permeability, and time-step size, require branch inputs of size $1$, giving $N_q=N_K=N_t=1$. Each scalar is broadcast to all collocation points during residual evaluation. The aforementioned architectural design, shown in Fig.~\ref{fig:methods_pifmionet_details}, is tailored for generalized examples introduced in Section~\ref{section:numerical_examples}. For other examples where the permeability $K_p$ and time increment $\Delta t$ are constant, we introduce a simplified PIF-MIONet, as shown in Fig.~\ref{fig:methods_pifmionet_simplified}. 

Regarding spatially varying input functions, such as strain or pressure, the input functions are represented by values sampled at nodes called sensors~\cite{lu_learning_2021}. Each branch encodes a single input function and may use its own sampling or projection method~\cite{jin_mionet_2022}. The number and distribution of input nodes may therefore differ between input functions. Evaluating the physics-informed PDE residual requires all field values at the collocation points. When these field values are obtained directly from the branch input functions, the node set for each spatially varying function must include all collocation points. Under this requirement, the collocation points constitute the minimum input set, with additional input points permitted for finer spatial resolution.

In this work, we add group normalization to the branch architecture. Our previous work~\cite{amin_i-fenn_2026} on DeepONets with IFENN has shown the efficacy of applying group normalization in branches where input fields suffer from error accumulation as simulations propagate through time increments. Further details on this approach are provided in Appendix~\ref{section:appendix:pif_mionet_training:stabilization}.

\subsection{Domain decomposition}
\label{section:pif_mionet_design:domain_decomposition}
\begin{figure}[hbt!]
\centering
\includegraphics[width=0.9\textwidth]{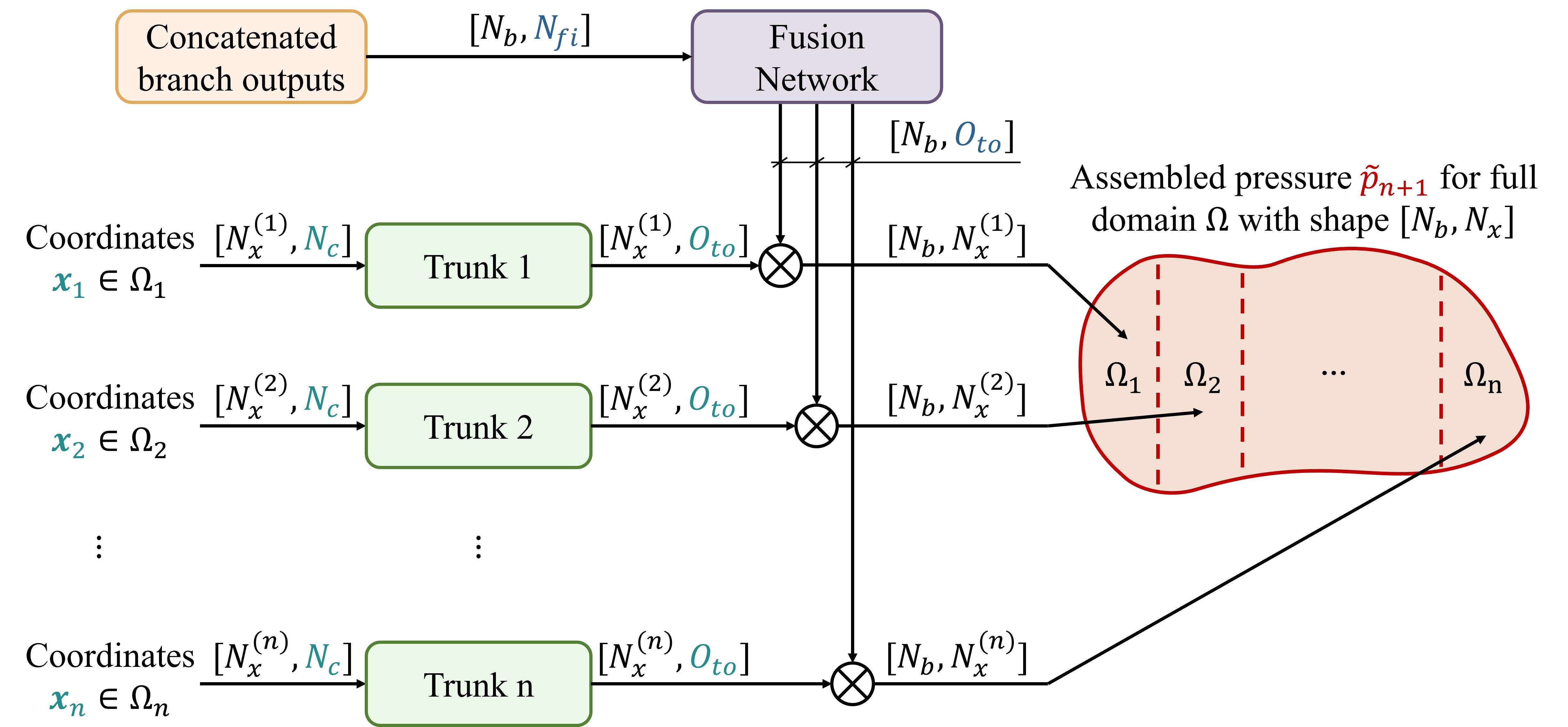}
\caption{PIF-MIONet with domain decomposition for efficient trunk utilization. The domain $\Omega$ is decomposed into different subdomains $\{ \Omega_1,\Omega_2,...,\Omega_n \}$, with each trunk responsible for encoding the coordinates of a specific subdomain. Outputs from each trunk are used to compute the pressure output of the corresponding subdomain.}
\label{fig:methods_pifmionet_domain_decomp}
\end{figure}

Several approaches use domain decomposition to make training more efficient~\cite{sharma_ensemble_2024, wu_physicsinformed_2024, yu_separable_2024}. We introduce domain decomposition into the PIF-MIONet architecture using multiple trunks, as shown in Fig.~\ref{fig:methods_pifmionet_domain_decomp}. 
The domain $\Omega$ is decomposed into a set of simpler subdomains $\{ \Omega_1,...,\Omega_n \}$.
For each subdomain $\Omega_k$, a dedicated trunk encodes the coordinates, which are then used to predict pressure within the subdomain. Inspired by continuity considerations for spatial domain decomposition~\cite{sharma_ensemble_2024}, we enforce continuity across common subdomain interfaces by averaging overlapping predictions. Each subdomain output is multiplied element-wise by a weighting tensor, then the weighted outputs are assembled to form the total output pressure field $p_{n+1}$. We compute the spatial derivatives required for the physics-informed loss after assembling the global pressure field, rather than independently for each subdomain. This ensures that the PDE residual is evaluated using the assembled pressure representation, including subdomain interfaces.

Weighting tensors enable efficient vectorized calculations and gradient computation for training. Each subdomain has a dedicated weighting tensor with weights of ones at the interior nodes, $1/2$ along interfaces, and $1/m_s$ at corners, where $m_s$ is the number of overlapping subdomains. This domain decomposition approach simplifies each trunk's task, allowing it to specialize in its assigned subdomain without requiring a single complex trunk for the whole domain. The efficacy of this approach for complex geometries is demonstrated in Section~\ref{section:numerical_examples:excav}.

\section{Training of the Physics-Informed Fusion-MIONet (PIF-MIONet)}
\label{section:pif_mionet_training}
All examples presented in this paper are purely physics-informed and exclude the data loss term (i.e., $\lambda_{d}=0$ in Eq.~\eqref{eq:loss_function_detailed}), except for one example, in Section~\ref{section:numerical_examples:excav}, that incorporates data loss to improve training accuracy and stability. Notably, even when the data loss term is omitted, we still utilize solutions from a fully coupled FEM solver to generate physically consistent network inputs, such as pressure and strain fields, which we collectively refer to as training data. In this section, we cover training data sampling, the adopted error metrics, and the computation of spatial derivatives.

\subsection{Physics-consistent input fields and sparse increment sampling}
\label{section:pif_mionet_training:sampling}
Training the PIF-MIONet requires physically consistent input fields for the pressure and strain branches, as illustrated in Fig.~\ref{fig:methods_pifmionet_details} and Fig.~\ref{fig:methods_pifmionet_simplified}. When training encompasses the complete loading history, the predicted fields at each loading increment can be used as inputs for the subsequent increment, enabling sequential propagation throughout the loading process. In this study, however, we adopt a computationally efficient sparse-sampling strategy, in which the training input fields at selected loading increments are obtained directly from fully coupled FEM simulations.

To this end, for a specific problem, we generate response data for multiple load histories using the FEM solver. The generated input/output histories are split into three groups: training, validation, and testing, where each load history is assigned exclusively to one of these groups. For each group, a few increments are sampled through a semi-random process. Specifically, a few increment indices are set, usually uniformly spaced over the time domain. Then, for each index, input/output data are sampled from multiple randomly selected load histories. 

In this sampling technique, we use disconnected, sparse increments for training rather than full continuous histories. This strategy reduces training cost and enables more efficient coverage of a wider range of simulations. In addition, this feature highlights the flexibility of the PIF-MIONet, which can be trained using sparse time increments instead of full histories. Finally, for the IFENN implementation, the framework is tested against complete histories from the testing group that the model optimizer has not seen.  

\subsection{Error metrics}
\label{section:pif_mionet_training:error}
During training, the loss metric $\mathcal{L}(\phi)$ is computed for both training and validation increments to monitor and detect overfitting. After concluding the training stage, different testing errors are computed as follows:
\begin{align}
        \label{eq:rel_error}
        \epsilon_{rel} &= {|y_{true} -y_{pred}|}/{(|y_{true}| + \epsilon_{tol})} 
    \\
        \label{eq:inc_error}
        L_2^{inc} &= {\left\| y_{true} - y_{pred} \right\|_2} / {\left\| y_{true} \right\|_2} 
    \\
        \label{eq:all_err}
        L_2^{all} &= {\left\| \{L_2^{inc=1},L_2^{inc=2},...,L_2^{inc=N_{inc}}\} \right\|_2}/{\sqrt{N_{inc}}}
\end{align}
\noindent
where $y_{true}$ is the fully coupled FEM response considered as the true reference value, $y_{pred}$ is the predicted response, $\epsilon_{rel}$ is the pointwise relative error for visual comparison at selected time steps, and $\epsilon_{tol} = 10^{-5} \max(|y_{true}|)$ is a tolerance for numerical stability. $L_2^{inc}$ measures the full-domain error at a single time increment, whereas $L_2^{all}$ represents the aggregated error across all sampled increments. 

\subsection{Spatial derivatives using FEM shape function (SF)}
\label{section:pif_mionet_design:sf_derivatives}
In PI-ML, automatic differentiation (AD) is commonly used to compute spatial derivatives required by the loss terms, providing a straightforward implementation, especially for mesh-free approaches~\cite{karniadakis_physics-informed_2021,rezaei_finite_2025}. However, AD can lead to unbounded errors~\cite{johnson_software-based_2025}, high computational cost~\cite{rezaei_finite_2025}, and instability~\cite{he_use_2023}, motivating the use of FEM shape function (SF) derivatives for more consistent, efficient, and accurate computations~\cite{he_use_2023, pantidis_i-fenn_2024}. SF-based derivatives require establishing a mesh for spatial discretization of the NN output; however, this introduces no additional meshing requirements to IFENN, because the framework already includes a discretized mesh for the FEM solver.

The residual forms in Eq.~\eqref{eq:residual_pde} and Eq.~\eqref{eq:residual_neumann} show that both first- and second-order spatial derivatives are required. To this end, we used a biquadratic quadrilateral element formulation to compute the derivatives. To the best of the authors' knowledge, this is the first implementation of 9-node element shape functions for computing spatial derivatives in operator learning within a PI-ML framework. We used the fact that our spatial discretizations are structured, with all prediction coordinates uniformly spaced. Each $3 \times 3$ set of nodes in the 2D space is grouped to form the response vector for a 9-node macro-element. The nodes are ordered lexicographically to enable efficient vectorized tensor product operations. More details on the adopted FEM formulas are provided in Appendix~\ref{section:appendix:pif_mionet_training:sf_derivatives}. Notably, for the 3D example in this paper, AD is the only tool used for computing spatial derivatives.

\FloatBarrier
\section{Numerical Examples}
\label{section:numerical_examples}
In this section, we present 2D and 3D examples that showcase the proposed framework's accuracy and generalizability. All the provided examples can be categorized under two main problems: (1) a poroelastic column with injection flux, tested in 2D and 3D domains, with varying material properties, (2) a 2D excavation problem with complex boundary conditions, subjected to a dewatering flux. The injection problem is purely physics-informed and excludes any data loss terms, while the dewatering problem incorporates a data loss term to address the domain's geometric complexities. 

For each example, a PIF-MIONet is trained as described in Section~\ref{section:pif_mionet_training} and integrated into IFENN for testing. Frameworks using deal.II and GEOS as mechanical field solvers are denoted IFENN-deal.II and IFENN-GEOS, respectively. Online IFENN execution requires only the trained model and target load history. Fully coupled FEM data ($y_{true}$) are used solely to monitor runtime errors and evaluate testing metrics (Eq.~\eqref{eq:rel_error}–Eq.~\eqref{eq:all_err}). Unless otherwise stated, the subcycle counts are $N_{sub}=2$ for IFENN-deal.II and $N_{sub}=1$ for IFENN-GEOS. All results use SI units, with displacement in m and pressure in N/m$^2$, unless otherwise specified. All simulations presented in the following sections are performed on an Ubuntu 24.04 workstation equipped with 128 GB of RAM, an Intel Xeon W3-2435 processor (8 physical cores), and an NVIDIA RTX A2000 GPU with 12 GB of dedicated memory. For consistent performance comparisons, both FEM and IFENN simulations are executed using 8 processes.

\subsection{2D injection example with constant \texorpdfstring{permeability $K_p$ and time increment $\Delta t$}{permeability and time increment}}
\label{section:numerical_examples:inj_flux_2d}

\subsubsection{Problem setup}
\label{section:numerical_examples:inj_flux_2d:setup}
\begin{figure}[hbt!]
\centering
\includegraphics[width=1.0\textwidth]{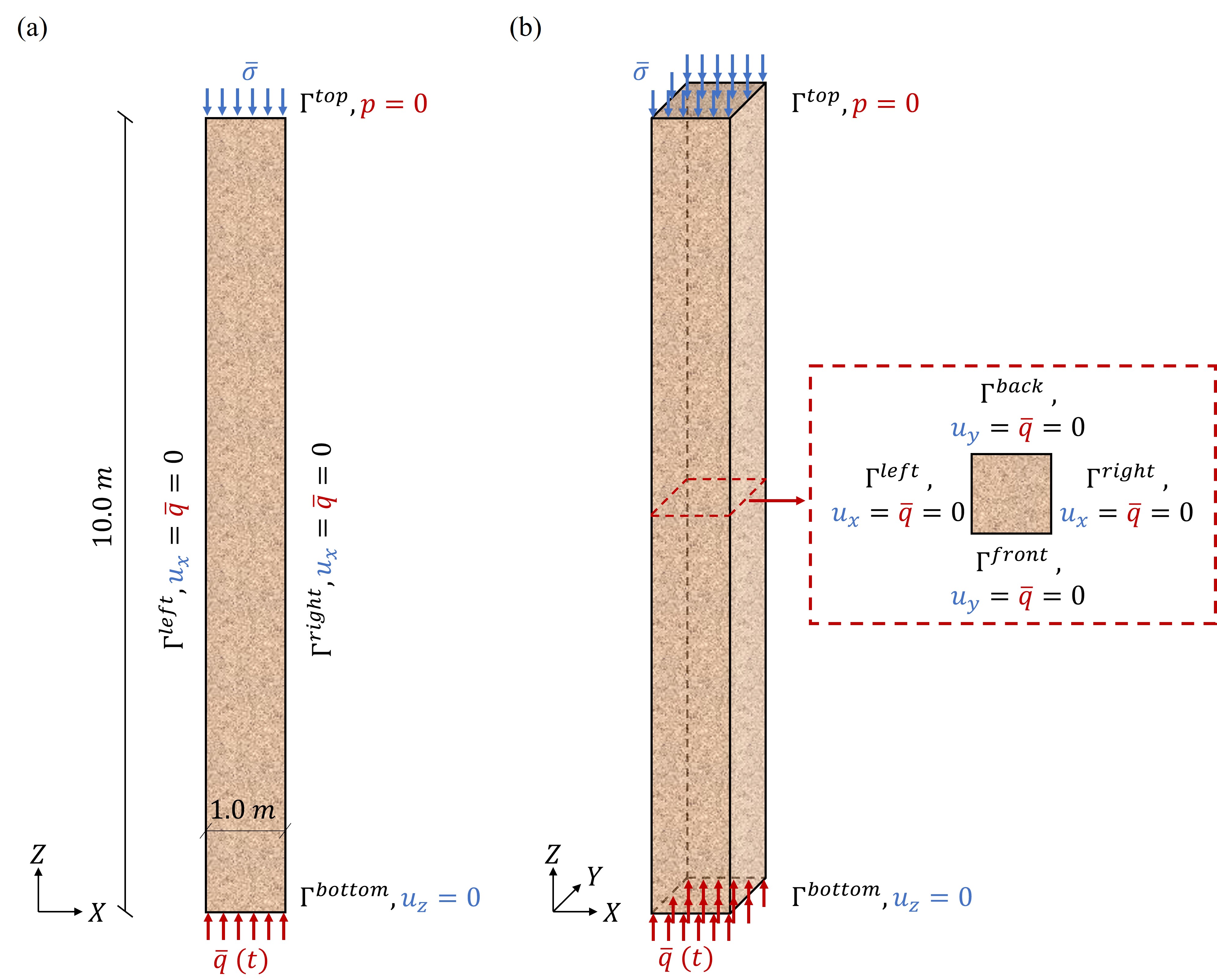}
\caption{Schematic representation of the geometry and boundary conditions
of the: (a) 2D injection problem, (b) 3D injection problem, with a square cross-section of width $1.0$ m.}
\label{fig:inj_setup}
\end{figure}

The first example is a 2D poroelastic column subjected to an injection flux. As shown in Fig.~\ref{fig:inj_setup}a, the top boundary ($\Gamma^{top}$) is fully drained ($p=0$) and subjected to a mechanical load of $\overline{\sigma} = 2000 $ N/m$^2$.  The bottom boundary ($\Gamma^{bottom}$) is mechanically restrained in the z-direction ($u_z=0$) and subjected to an injection flux ($\overline{q}(t)$), which is uniform along the boundary and randomly fluctuating in time. The left and right boundaries ($\Gamma^{left}$ and $\Gamma^{right}$) are impermeable and mechanically restrained in x-direction ($u_x=0$).

The gravitational forces are not considered in this example; thus, the behavior is mainly driven by the mechanical load $\overline{\sigma}$ and the injection flux $\overline{q}(t)$. Given these two driving forces, this problem setup is suitable for studying the coupled hydro-mechanical interactions, which have been incorporated into multiple investigations, including numerical stability\cite{kim_stability_2009} and damage in porous media~\cite{mobasher_non-local_2017}. In the case of a zero injection flux, the problem reduces to the classic Terzaghi consolidation problem, found in the literature~\cite{white_stabilized_2008}. 

Inspired by different examples in literature~\cite{amin_i-fenn_2026, li_stabilized_2018}, the poroelastic medium properties are set as $K_s=55556$ MPa, $\lambda=8.375$ MPa, $\mu=5.58$ MPa, $K_f=2200$ MPa, $\phi_b = 0.4$, $\mu_f = 1.0 \times 10^{-3}$ Pa.s, and $K_p = 1.0 \times 10^{-13}$ m$^2$. The fluctuating injection flux is set to have a maximum amplitude of $ 5.0 \times 10^{-6} $ m/s, and the duration of the simulation is set to $5.0 \times 10^{7}$ s, split over $50$ time steps of increment size $\Delta t =  1.0\times 10^6 $ s. 

\subsubsection{PIF-MIONet training}
\label{section:numerical_examples:inj_flux_2d:training}
As explained in Section~\ref{section:pif_mionet_training:sampling}, the adopted training/sampling strategy of the PIF-MIONet requires datasets of physically consistent input fields. For that purpose, a dataset of $100$ Gaussian random injection flux histories and their fully coupled FEM responses is divided into $70$ training, $20$ validation, and $10$ testing load cases (LCs). Increments are sampled as described in Section~\ref{section:pif_mionet_training:sampling}. Training uses $72$ of $3500$ available time increments ($\approx 2\%$), while validation uses $24$ of $1000$. 

During training, the input fields at the sampled increments are supplied to the PIF-MIONet, while the optimizer minimizes the physics-based loss function. Pure physics-informed training is performed for this problem, considering $5$ loss terms, without a data loss term. The first loss term, $\mathcal{L}_1^{pde}$, is computed based on the PDE residual term $\boldsymbol{R}_P$, defined in Eq.~\eqref{eq:residual_pde}. The term $\mathcal{L}_2^{top}$ is the residual of the Dirichlet boundary condition over the boundary $\Gamma^{top}$ ($R_D$ in Eq.~\eqref{eq:residual_dirich}). $\mathcal{L}_3^{bottom}$ is satisfying the injection flux $\overline{q}(t)$ as a Neumann boundary condition over $\Gamma^{bottom}$ ($R_N$ in Eq.~\eqref{eq:residual_neumann}). Finally, $\mathcal{L}_4^{left}$ and $\mathcal{L}_5^{right}$ are considered for satisfying the zero Neumann boundary conditions on $\Gamma^{left}$ and $\Gamma^{right}$, respectively. 

To determine the optimal training configuration and identify ineffective approaches for the problem at hand, we conduct an extensive parametric study that considers multiple training variations, as presented in Appendix~\ref{section:numerical_examples_appendix:inj_flux_2d:hyperparameters}. Our findings indicate that the best results were obtained using shape function derivatives (SF), GradNorm adaptive weights (GN), dimensionless scaling (DS), and the L2 norm $N(.)=\left\|. \right\|_2$ for computing the norm for each loss term. These choices are referred to as ``SF-GN-DS-L2".

The trained model follows the architecture shown in Fig.~\ref{fig:methods_pifmionet_simplified}. A structured grid of equally spaced nodes ($41 \times 5 =205$ points) over the domain is selected to represent the input nodes for strain and pressure fields. The input/output sizes, hidden layers, and group normalization details of each MLP network are provided in Table~\ref{tab:inj_pifmionet_arch}, resulting in a total of $4592$ trainable parameters. Training was performed over $40000$ epochs using an Adam optimizer with a custom learning rate scheduler, requiring a total training time of $35$ minutes.
More details on the custom learning rate are provided in Appendix~\ref{section:numerical_examples_appendix:inj_flux_2d:SF-GN-DS-L2}.

\begin{table}[hbt!]
\renewcommand{\arraystretch}{1.2}
\centering
\caption{Hyperparameters of the PIF-MIONet model trained for the 2D injection example}

\begin{threeparttable}
\begin{tabular}{ccccc}
\hline
Network     & Input size & Sizes of hidden layers\tnote{a} &  $N_{ch}$\tnote{b}  & Output size \\
\hline
Branch \# 1 &$N_q=$1               & [8,8]   & - & 8 \\
Branch \# 2 &$N_{\varepsilon}=$205 & [8]     & 4 & 8 \\
Branch \# 3 &$N_p=$205             & [8]     & 4 & 8 \\
Fusion      &$N_{fi}=$24           & [16,16] & - & 8 \\
Trunk       &$N_c=$2               & [8,8]   & - & 8 \\
\hline
\end{tabular}

\begin{tablenotes}
\footnotesize
\item[a] The length of the sizes array is the number of hidden layers.
\item[b] Group normalization using $N_{ch}$ channels is implemented on the output of the input layer. ``-" means no group normalization is considered. 
\end{tablenotes}
\vspace{-7pt}
\rule{\linewidth}{0.1pt}
\end{threeparttable}

\label{tab:inj_pifmionet_arch}
\end{table}

\begin{figure}[hbt!]
\centering
\includegraphics[width=1.0\textwidth]{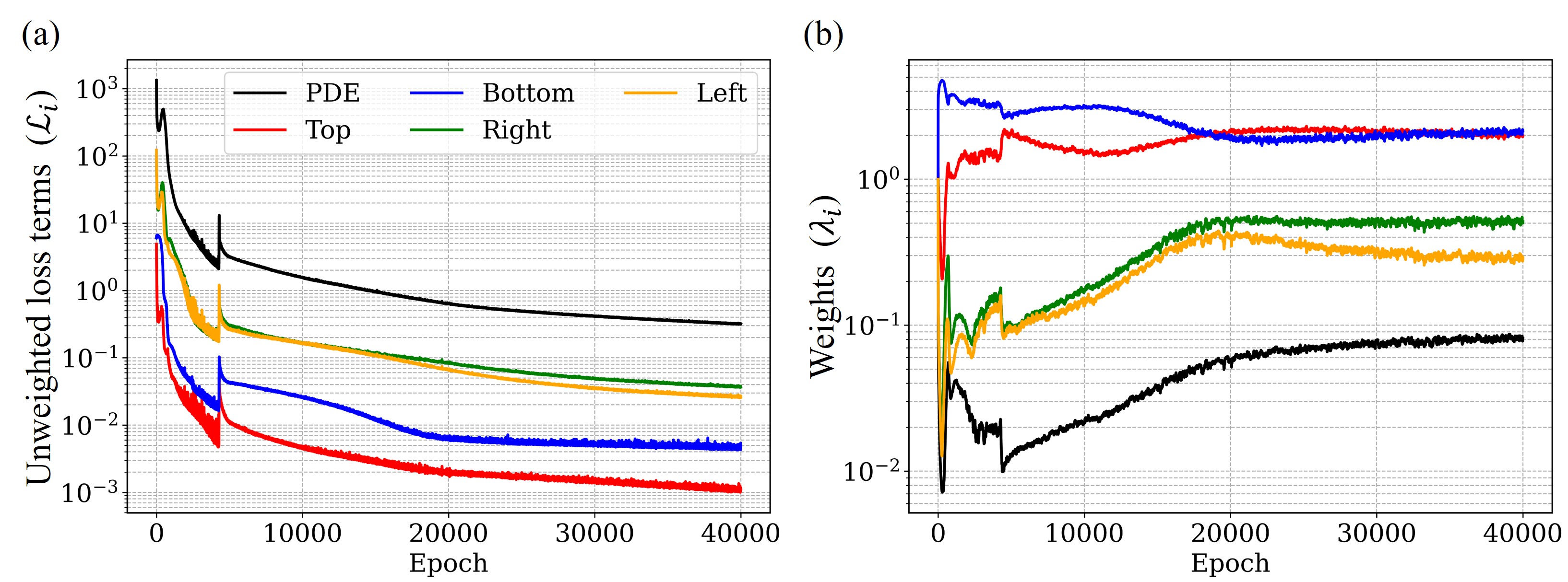}
\caption{Training metrics for the 2D injection example: (a) unweighted loss terms $\mathcal{L}_i$, (b) the corresponding adaptive weights $\lambda_i$. Only PDE residual and boundary condition losses are included; no data loss is used.}
\label{fig:inj_2d_terms}
\end{figure}

Fig.~\ref{fig:inj_2d_terms} shows the unweighted training loss terms and the development of their adaptive weights across epochs. The plots show how GradNorm adaptively updates the weights to ensure a balanced reduction across all loss terms. All loss terms dropped more than three orders of magnitude from their initial value, highlighting the efficiency of the training setup using SF-GN-DS-L2. We also tested training for another variation, where ReLoBRaLo (RB) is used instead of GradNorm (i.e., training combination SF-RB-DS-L2). The results showed training instability and higher error values. More details on the training using SF-RB-DS-L2 are provided in Appendix~\ref{section:numerical_examples_appendix:inj_flux_2d:SF-RB-DS-L2}.

\subsubsection{IFENN accuracy and portability}
\label{section:numerical_examples:inj_flux_2d:ifenn}
\begin{figure}[hbt!]
\centering
\includegraphics[width=1.0\textwidth]{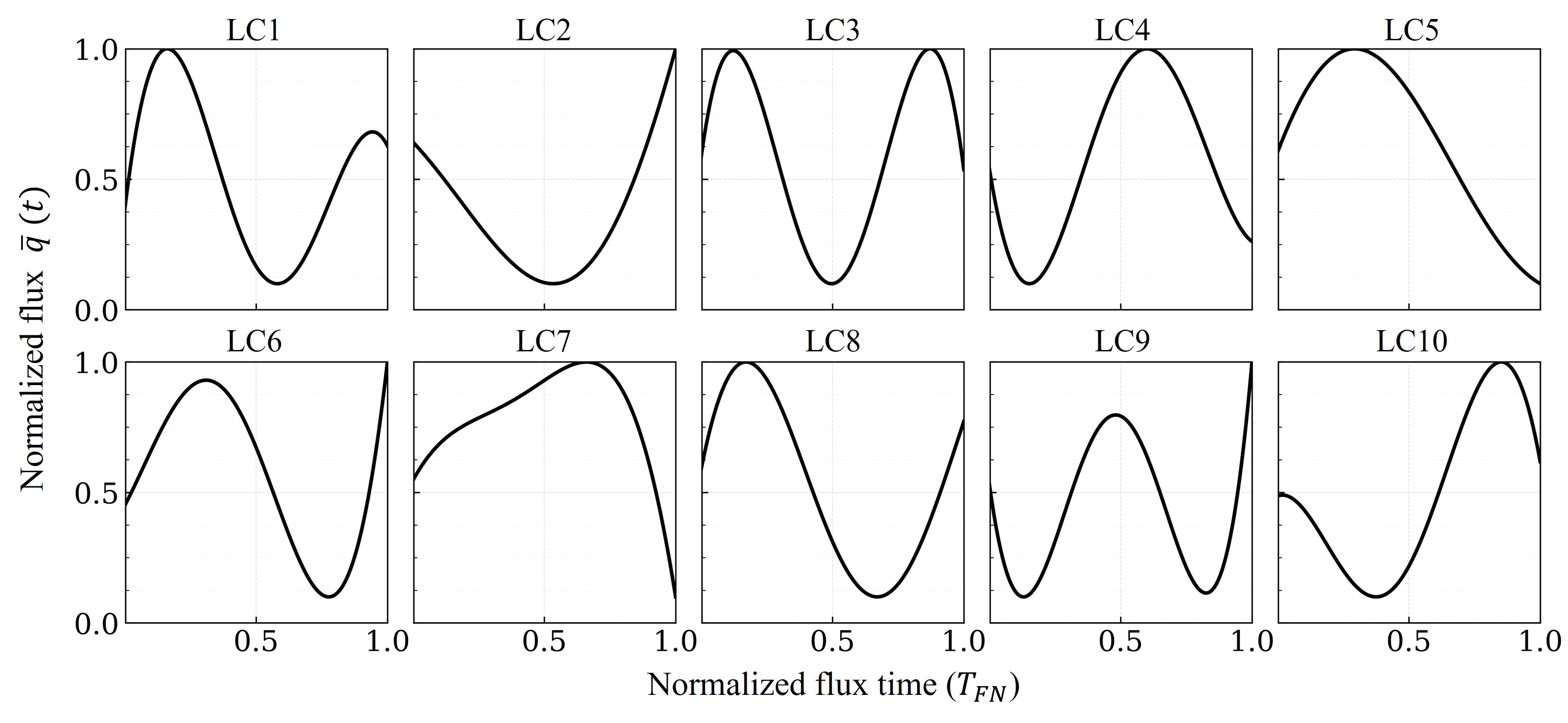}
\caption{Normalized flux amplitude versus normalized flux time ($T_{FN}$) for all testing load cases (LCs) of the injection problem. Flux and time are normalized by their respective maximum values.}
\label{fig:inj_flux_testing_lcs}
\end{figure}

\begin{figure}[hbt!]
\centering
\includegraphics[width=1.0\textwidth]{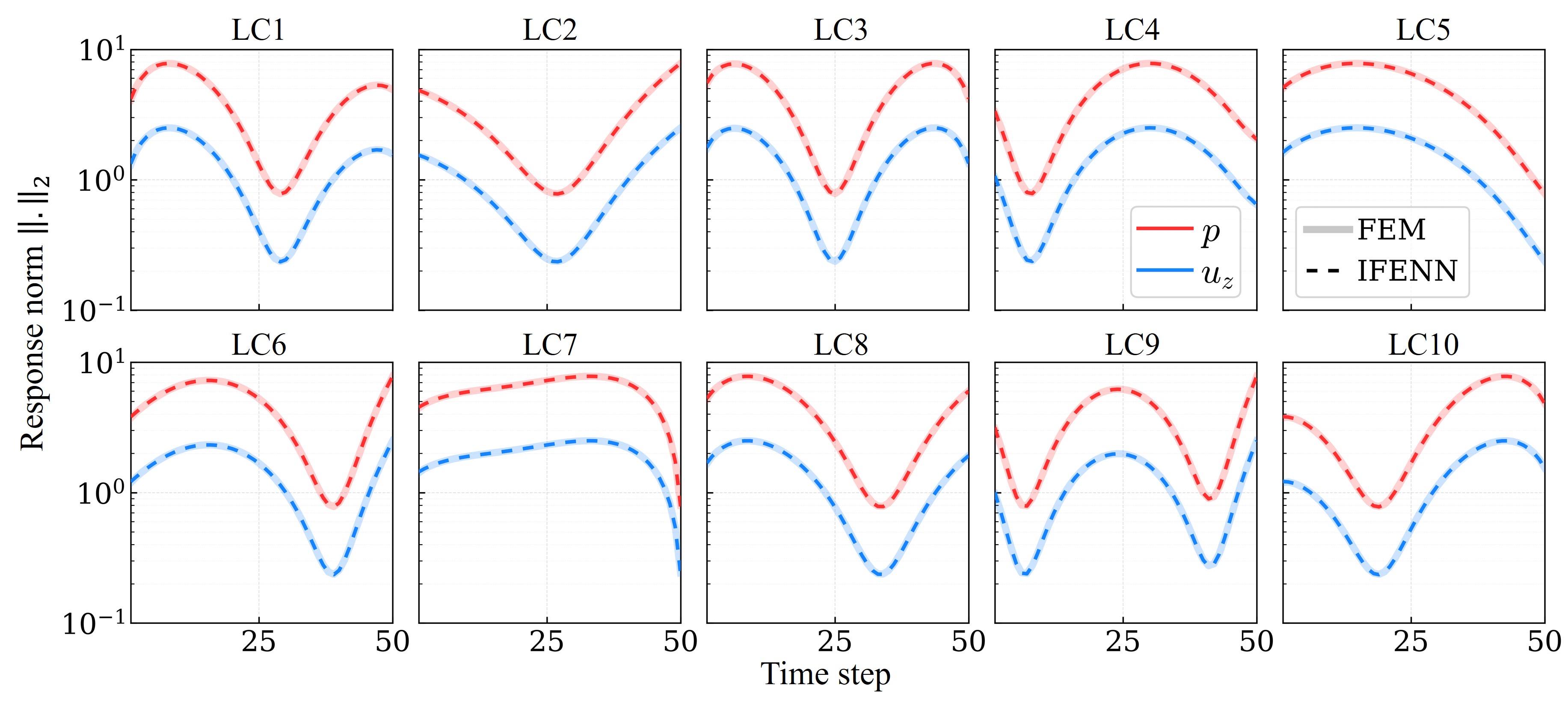}
\caption{Pressure norm ${\left\| p \right\|_2}$ and displacement norm ${\left\| u_z \right\|_2}$ obtained using the fully coupled FEM solver and IFENN-deal.II for the 2D injection problem. Response norm units are MPa and m for pressure and displacement, respectively.}
\label{fig:inj_ifenn_dealii_2D_Norms}
\end{figure}

After being trained on sparse randomly sampled time steps, the model is integrated into the IFENN-deal.II framework and tested on $10$ load cases (LCs), shown in Fig.~\ref{fig:inj_flux_testing_lcs}, that were unseen by the training optimizer. Given the problem setup, the response is effectively one-dimensional; hence, we focus only on the pressure field ($p$) and displacement in the z-direction ($u_z$). Fig.~\ref{fig:inj_ifenn_dealii_2D_Norms} depicts the norm for the pressure (${\left\| p \right\|_2}$) and displacement (${\left\| u_z \right\|_2}$) fields, per time increment, obtained using the fully coupled FEM solver ($y_{true}$) and IFENN-deal.II ($y_{pred}$). The plots show consistent accuracy, with the IFENN-deal.II response closely matching the fully coupled solver response for all testing cases. In addition, both pressure and displacement responses follow the injection flux development (in Fig.~\ref{fig:inj_flux_testing_lcs}) across all time increments and load cases.

\begin{figure}[hbt!]
\centering
\includegraphics[width=1.0\textwidth]{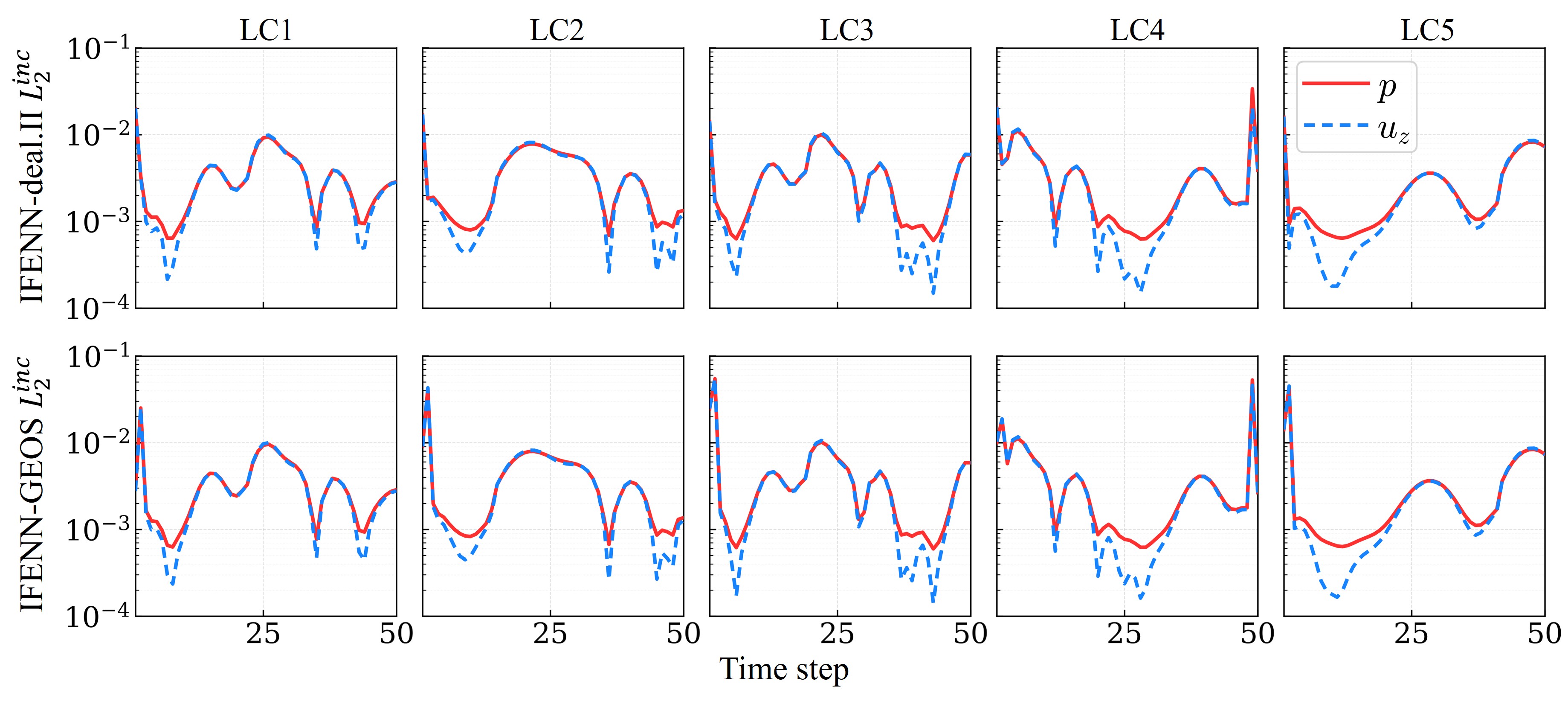}
\caption{Testing error $L_2^{inc}$ of pressure $(p)$ and displacement in the z-direction $(u_z)$ for testing load cases LC1-LC5 of the 2D injection problem, using IFENN-deal.II and IFENN-GEOS.}
\label{fig:inj_2D_ifenn_dealii_vs_geos_1_5}
\end{figure}

To further assess the accuracy and the framework portability, the simulations were repeated using IFENN-GEOS. For this purpose, a GEOS-compatible three-dimensional mesh consisting of a single layer of elements in the out-of-plane direction was employed, with appropriate boundary conditions imposed to enforce plane-strain conditions. Fig.~\ref{fig:inj_2D_ifenn_dealii_vs_geos_1_5} shows the testing error ($L_2^{inc}$ defined in Eq.~\eqref{eq:inc_error}) for both $p$ and $u_z$ fields, computed for IFENN-deal.II and IFENN-GEOS responses for load cases LC1-LC5. The error plots show low error magnitudes, ranging from $1\times10^{-2}$ to $1\times10^{-4}$, with the errors in both fields ($p$ and $u_z$) relatively close. Errors for load cases LC6-L10 have a consistent behavior as shown in the Appendix Fig.~\ref{fig:inj_2D_ifenn_dealii_vs_geos_6_10}.

The fluctuations in the error metric $L_2^{inc}$ can be related to the fluctuations in the injection flux $\overline{q}(t)$. When the injection flux is at its lowest magnitude, the true pressure norm ${\left\| y_{true} \right\|_2}$ drops to values close to zero, magnifying the relative error computation of $L_2^{inc}$ (see Eq.~\eqref{eq:inc_error}). It's worth emphasizing that these results are derived from a network trained solely on 72 non-consecutive time increments in a purely physics-informed fashion, without data in the loss function, and evaluated against previously unseen consecutive time series.

In addition to low error magnitudes, IFENN-deal.II and IFENN-GEOS errors are nearly identical. One trained PIF-MIONet is integrated with two different solvers, providing almost the same response, highlighting the portability of the IFENN framework. More results highlighting the achieved accuracy and portability for IFENN-deal.II and IFENN-GEOS are provided in Appendix~\ref{section:numerical_examples_appendix:inj_flux_2d:ifenn_dealii_geos_2d}.

\subsubsection{IFENN efficiency}
\label{section:numerical_examples:inj_flux_2d:efficiency}
The efficiency of the IFENN framework has been studied before in the literature for different problems, including thermoelasticity~\cite{amin_i-fenn_2026} and phase-field fracture modeling~\cite{pantidis_integrated_2026}. In this work, we investigate the IFENN framework's efficiency for poroelasticity problems for the first time. This study differs from previous work due to the numerical FEM stability requirements of poroelasticity problems. As discussed in Section~\ref{section:poroelasticity:weak}, a proper FEM analysis can be done using biquadratic-displacement/bilinear-pressure (U2P1) elements or bilinear-displacement/bilinear-pressure (U1P1) elements with a stabilization algorithm. 

\begin{figure}[hbt!]
\centering
\includegraphics[width=.7\textwidth]{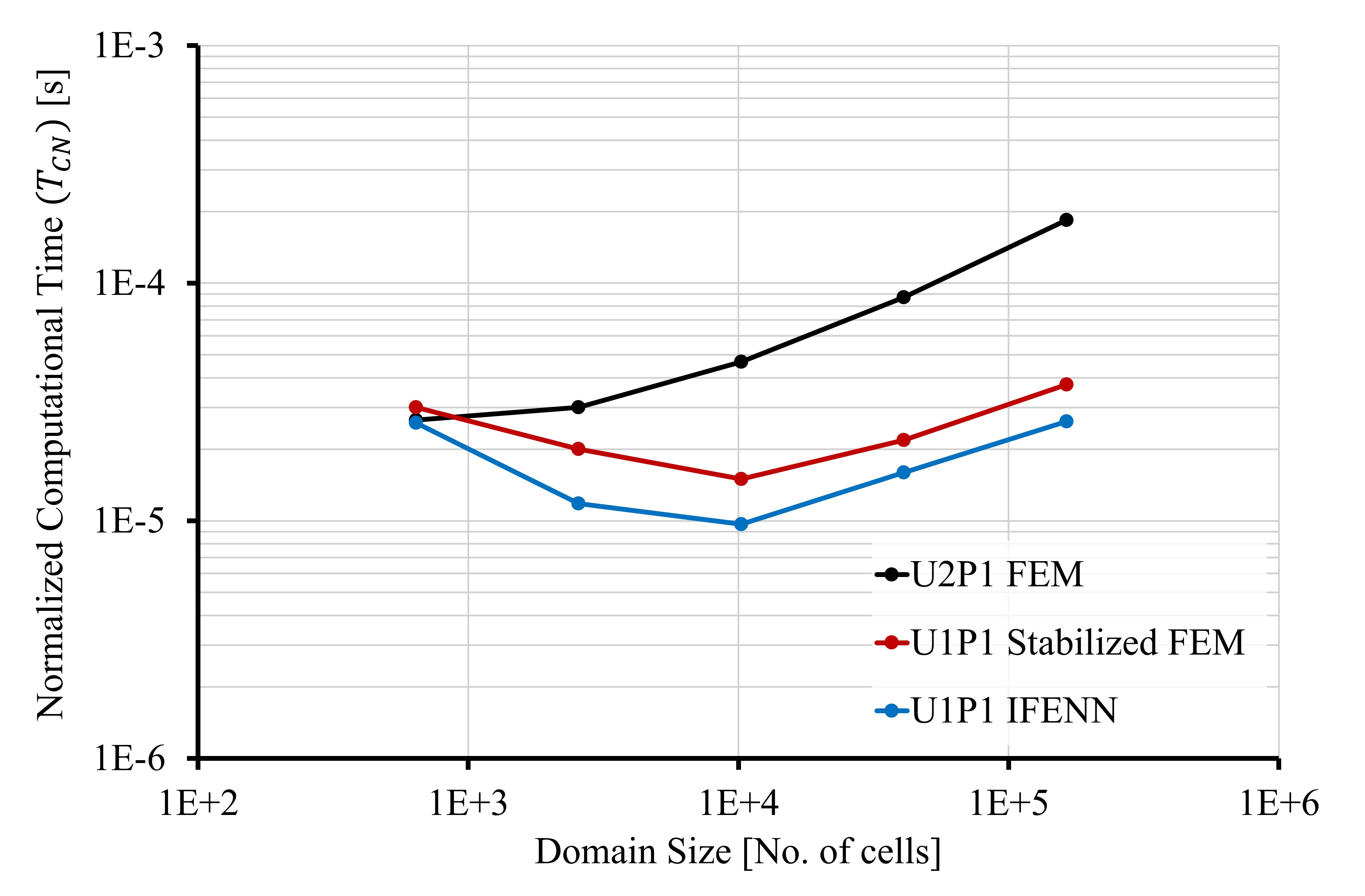}
\caption{Efficiency analysis: Normalized computational time ($T_{CN}$) for staggered coupled FEM solvers and IFENN-deal.II. $T_{CN}$ is computed as the total computational time divided by (the number of time increments $N_{inc}$ $\times$ the number of degrees of freedom $N_{DOF}$).}
\label{fig:inj_efficiency}
\end{figure}

For this study, we implement two staggered FEM solvers. Specifically, one staggered solver is implemented for the unstabilized weak form (Eq.~\eqref{eq:poro_w_blm_time} and Eq.~\eqref{eq:poro_w_cont_time}) using U2P1 elements, and another for the stabilized weak form (Eq.~\eqref{eq:poro_w_blm_time} and Eq.~\eqref{eq:poro_w_cont_time_stab}) using U1P1 elements. We compare the U1P1-IFENN efficiency against these two options, considering different mesh resolutions. The results are presented in Fig.~\ref{fig:inj_efficiency}, showing normalized computational time versus domain size. The normalized computational time ($T_{CN}$) is computed as the total computational time divided by ($N_{inc}$ $\times$ $N_{DOF}$), where $N_{inc}$ is the number of time increments and $N_{DOF}$ is the number of degrees of freedom.  

The results show that for small domain sizes, the cost of IFENN per DOF is comparable to the FEM solvers. This comparable performance at small domain sizes can be attributed to the computing overhead from the debugging features added to the IFENN framework to compute runtime errors. However, with mesh refinement, the IFENN framework shows better performance compared to the FEM solvers. With this consistently enhanced performance, IFENN shows potential scalability compared to other solvers, especially for large-scale industrial problems. 

It is worth noting that the computational time for IFENN did not include the training time of the PIF-MIONet. The model is trained once and can then be used for any number of simulations, with different flux histories and mesh-refinement setups. Therefore, as the number of required simulations increases, especially with fine meshes, training becomes relatively insignificant compared to the total computational time required.   

\FloatBarrier
\subsection{2D injection example with variable \texorpdfstring{permeability $K_p$ and time increment $\Delta t$}{permeability and time increment}}
\label{section:numerical_examples:inj_flux_2d_k_dt}
\subsubsection{PIF-MIONet training}

The PIF-MIONet is tested in a new generalization setup, where the model encodes variable material properties and time discretizations. In this setup, we train the model on variable permeability ($K_p$) and variable time step size ($\Delta t$). Variations in $K_p$ and $\Delta t$ are organized into three permeability groups, each with three subgroups corresponding to different time discretizations, as listed in Table~\ref{tab:inj_group_prop}. All groups are subjected to the same injection histories; however, the injection rates (total time) are adjusted relative to the permeability $K_p$ values. For subgroups with the same $K_p$, the number of time increments is set so that the total time is the same. This wide range of permeability values under the same injection magnitudes is expected to result in deformations that exceed the linear poroelasticity assumptions. However, for consistency throughout this paper, all simulations assume linear elastic behavior.     

\begin{table}[hbt!]
    \centering
    \caption{Different group properties for the generalized PIF-MIONet model trained for variable $K_p$ and $\Delta t$.}
    
    \renewcommand{\arraystretch}{1.2}
    \setlength{\tabcolsep}{12pt}

    \begin{tabular}{@{}lccc@{}}
        \toprule

        Group Number
        & G1 & G2 & G3 \\

        \midrule

        Permeability ($K_p$)
        & $1 \times 10^{-14}$
        & $1 \times 10^{-13}$
        & $1 \times 10^{-12}$ \\

        Total time
        & $5.0 \times 10^{8}$
        & $5.0 \times 10^{7}$
        & $5.0 \times 10^{6}$ \\

        Pressure scaling factor ($p^*$)
        & $5.0 \times 10^{6}$
        & $5.0 \times 10^{5}$
        & $5.0 \times 10^{4}$ \\

        Strain scaling factor ($\varepsilon^*$)
        & $2.5 \times 10^{-1}$
        & $2.5 \times 10^{-2}$
        & $2.5 \times 10^{-3}$ \\

        \addlinespace[3pt]
        \midrule

        Number of increments ($N_{\mathrm{inc}}$)
        & \multicolumn{3}{c}{Increment size ($\Delta t$)} \\

        \cmidrule(lr){2-4}

        \multicolumn{1}{c}{500}
        & $1 \times 10^{6}$
        & $1 \times 10^{5}$
        & $1 \times 10^{4}$ \\

        \multicolumn{1}{c}{50}
        & $1 \times 10^{7}$
        & $1 \times 10^{6}$
        & $1 \times 10^{5}$ \\

        \multicolumn{1}{c}{5}
        & $1 \times 10^{8}$
        & $1 \times 10^{7}$
        & $1 \times 10^{6}$ \\

        \bottomrule
    \end{tabular}
    \label{tab:inj_group_prop}
\end{table}

For this generalized problem, two branches for permeability $K_p$ and time increment size $\Delta t$ are incorporated into the PIF-MIONet architecture, as shown in Fig.~\ref{fig:methods_pifmionet_details}. The training in this example is purely physics-informed, with the data loss term excluded from the loss function. However, the sparse increment training strategy, explained in Section~\ref{section:pif_mionet_training:sampling}, requires datasets of physically consistent input fields. For this purpose, training, validation, and testing datasets are sampled from all the groups listed in Table~\ref{tab:inj_group_prop}. As a result, the magnitudes of pressure and strain fields are expected to vary by orders of magnitude. Globally normalizing all sampled increments could harm the network's ability to encode some inputs. For example, normalizing pressure inputs for increments spanning magnitudes $1.0\times10^4$ to $1.0\times10^6$ to the range $[0,1]$ would compress the variations in pressure fields with small magnitudes. Hence, we adopt a per-sample normalization strategy, where each input/output for a specific increment is scaled using a predefined scaling factor.

The per-sample normalization strategy is applied only to the pressure and strain fields for each increment. Other parameters, such as flux, permeability, and time increment size, are normalized globally, enabling the network to treat these inputs as continuous fields and accept any value. The scaling factors for pressure ($p^*$) and strain ($\varepsilon^*$) are provided in Table~\ref{tab:inj_group_prop}. Notably, the pressure scaling factor ($p^*$) is the same factor utilized in the dimensionless scaling of the residual forms (see Section~\ref{section:pif_mionet_design:dim}). For brevity, more details on the training process are provided in the Appendix~\ref{section:numerical_examples_appendix:inj_flux_2d_k_dt}. 

\subsubsection{IFENN-deal.II results for variable \texorpdfstring{permeability $K_p$ and time increment $\Delta t$}{permeability and time increment}}
\begin{figure}[hbt!]
\centering
\includegraphics[width=1.0\textwidth]{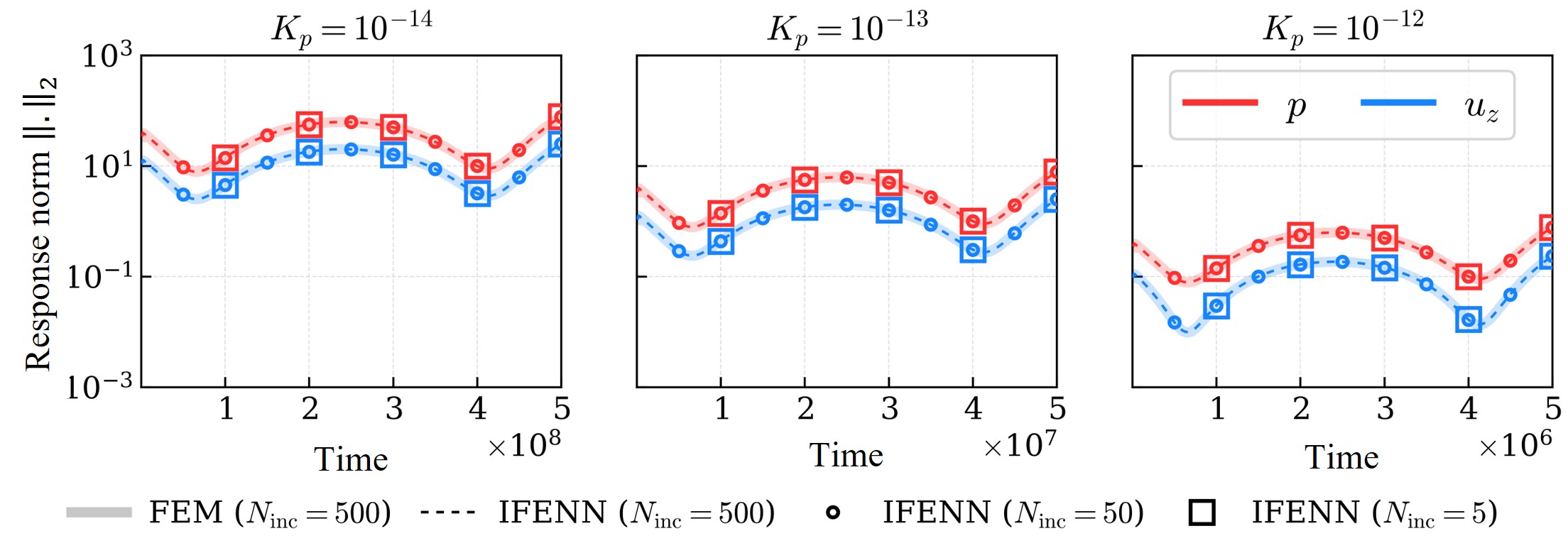}
\caption{Pressure norm ${\left\| p \right\|_2}$ and displacement norm ${\left\| u_z \right\|_2}$ obtained using the fully coupled FEM solver and IFENN-deal.II for 9 different setups (3 permeability values $\times$ 3 time increment values) in the 2D injection problem, subjected to the 9th testing load case. Response norm units are MPa and m for pressure and displacement, respectively.}
\label{fig:inj_gen_3x1_LC9}
\end{figure}

The trained model is integrated into the IFENN-deal.II framework and tested for the three groups (nine subgroups) listed in Table~\ref{tab:inj_group_prop}. The 9th testing load case showed high fluctuation and reached the maximum amplitude at the last time increment (see Fig.~\ref{fig:inj_flux_testing_lcs}), and was therefore selected for evaluation, as shown in Fig.~\ref{fig:inj_gen_3x1_LC9}.

Within each group in Fig.~\ref{fig:inj_gen_3x1_LC9}, the pressure norm $\|p\|_2$ and displacement norm $\|u_z\|_2$ remain consistent across subgroups with different time discretizations. In particular, reducing the total number of increments $N_{inc}$ from $500$ to $5$ preserves end-of-simulation accuracy. Comparisons across G1-G3 further demonstrate that the same trained PIF-MIONet accurately predicts pressure fields spanning two to three orders of magnitude. This pressure range (and the corresponding strains) is handled by the physics-based scaling factors ($p^*$ and $\varepsilon^*$ in Table~\ref{tab:inj_group_prop}), while the PIF-MIONet captures the remaining dependence on permeability $K_p$ and time increment $\Delta t$. Similar observations hold for the 7th testing load case, in Appendix Fig.~\ref{fig:inj_gen_3x1_LC7}.

\subsection{3D injection example}
\label{section:numerical_examples:inj_flux_3D}

\subsubsection{Problem setup}
The previous sections showed IFENN's performance for the 2D injection problem. Here, we extend the same problem in Section~\ref{section:numerical_examples:inj_flux_2d} to a 3D domain to test whether the proposed framework can handle volumetric inputs, three-dimensional pressure derivatives, and additional boundary-condition terms. Although the resulting pressure response is expected to exhibit a one-dimensional behavior, the problem is formulated and solved entirely in three dimensions. Fig.~\ref{fig:inj_setup}b depicts the 3D domain for the problem and its boundary conditions. The 3D domain is discretized using a structured mesh of eight-node hexahedral elements, each with an edge length of 1/8 m in all three spatial directions.

\subsubsection{PIF-MIONet training}
For this 3D setup, a 3D-compatible PIF-MIONet is trained using pressure and strain-trace fields defined over the domain, with a trunk input containing three coordinate components. The 3D problem preserves the main loading and flow configuration of the 2D example, while adding the front and back boundaries, $\Gamma^{front}$ and $\Gamma^{back}$, in the $y$ direction. These boundaries are treated analogously to $\Gamma^{left}$ and $\Gamma^{right}$. Thus, the front and back faces are impermeable for the fluid flow problem and mechanically restrained in the $y$ direction. Since the 3D mesh uses the same size and resolution in the $x$ and $y$ directions, the left/right and front/back boundary responses are expected to perform similarly.

For the 3D training study, the original 2D deal.II simulations were used to construct a controlled pseudo-3D dataset. The 2D coordinate grid was extruded in the $y$ direction, and the pressure and strain-trace histories were duplicated through the added depth direction. This pseudo-3D dataset provides physically consistent pressure and strain-trace histories for the branch inputs. Notably, this dataset is used exclusively to provide network inputs during training and does not contribute to any data-driven loss term.

\begin{figure}[hbt!]
\centering
\includegraphics[width=1.0\textwidth]{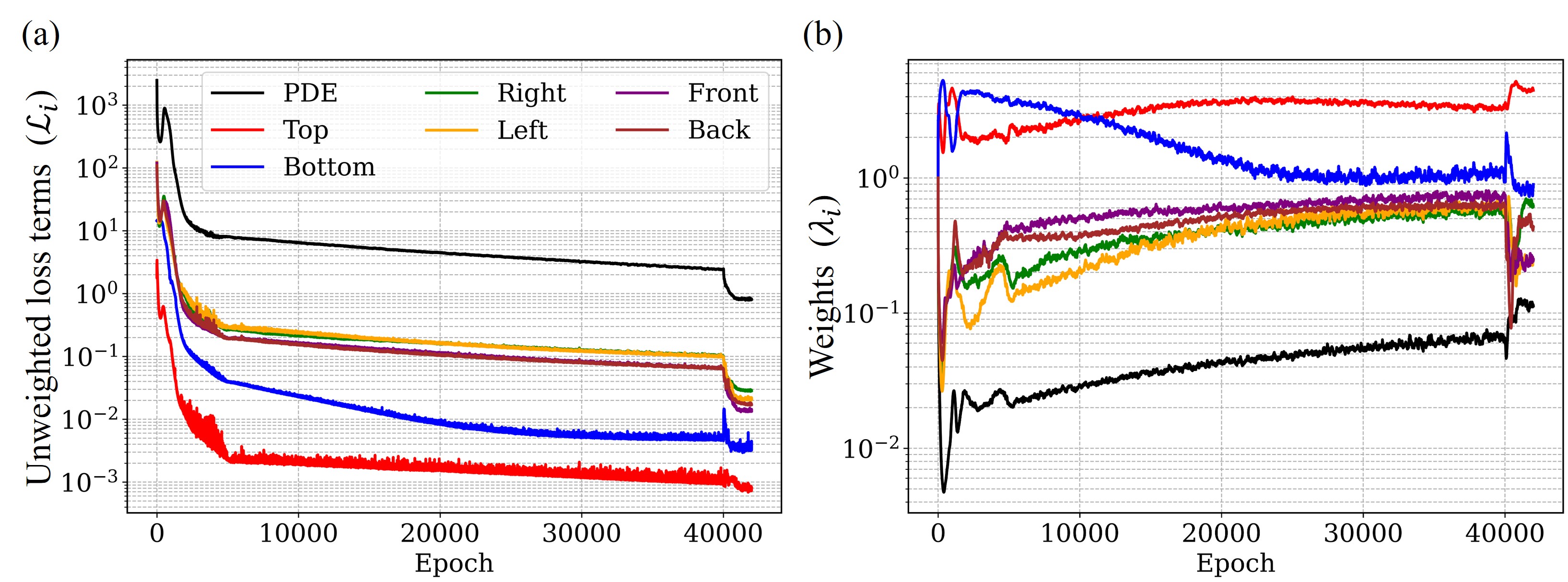}
\caption{Training metrics for the 3D injection example: (a) unweighted loss terms $\mathcal{L}_i$, (b) the corresponding adaptive weights $\lambda_i$. Only PDE residual and boundary condition losses are included; no data loss is used.}
\label{fig:inj_3d_terms}
\end{figure}

The 3D physics-informed loss extends the 2D loss definition by adding two boundary terms for the front and back faces. The two additional terms, $\mathcal{L}_6^{front}$ and $\mathcal{L}_7^{back}$, enforce the zero Neumann conditions in the $y$ direction. The pressure Laplacian in the PDE residual ($\boldsymbol{R}_P$ in Eq.~\eqref{eq:residual_pde}) is also extended to consider the $y$ direction. Automatic differentiation is used to compute the spatial derivatives. To enforce the Neumann boundary conditions, pressure gradients are evaluated along the discretized domain in the $x$, $y$, and $z$ directions at their corresponding boundary faces. More details on the training process and the adopted custom learning rate are provided in Appendix~\ref{section:numerical_examples_appendix:inj_flux_3d}. The monitored loss terms for the 3D PIF-MIONet and their adaptive weights are provided in Fig.~\ref{fig:inj_3d_terms}. The plots show that the front/back terms behave similarly to the left/right terms, which is expected from the symmetry between the $x$ and $y$ directions.

\subsubsection{IFENN-GEOS results for the 3D injection example}
\begin{figure}[hbt!]
\centering
\includegraphics[width=1.0\textwidth]{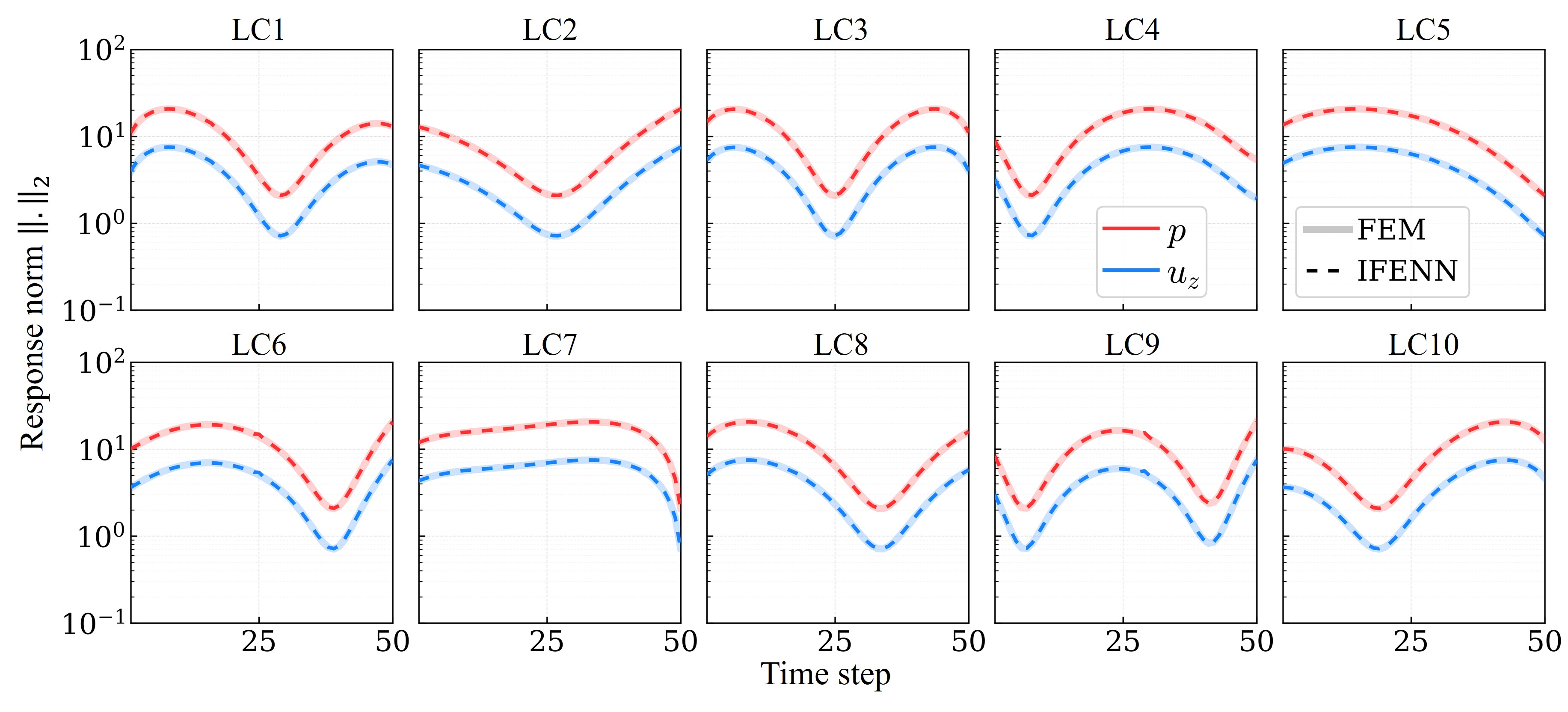}
\caption{Pressure norm ${\left\| p \right\|_2}$ and displacement norm ${\left\| u_z \right\|_2}$ obtained using the fully coupled FEM solver and IFENN-GEOS for the 3D injection problem. Response norm units are MPa and m for pressure and displacement, respectively.}
\label{fig:inj_ifenn_geos_3D_Norms}
\end{figure}

\begin{figure}[hbt!]
\centering
\includegraphics[width=1.0\textwidth]{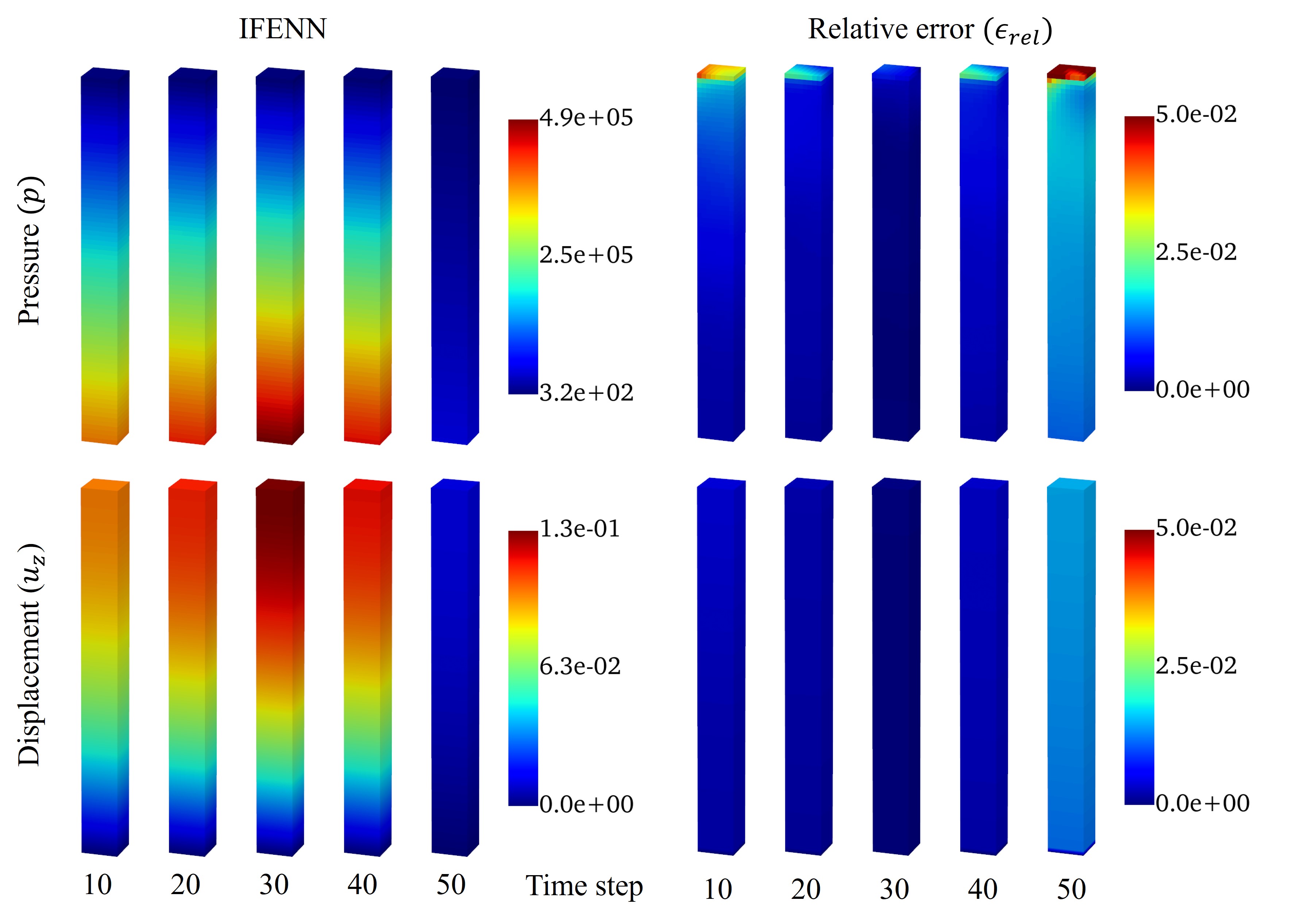}
\caption{Pressure $p$ and displacement $u_z$ obtained using IFENN-GEOS and the corresponding relative error ($\epsilon_{rel}$) for the 3D injection example at different time steps of the 7th testing load case.}
\label{fig:inj_ifenn_geos_3D_LC7}
\end{figure}

Fig.~\ref{fig:inj_ifenn_geos_3D_Norms} compares the response norms obtained using IFENN-GEOS and the fully coupled FEM solver for the 3D problem. Their close agreement demonstrates the framework's accuracy in 3D, consistent with the 2D results in Fig.~\ref{fig:inj_ifenn_dealii_2D_Norms}. A sample load case is presented in Fig.~\ref{fig:inj_ifenn_geos_3D_LC7}, where the pressure and displacement fields and their relative errors are shown. The pressure distribution is uniform in the $x$- and $y$-directions, consistent with zero-flux conditions on the front, back, left, and right boundaries. The largest relative errors occur in regions with near-zero field values, while errors remain well below $5\%$ elsewhere.

Overall, the results show that the framework remains stable after introducing volumetric fields and the additional $y$-direction boundary terms, both in terms of offline training as well as online implementation. This example demonstrates that IFENN can be extended from 2D to 3D domains and coupled with GEOS for volumetric poromechanical simulations. Additional results showing the relative errors and predicted fields for different load cases are provided in Appendix~\ref{section:numerical_examples_appendix:inj_flux_3d:results}.

\subsection{Excavation problem}
\label{section:numerical_examples:excav}

\subsubsection{Problem setup}
\label{section:numerical_examples:excav:setup}

The last example is an excavation problem subjected to a dewatering flux. This example is inspired by geotechnical engineering problems in the literature~\cite{potts_finite_2001, li_finite_2004} and was previously investigated by the authors using a data-driven training approach~\cite{amin_i-fenn_2026}. We revisit this problem with a physics-based approach, discussing the challenges posed by the complex boundary conditions and highlighting the advantages of using PI-ML for this type of problem. 

\begin{figure}[hbt!]
\centering
\includegraphics[width=.8\textwidth]{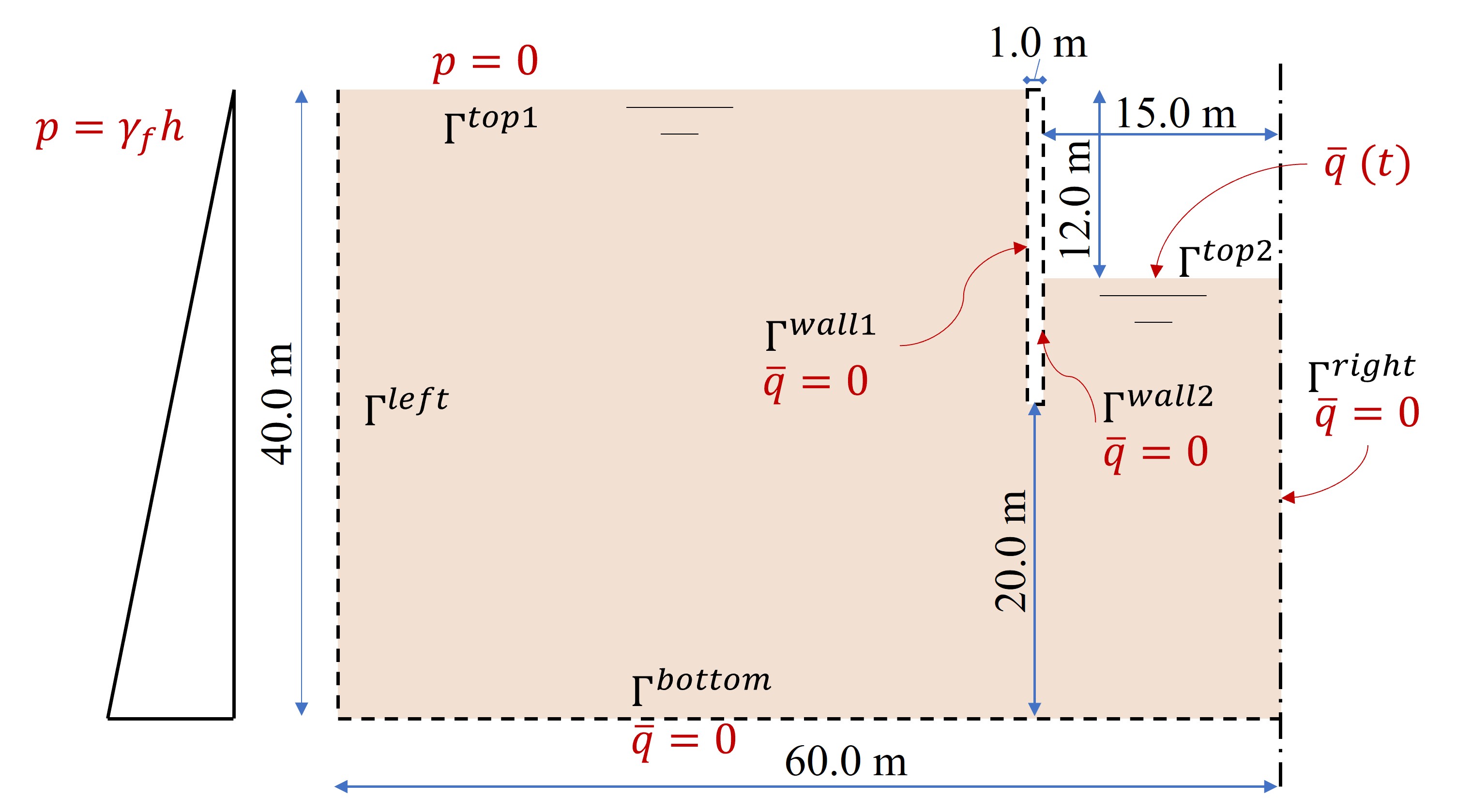}
\caption{Schematic representation of the geometry and boundary conditions
of the excavation problem. For training considerations, the boundaries are grouped into two subsets:  $\Gamma^{base}= \left\{\Gamma^{top1},\Gamma^{top2},\Gamma^{left},\Gamma^{bottom}   \right\}$ and $\Gamma^{comp}= \left\{\Gamma^{wall1},\Gamma^{wall2},\Gamma^{right} \right\}$.}
\label{fig:excav_setup}
\end{figure}

The problem geometry is shown in Fig.~\ref{fig:excav_setup}, which depicts a dewatering flux $\overline{q}(t)$ and the different boundary conditions. The material properties are set as $K_s=55556$ MPa, $\lambda=8.375$ MPa, $\mu=5.58$ MPa, $K_f=2200$ MPa, $\phi_b = 0.4$, $\mu_f = 1.0 \times 10^{-3}$ Pa.s, and $K_p =  1.0 \times 10^{-11}$ m$^2$. The simulation time is set to $3\times10^{6}$ seconds, with a total number of increments $N_{inc}=60$.  

\subsubsection{PIF-MIONet design and training options}
\label{section:numerical_examples:excav:training}
A total of $100$ Gaussian random histories of dewatering flux $\overline{q}(t)$ are generated and then used to prepare the corresponding input/output pairs. The dataset generated for the $100$ simulations is split into subsets of $70$, $20$, and $10$ for training, validation, and testing, respectively. $33$ increments are sampled from the total training simulations ($4200$ time increments), and $33$ increments are sampled from the validation simulations ($1200$ time increments). 

The given model needs to account for $7$ boundary conditions, as shown in Fig.~\ref{fig:excav_setup}. Initial training trials showed that excluding the three boundary parts $\Gamma^{right}$, $\Gamma^{wall2}$, and $\Gamma^{wall1}$ can enhance the training accuracy. Hence, we divide the boundary conditions into two subsets: $\Gamma^{base}= \left\{\Gamma^{top1},\Gamma^{top2},\Gamma^{left},\Gamma^{bottom} \right\}$ and $\Gamma^{comp}= \left\{\Gamma^{wall1},\Gamma^{wall2},\Gamma^{right} \right\}$. For completeness, we examine both cases, excluding and including $\Gamma^{comp}$. For clarity in the following discussions, the weighted sums of the loss terms associated with the boundary subsets $\Gamma^{base}$ and $\Gamma^{comp}$ are denoted by $\mathcal{L}^{base}=\sum_{\Gamma^i\in\Gamma^{base}}\lambda_i\,\mathcal{L}_i$ and $\mathcal{L}^{comp}=\sum_{\Gamma^i\in\Gamma^{comp}}\lambda_i\,\mathcal{L}_i$, respectively. 
The adaptive weights associated with the boundary conditions at $\Gamma^{comp}$ are collected in the vector $\boldsymbol{\lambda}_{comp} = \{\lambda_{wall1}, \lambda_{wall2}, \lambda_{right}\}$. The inclusion or exclusion of these boundary conditions is controlled by their corresponding weights, with zero weights indicating that the associated loss terms are omitted from training.

For this complex geometry, evaluating spatial derivatives using automatic differentiation led to unstable training that failed to converge (model M1-AD in Table~\ref{tab:excav_training_summary_bcs}); a possible reason for this behavior is discussed in Appendix~\ref{section:numerical_examples_appendix:excav:param}. Guided by this remark and the findings from previous examples, we adopt shape function derivatives (SF), GradNorm adaptive weighting (GN), dimensionless scaling (DS), and the $L_2$ norm, $N(\cdot)=\|\cdot\|_2$, for evaluating individual loss terms. This combination, denoted ``SF-GN-DS-L2'', is used by default for all subsequent architectural and training configurations. 

\begin{figure}[hbt!]
\centering
\includegraphics[width=.8\textwidth]{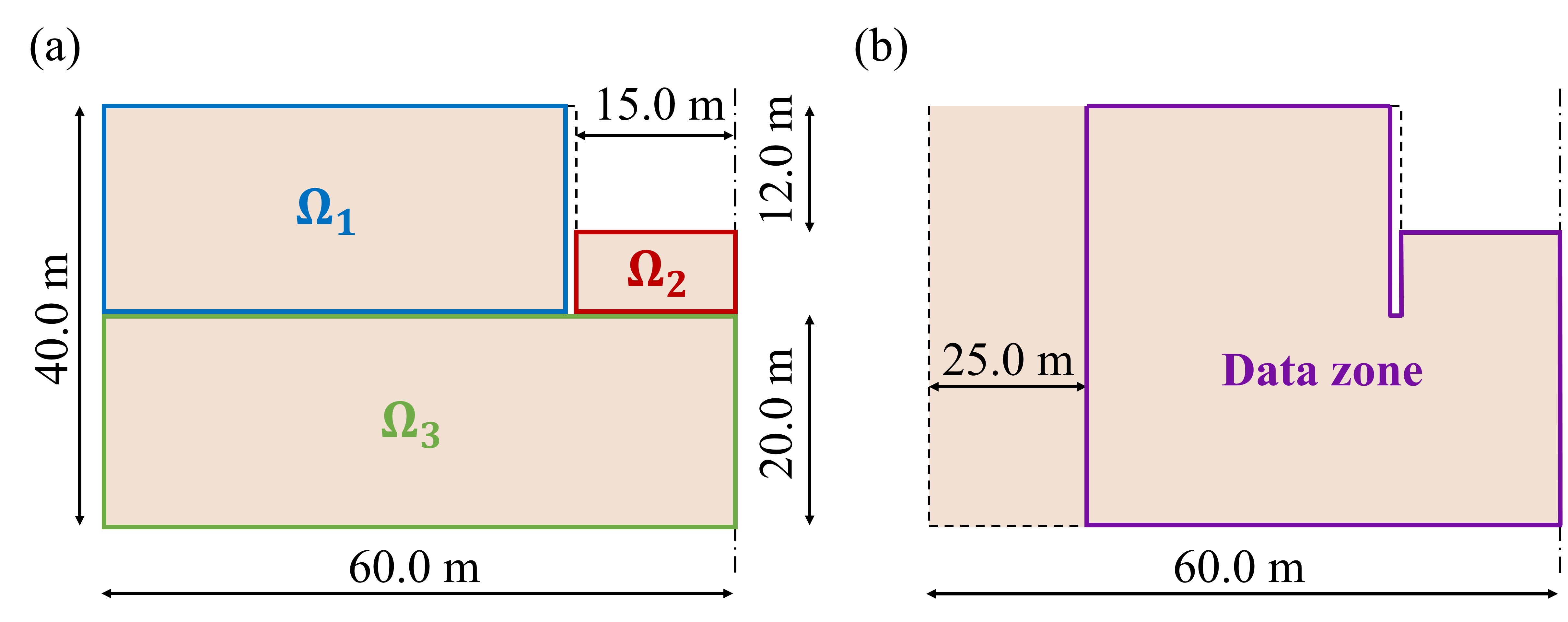}
\caption{Schematic representation of: (a) decomposed domain into $3$ subdomains for training using domain decomposition, (b) the boundaries of the data zone, utilized in sampling nodes for the data loss term.}
\label{fig:excav_zones}
\end{figure}

We conducted several training trials to evaluate PIF-MIONet architectures with single and multiple trunks, illustrated in Figs.~\ref{fig:methods_pifmionet_simplified} and~\ref{fig:methods_pifmionet_domain_decomp}, respectively. For the multiple-trunk architecture, the domain is divided into three subdomains (Fig.~\ref{fig:excav_zones}a), each with a dedicated trunk to encode its spatial coordinates. Table~\ref{tab:excav_pifmionet_arch_3_trunk} lists the details of the three-trunk architecture utilized for this problem. Details of the single-trunk architecture are provided in the Appendix Table~\ref{tab:excav_pifmionet_arch_1_trunk}.

\begin{table}[hbt!]
\renewcommand{\arraystretch}{1.2}
\centering
\caption{Hyperparameters of the three-trunk PIF-MIONet model trained for the 2D excavation example}

\begin{threeparttable}
\begin{tabular}{ccccc}
\hline
Network     & Input size & Sizes of hidden layers\tnote{a} &  $N_{ch}$\tnote{b}  & Output size \\
\hline
Branch \# 1 &$N_q=$1               & [64,64]        & - & 64 \\
Branch \# 2 &$N_{\varepsilon}=$499 & [64,64]        & - & 64 \\
Branch \# 3 &$N_p=$499             & [64,64]        & - & 64 \\
Fusion      &$N_{fi}=$192          & [192]          & - & 256 \\
Trunk  \#1  &$N_c=$2               & [128,128,128]  & - & 256 \\
Trunk  \#2  &$N_c=$2               & [128,128,128]  & - & 256 \\
Trunk  \#3  &$N_c=$2               & [128,128,128]  & - & 256 \\
\hline
\end{tabular}

\begin{tablenotes}
\footnotesize
\item[a] The length of the sizes array is the number of hidden layers.
\item[b] Group normalization using $N_{ch}$ channels is implemented on the output of the input layer. ``-" means no group normalization is considered. 
\end{tablenotes}
\vspace{-7pt}
\rule{\linewidth}{0.1pt}
\end{threeparttable}

\label{tab:excav_pifmionet_arch_3_trunk}
\end{table}

Another aspect that was tested during the training trials is incorporating the data loss term ($\mathcal{L}^{data}$), with its corresponding adaptive weight denoted by ($\lambda_{data}\equiv\lambda_d$), as introduced in Eq.~\eqref{eq:loss_function_detailed}. Notably, this is the only example where a data loss term is added to the physics-informed loss function. The collocation points used for enforcing the data term are sampled from the data zone depicted in Fig.~\ref{fig:excav_zones}b. 

Appendix~\ref{section:numerical_examples_appendix:excav:param} presents the detailed parametric results and descriptions of all loss terms. The best-performing configuration, a PIF-MIONet with three trunks, is adopted as the reference model. Its loss function combines PDE and data losses with boundary condition terms on $\Gamma^{base}$, while excluding those on $\Gamma^{comp}$. Training uses the Adam optimizer for $30000$ epochs at a constant learning rate of $1.0\times10^{-5}$.

\begin{figure}[hbt!]
\centering
\includegraphics[width=1.0\textwidth]{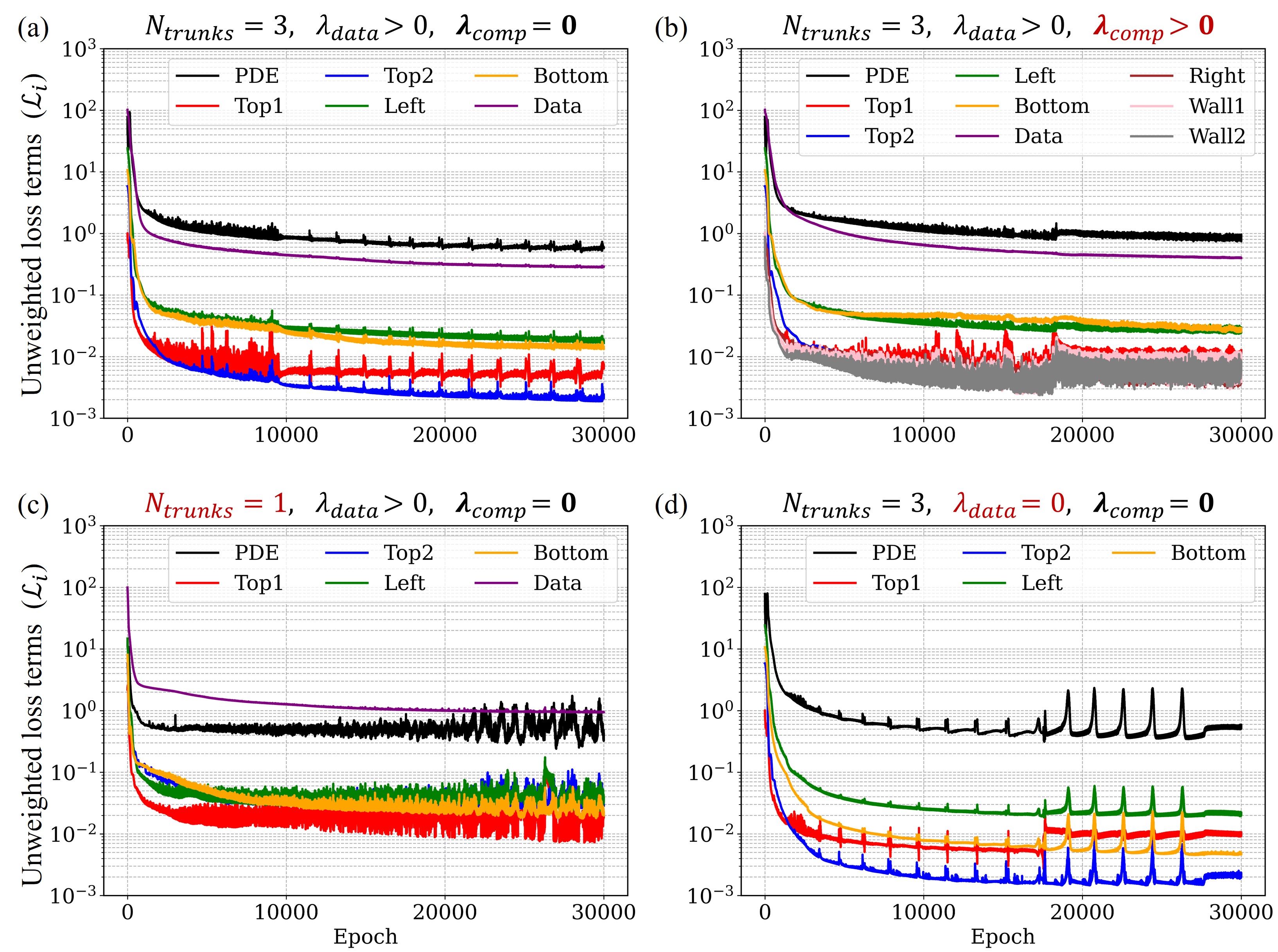}
\caption{Training unweighted loss terms $\mathcal{L}_i$ of the excavation problem under four training configurations: 
(a) A reference model of three trunks with
$\mathcal{L}^{data}$ ($\lambda_{data}>0$)
and without $\mathcal{L}^{comp}$
($\boldsymbol{\lambda}_{comp}=\boldsymbol{0}$),
(b) three trunks with both $\mathcal{L}^{data}$
($\lambda_{data}>0$) and $\mathcal{L}^{comp}$
($\boldsymbol{\lambda}_{comp}>\boldsymbol{0}$),
(c) one trunk with $\mathcal{L}^{data}$
($\lambda_{data}>0$) and without
$\mathcal{L}^{comp}$
($\boldsymbol{\lambda}_{comp}=\boldsymbol{0}$),
and (d) three trunks without either term
($\lambda_{data}=0$ and
$\boldsymbol{\lambda}_{comp}=\boldsymbol{0}$).
}
\label{fig:excav_terms_4_models}
\end{figure}

\begin{figure}[hbt!]
\centering
\includegraphics[width=1.0\textwidth]{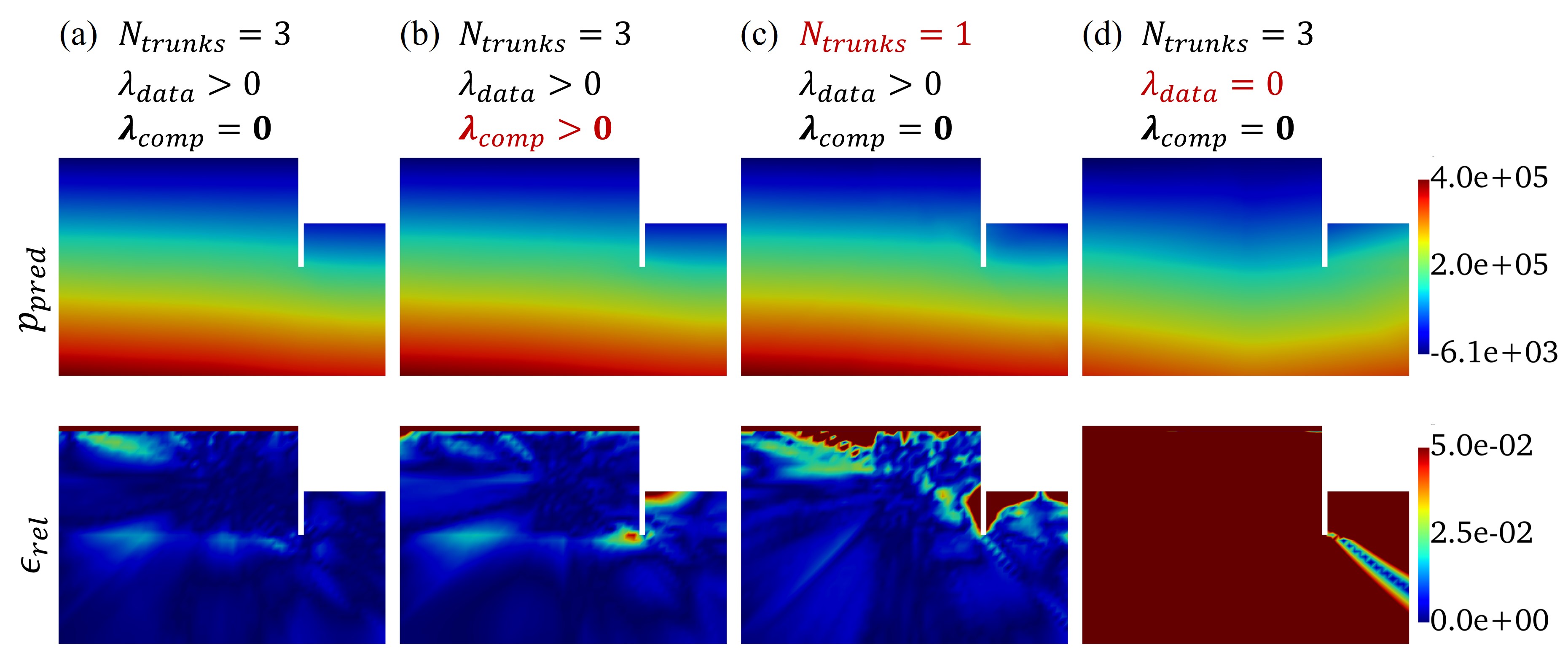}
\caption{Predicted pressure fields $p_{pred}$ (top row) and relative errors $\epsilon_{rel}$ (bottom row) for the 30th increment of the 9th testing load case of the excavation example. Models from left to right: 
(a) A reference model of three trunks with
$\mathcal{L}^{data}$ ($\lambda_{data}>0$)
and without $\mathcal{L}^{comp}$
($\boldsymbol{\lambda}_{comp}=\boldsymbol{0}$),
(b) three trunks with both $\mathcal{L}^{data}$
($\lambda_{data}>0$) and $\mathcal{L}^{comp}$
($\boldsymbol{\lambda}_{comp}>\boldsymbol{0}$),
(c) one trunk with $\mathcal{L}^{data}$
($\lambda_{data}>0$) and without
$\mathcal{L}^{comp}$
($\boldsymbol{\lambda}_{comp}=\boldsymbol{0}$),
and (d) three trunks without either term
($\lambda_{data}=0$ and
$\boldsymbol{\lambda}_{comp}=\boldsymbol{0}$).}
\label{fig:excav_ifenn_dealii_LC9_T30_different_models}
\end{figure}

The unweighted loss terms for the reference model and three alternative architectural and training configurations are compared in Fig.~\ref{fig:excav_terms_4_models}. Their IFENN-deal.II pressure predictions and relative errors for the 9th testing load case are presented in Fig.~\ref{fig:excav_ifenn_dealii_LC9_T30_different_models}, with the testing load histories shown in Fig.~\ref{fig:excav_flux_testing_lcs}. The following sections examine how different choices affect training and predictive performance.

\paragraph{Effect of including all boundary conditions:}
Including $\mathcal{L}^{comp}$ to enforce all boundary conditions produced oscillations in the training losses (Fig.~\ref{fig:excav_terms_4_models}b) more than the reference model (Fig.~\ref{fig:excav_terms_4_models}a). This configuration also yielded higher local errors near the additional boundaries, $\Gamma^{wall1}$ and $\Gamma^{wall2}$ (Fig.~\ref{fig:excav_ifenn_dealii_LC9_T30_different_models}b). 
Notably, excluding the boundary conditions associated with $\Gamma^{comp}$ is intended to improve optimization stability rather than modify the underlying physical problem.

\paragraph{Effect of the number of trunks:}
An additional model is trained with a single trunk comprising three hidden layers with $256$ neurons per layer, as detailed in the Appendix Table~\ref{tab:excav_pifmionet_arch_1_trunk}. The three-trunk and single-trunk models have $374848$ and $373696$ trainable parameters, respectively, ensuring comparable model sizes. Their final validation errors are $L_2^{all}=3.7\times10^{-3}$ and $L_2^{all}=1.2\times10^{-2}$, respectively.

The single-trunk model also exhibits larger oscillations in the PDE and boundary condition losses than the reference model (Fig.~\ref{fig:excav_terms_4_models}c). This behavior suggests greater difficulty representing the solution across the complex domain using a single set of basis functions. Pressure predictions for the single-trunk model also show higher errors near the wall and top boundaries (Fig.~\ref{fig:excav_ifenn_dealii_LC9_T30_different_models}c). These results highlight the benefit of the proposed PIF-MIONet architecture with domain decomposition, which achieves higher accuracy while maintaining a comparable total parameter count.

\paragraph{Effect of data loss on physics-informed training:}
Initial training experiments without the data loss term exhibited oscillatory convergence (Fig.~\ref{fig:excav_terms_4_models}d) and large prediction errors (Fig.~\ref{fig:excav_ifenn_dealii_LC9_T30_different_models}d). These difficulties may be associated with singular behavior near the sharp corner at the impermeable wall tip, where localized strain concentrations are observed (Appendix, Fig.~\ref{fig:excav_fem_strain_LC9_T30}). Such geometric singularities challenge physics-informed neural networks, motivating approaches such as response enrichment~\cite{hu_solving_2024} and adaptive collocation sampling~\cite{mao_physics-informed_2023,gao_failure-informed_2023}.

\begin{figure}[hbt!]
\centering
\includegraphics[width=1.0\textwidth]{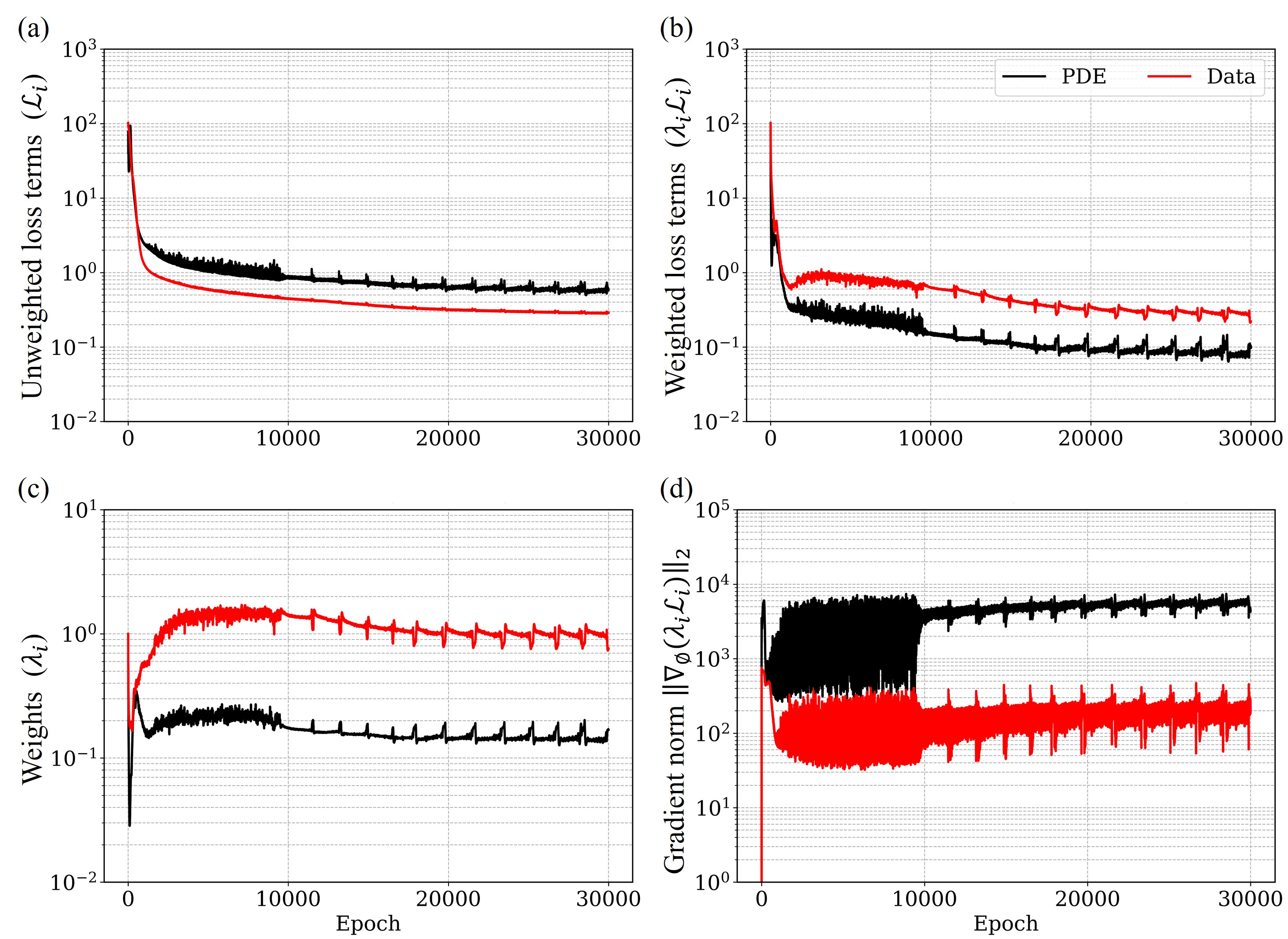}
\caption{Training metrics for the reference PIF-MIONet with three trunks for the excavation example: (a) unweighted loss terms $\mathcal{L}_i$, (b) weighted loss terms $\lambda_i\mathcal{L}_i$, (c) adaptive weights $\lambda_i$, and (d) weighted gradient norms $\|\nabla_{\phi}(\lambda_i\mathcal{L}_i)\|_2$. For clarity, only PDE and data terms are represented in this plot.}
\label{fig:excav_training_pde_data}
\end{figure}

In this study, we address these training difficulties for complex geometries by augmenting the loss function with a supplementary data term ($\mathcal{L}^{data}$). Notably, this is the only example in this study in which a supplementary data loss term is introduced to enhance training. Accordingly, we investigate three key questions: (1) how the data term affects training stability and prediction accuracy, (2) whether it dominates the optimization at the expense of the physics-informed constraints, and (3) how the resulting approach compares with a purely data-driven model in terms of data requirements. Comparing Figs.~\ref{fig:excav_terms_4_models}a and~\ref{fig:excav_terms_4_models}d, alongside Figs.~\ref{fig:excav_ifenn_dealii_LC9_T30_different_models}a and~\ref{fig:excav_ifenn_dealii_LC9_T30_different_models}d, demonstrates that incorporating the data term stabilizes training and substantially reduces prediction errors, addressing the first question. To address the second question, we isolate the PDE and data loss terms of the reference PIF-MIONet for comparison, as shown in Fig.~\ref{fig:excav_training_pde_data}.

Fig.~\ref{fig:excav_training_pde_data}a shows an overall decrease in both unweighted losses, indicating simultaneous improvements in PDE and data terms. Although the unweighted data loss is smaller than the PDE loss, this order reverses after weighting (Fig.~\ref{fig:excav_training_pde_data}b), reflecting the larger adaptive weight assigned to the data term by GradNorm (Fig.~\ref{fig:excav_training_pde_data}c). However, a larger contribution to the scalar loss does not necessarily imply a larger gradient contribution. Indeed, the weighted PDE gradient norms are approximately one to two orders of magnitude larger than those of the data term over most of training (Fig.~\ref{fig:excav_training_pde_data}d). Thus, despite its larger weighted loss, the data term does not dominate the PDE term in gradient magnitude during training backpropagation, and the PDE residual continues to decrease.

To address the third question, we compare the data requirements of the proposed approach with those of a purely data-driven model~\cite{amin_i-fenn_2026}. The proposed physics-informed, data-assisted approach utilizes 33 sparsely sampled, nonconsecutive time increments. In contrast, the data-driven model was trained using 28,800 time increments from continuous histories. Thus, the proposed approach utilizes approximately 0.1\% of the training increments used in the data-driven model. This highlights the potential of physics-informed training to substantially reduce data requirements. Future research could focus on optimizing the count and spatial distribution of data points. Additional parametric studies exploring alternative training configurations are presented in Appendix~\ref{section:numerical_examples_appendix:excav:param}.

\subsubsection{IFENN-deal.II using multiple subcycles}
\begin{figure}[hbt!]
\centering
\includegraphics[width=1.0\textwidth]{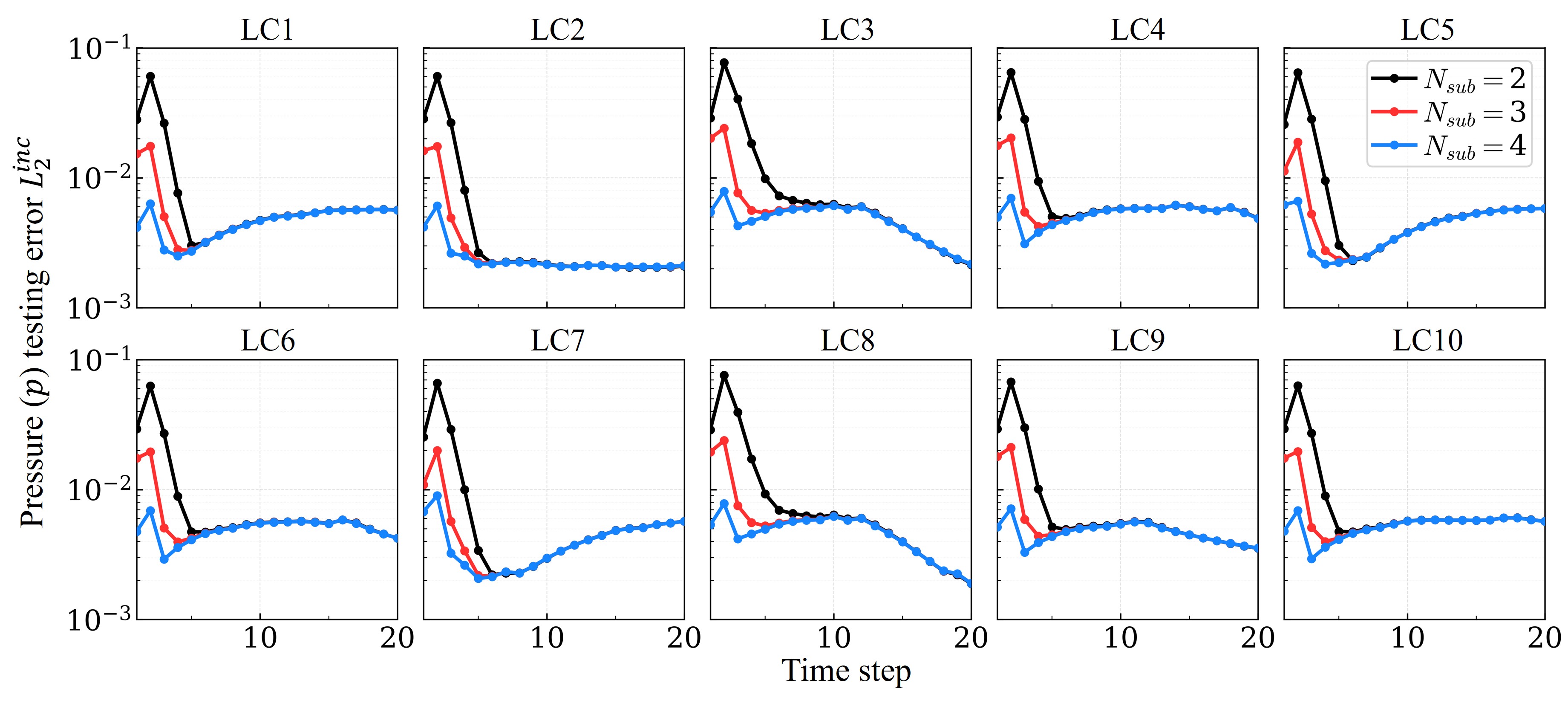}
\caption{Normalized testing error $L_2^{inc}$ of pressure $(p)$ for the first 20 increments of all testing load cases (LCs) using IFENN-deal.II with variations of predictor-multi-corrector subcycles ($N_{sub}=2,3,4$), for the excavation problem.}
\label{fig:excav_2_3_4_subcylces_p}
\end{figure}

Initial IFENN-deal.II results, using the trained three-trunk reference model, showed that using the predictor-multi-corrector approach with a total number of subcycles $N_{sub}=2$ resulted in high error magnitudes, especially at the initial time steps. Further investigation showed that increasing the number of subcycles ($N_{sub}$) can greatly reduce these errors. Fig.~\ref{fig:excav_2_3_4_subcylces_p} provides the pressure $L_2^{inc}$ error for the testing load cases (shown in Fig.~\ref{fig:excav_flux_testing_lcs}), utilizing different numbers of subcycles ($N_{sub}=2,3,4$). The error estimates for displacement ($u_z, u_x$) are depicted in Appendix~\ref{section:numerical_examples_appendix:excav:subcycles}, providing consistent behavior. 

In the predictor-multi-corrector scheme, the predictor subcycle uses known strain rate values from preceding increments, improving the initial strain-rate estimate. However, for the first time increment, the initial estimate of the strain rate trace is set to zero, as there is no prior knowledge from preceding increments. This is clearly reflected in the results shown in Fig.~\ref{fig:excav_2_3_4_subcylces_p}. Increasing the subcycle count ($N_{sub}$) improves predictions and can reduce the error $L_2^{inc}$ by an order of magnitude during approximately the first five increments. At later increments, two subcycles ($N_{sub}=2$) provide satisfactory accuracy, and no improvement occurs by increasing the subcycles, because the predictor–multi-corrector algorithm already starts with good estimates informed by preceding increments.

\begin{figure}[hbt!]
\centering
\includegraphics[width=1.0\textwidth]{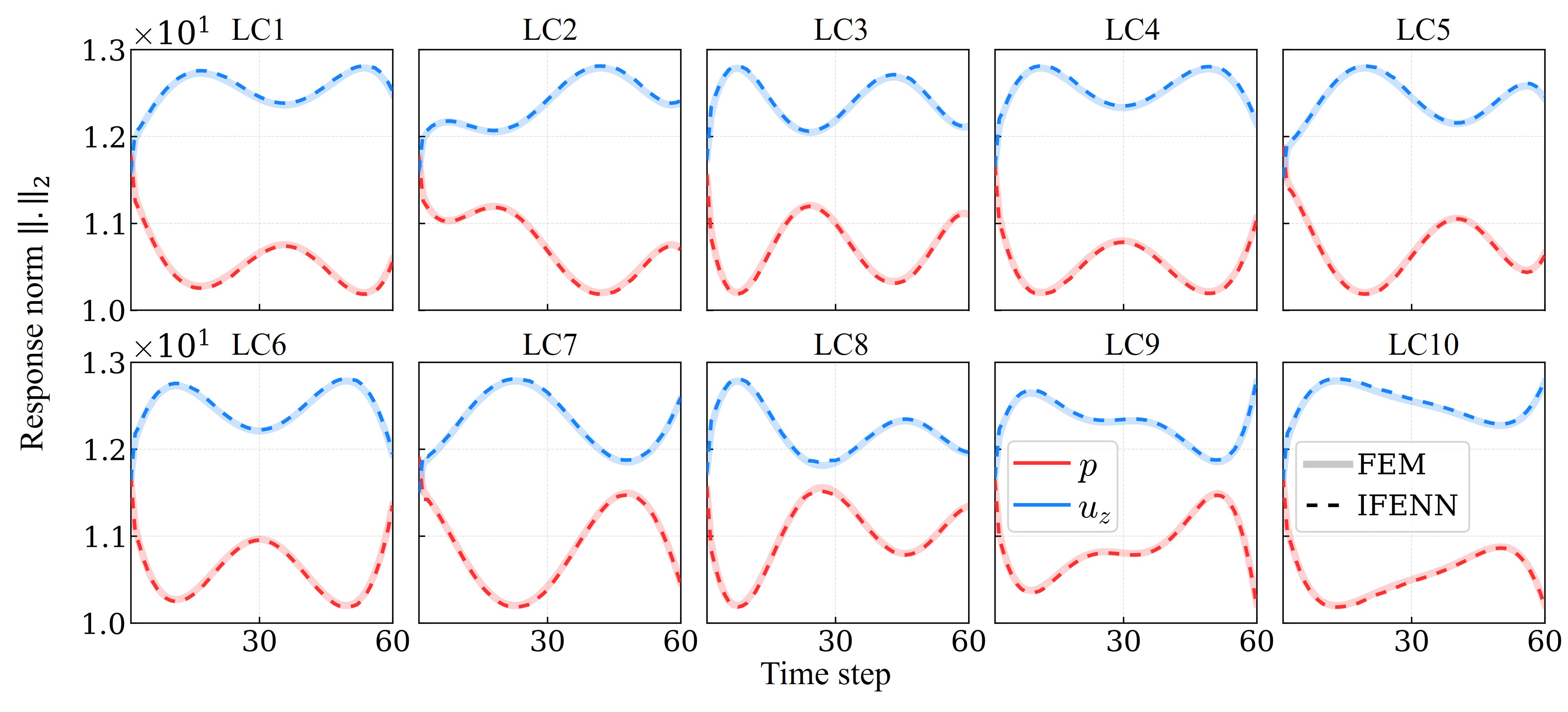}
\caption{Pressure norm ${\left\| p \right\|_2}$ and displacement norm ${\left\| u_z \right\|_2}$ obtained using the fully coupled FEM solver and IFENN-deal.II with four predictor-multi-corrector subcycles ($N_{sub}=4$), for all testing load cases (LCs) of the excavation problem. Response norm units are MPa and m for pressure and displacement, respectively.}
\label{fig:excav_ifenn_4SC_Norms}
\end{figure}

Using four subcycles ($N_{sub}=4$), the predicted pressure and vertical displacement norms closely match those obtained with the fully coupled FEM solver across all load cases and time increments, as shown in Fig.~\ref{fig:excav_ifenn_4SC_Norms}. Unlike the injection examples, the excavation problem exhibits opposing trends in the pressure and vertical displacement norms. As dewatering flux increases, the pressure norm decreases, while the displacement norm increases. The close agreement with the FEM results demonstrates that IFENN captures this distinct response, supporting its applicability to different poroelasticity problems. Additional results demonstrating prediction accuracy are provided in Appendix~\ref{section:numerical_examples_appendix:excav}.

\section{Discussion and conclusion}
\label{section:discussion_conclusion}
\subsection{Discussion}
\label{section:discussion_conclusion:discussion}
PIF-MIONet models were tested using IFENN-deal.II and IFENN-GEOS, yielding accurate, consistent predictions across all load cases for the 2D and 3D injection problems (Figs.~\ref{fig:inj_ifenn_dealii_2D_Norms}, \ref{fig:inj_2D_ifenn_dealii_vs_geos_1_5}, and \ref{fig:inj_ifenn_geos_3D_Norms}). These results demonstrate the portability of a single trained network across FEM solvers within IFENN. Innovative techniques were adopted to successfully introduce IFENN-GEOS for the first time. Further development could enable any trained physics-informed model to be plugged in as a solver within IFENN-GEOS without changes to the integration workflow. This would allow direct access to GEOS features for complex, industrial-scale applications.

The generalized PIF-MIONet was trained for variable permeability $K_p$ and time increments $\Delta t$ using predefined scaling factors to normalize pressure and strain fields per sample. Requiring predefined scaling factors for these fields remains a limitation of the current implementation. Nevertheless, this limitation could be addressed by an algorithm that automatically classifies input fields and selects suitable scaling factors. Alternatively, future work could explore encoding methods for inputs spanning multiple orders of magnitude to improve PIF-MIONet generalization.

The injection problem was purely physics-informed and required no data loss term, whereas the excavation problem required one to improve training accuracy under more complex boundary conditions. This used only $0.1\%$ of the labeled data increments employed in data-driven training~\cite{amin_i-fenn_2026}, highlighting the reduced data requirements of physics-informed training. The results also indicate that the data term regularizes training without dominating the optimization process. Future work could further optimize the size of the incorporated data or explore eliminating the need for it.

\subsection{Conclusion}
\label{section:discussion_conclusion:conclusion}
This work introduced PIF-MIONet for physics-informed learning in coupled multiphysics problems and demonstrated its integration into IFENN through 2D and 3D poroelasticity examples. The main contributions and findings are:

\begin{itemize}
\item The fusion design couples an arbitrary number of input fields, supporting extensions to additional interacting physics.

\item Dedicated subdomain trunks, through domain decomposition, improve accuracy in complex geometries while maintaining a comparable trainable parameter count.

\item Training on sparse, disconnected increments from varied load cases enables prediction of unseen histories, with relative errors below $5\%$ in non-trivial zones.

\item The injection examples are trained without a data loss term. For the excavation example with geometric singularities, incorporating a supplementary data loss improves training stability and accuracy while requiring approximately 0.1\% of the supervised training increments used by the previously reported data-driven approach. 

\item A single trained PIF-MIONet is portable across FEM solvers within IFENN and produces consistent predictions.

\item The first IFENN-GEOS integration extends IFENN to an industrial simulation environment, with predictions independently verified against deal.II FEM results.

\item IFENN extends its computational efficiency gains to poroelasticity, exhibiting favorable scaling relative to fully coupled FEM as domain size increases.

\end{itemize}

These findings support further development of IFENN for complex subsurface and multiphysics applications involving millions of degrees of freedom.

\section*{Acknowledgments}
This work was partially supported by the Sand Hazards and Opportunities for Resilience, Energy, and Sustainability (SHORES) Center, funded by Tamkeen under the NYUAD Research Institute Award CG013. The authors also acknowledge the support of the NYUAD Center for Research Computing, which provided resources, services, and staff expertise that contributed to this work.
Portions of this work were performed under the auspices of the U.S. Department of Energy by Lawrence Livermore National Laboratory under Contract DE-AC52-07NA27344.

\section*{Data Availability}
The data and source code supporting the findings of this study will be made publicly available upon publication of the article.

\appendix
\numberwithin{equation}{section}
\counterwithin{figure}{section}
\counterwithin{table}{section}

\makeatletter
\renewcommand{\theHsection}{\Alph{section}}
\makeatother

\section{Additional details on the PIF-MIONet training}
\label{section:appendix:pif_mionet_training}

\subsection{Adaptive weights}
\label{section:appendix:pif_mionet_training:adaptive}

Physics-informed training can suffer from numerical stiffness when high-gradient terms dominate the training process, adversely affecting convergence~\cite{wang_understanding_2021, haghighat_physics-informed_2022}. To overcome this numerical stiffness, researchers use adaptive weighting algorithms to dynamically balance the loss terms during training~\cite{bischof_multi-objective_2025, chen_gradnorm_2017, heydari_softadapt_2019, nath_amore_2026, wang_understanding_2021}. In this paper, we tested two adaptive weight algorithms that use fundamentally different strategies. The first adaptive weight algorithm, GradNorm, was introduced by Chen et al. ~\cite{chen_gradnorm_2017} and shows improved accuracy and reduced overfitting. In addition, it has been utilized in the literature for poroelasticity problems~\cite{haghighat_physics-informed_2022}. Before proceeding with the algorithm details, we list a group of preliminary definitions as follows:
\begin{align}
    \label{eq:grad_norm_per_term}
    G_i^\phi &= \left\| \nabla_\phi (\lambda_{i}  \mathcal{L}_i)  \right\|_2
    \\
    \label{eq:grad_norm_mean}
    \overline{G}^\phi &= E_{\forall i}[G_i^\phi]
    \\
    \label{eq:loss_ratio}
    \tilde{\mathcal{L}}_i &= \mathcal{L}_i/\mathcal{L}_i^0
    \\
    \label{eq:rel_inv_train_rate}
    r_i &= \tilde{\mathcal{L}}_i / E_{\forall i}[\tilde{\mathcal{L}}_i]
\end{align}
\noindent
where $G_i^\phi$ is the norm of the gradient of the weighted loss component $\lambda_{i}  \mathcal{L}_i$, with respect to the model parameters $\phi$. $\overline{G}^\phi$ is the average norm across all $G_i^\phi$ values, $\mathcal{L}_i^0$ is the initial loss value $\mathcal{L}_i$ at the first training epoch, $\tilde{\mathcal{L}}_i$ is the inverse training rate of term $i$, and $r_i$ is the relative inverse training rate. To save computational cost, the gradients' norm $(G_i^\phi)$ can be computed for a subset of the model weights~\cite{chen_gradnorm_2017}. However, because of the intricate architecture of the PIF-MIONet, we compute $G_i^\phi$ over all trainable parameters. 

In the first epoch, all the weights are set to unity. Then for each subsequent epoch $\mathcal{E}$, the new weight $\lambda_i^\mathcal{E}$ can be updated as follows~\cite{chen_gradnorm_2017,haghighat_physics-informed_2022}:
\begin{align}
    \label{eq:grad_norm_euler_update}
    \hat{\lambda}_i^\mathcal{E} &= \lambda_i^{\mathcal{E}-1} \dfrac{\overline{G}^\phi}{G_i^\phi} r_i^{\alpha_g}
    \\
    \label{eq:grad_norm_weight_update}
    \lambda_i^\mathcal{E} &= (1-\beta_g) \lambda_i^{\mathcal{E}-1} + \beta_g \hat{\lambda}_i^\mathcal{E}
\end{align}
\noindent
where $\alpha_g$ and $\beta_g$ are GradNorm tuning parameters that can manipulate the algorithm's behavior. $\alpha_g$ controls the strength of the penalty on terms with a slower learning rate. $\beta_g$ is a smoothing factor $(0<\beta_g\leq1)$ that prevents high fluctuations in the updated weights. Finally, it is important to note that after computing the new weights using Eq.~\eqref{eq:grad_norm_weight_update}, the weights are normalized to have a mean of unity, to decouple the adaptive weights algorithm from the global learning rate of the optimizer~\cite{chen_gradnorm_2017}.  

The second adaptive weights algorithm considered in this work is Relative Loss Balancing with Random Lookback (ReLoBRaLo), introduced by Bischof and Kraus~\cite{bischof_multi-objective_2025}. ReLoBRaLo avoids using the gradient statistics, unlike GradNorm, and only tracks loss statistics for adaptive weight computation. The algorithm depends on two functions that can be defined as:
\begin{align}
    \label{eq:relobralo_bal_method}
    \lambda_i^{bal}(\mathcal{E}, \mathcal{E}^\prime) &= N_{terms} \dfrac{\text{exp}\bigl(\dfrac{\mathcal{L}_i^\mathcal{E}}{\theta_r \mathcal{L}_i^{\mathcal{E}^\prime}}\bigr)}{\sum_{j=1}^{N_{terms}}\text{exp}\bigl(\dfrac{\mathcal{L}_j^\mathcal{E}}{\theta_r \mathcal{L}_j^{\mathcal{E}^\prime}}\bigr)}
    \\
    \label{eq:relobralo_hist_method}
    \lambda_i^{hist}(\mathcal{E}) &= \rho_r \lambda_i^{\mathcal{E}-1} + (1-\rho_r) \lambda_i^{bal}(\mathcal{E}, 0)
    \end{align}
\noindent
where $\mathcal{L}_i^\mathcal{E}$ is the loss of term $i$ at epoch $\mathcal{E}$, $\theta_r$ is a temperature parameter that controls the magnitudes of the scalings, and $\rho_r$ is a Bernoulli random variable $(\rho_r \in \{0,1\})$, with an expected value $E[\rho_r]$ close to unity. 
Utilizing the previous functions, the new ReLoBRaLo weight $\lambda_i^\mathcal{E}$ can be computed as: 
\begin{align}
    \label{eq:relobralo_update}
    \lambda_i^\mathcal{E} &=  \alpha_r \lambda_i^{hist}(\mathcal{E}) + (1-\alpha_r) \lambda_i^{bal}(\mathcal{E}, \mathcal{E}-1)
\end{align}
\noindent
where $\alpha_r$ is the exponential decay rate controlling the relative contribution of previous weights and the new balancing weights.

\subsection{Model stabilization}
\label{section:appendix:pif_mionet_training:stabilization}
\begin{figure}[hbt!]
\centering
\includegraphics[width=0.5\textwidth]{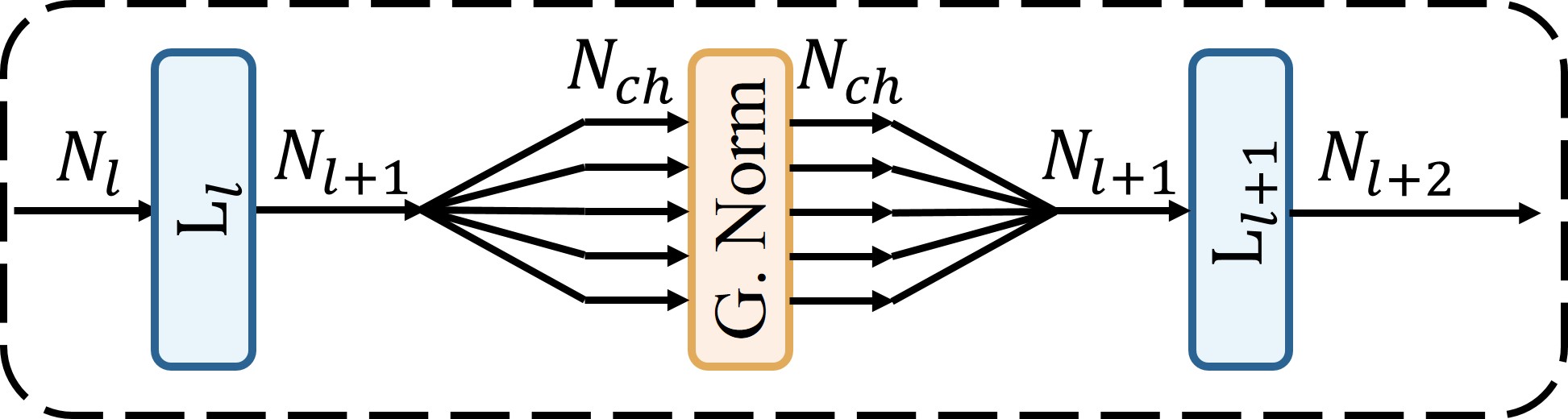}
\caption{A schematic representation of the stabilization approach, implemented between two linear feedforward layers. A layer $L_l$ has an input size $N_l$ and an output size $O_l=N_{l+1}$.}
\label{fig:methods_pifmionet_stabilization}
\end{figure}

As depicted in Fig.~\ref{fig:methods_pifmionet_stabilization}, group normalization between two layers ($L_l$ and $L_{l+1}$) is done by splitting the output vector of size $N_{l+1}$ into $N_{ch}$ channels, where group normalization is done for each channel, and then the outputs are reshaped again into a unified vector of shape $N_{l+1}$. 

\subsection{Shape-function derivatives for FEM biquadratic elements}
\label{section:appendix:pif_mionet_training:sf_derivatives}

For a biquadratic quadrilateral, with 9 nodes, the shape functions and their derivatives in natural coordinates are listed in Table~\ref{tab:q9_sf_derivatives}. Instead of numbering perimeter or corner nodes first, a lexicographic ordering of nodes is used to enable efficient vectorized tensor-product operations.

\begin{table}[htbp]
\centering
\caption{Shape functions and their natural-coordinate derivatives.}
\label{tab:q9_sf_derivatives}
\renewcommand{\arraystretch}{1.25}
\setlength{\tabcolsep}{4pt}
\small

\resizebox{\linewidth}{!}{%
\begin{tabular}{c *{6}{r@{}l}}
\toprule
\textbf{Node}
& \multicolumn{2}{c}{\(N_n\)}
& \multicolumn{2}{c}{\(\partial N_n/\partial\xi\)}
& \multicolumn{2}{c}{\(\partial N_n/\partial\eta\)}
& \multicolumn{2}{c}{\(\partial^2N_n/\partial\xi^2\)}
& \multicolumn{2}{c}{\(\partial^2N_n/\partial\eta^2\)}
& \multicolumn{2}{c}{\(\partial^2N_n/\partial\xi\partial\eta\)}
\\
\midrule

1
& \(\xi\eta(\xi-1)(\eta-1)\) & \(/4\)
& \(\eta(2\xi-1)(\eta-1)\) & \(/4\)
& \(\xi(\xi-1)(2\eta-1)\) & \(/4\)
& \(\eta(\eta-1)\) & \(/2\)
& \(\xi(\xi-1)\) & \(/2\)
& \((2\xi-1)(2\eta-1)\) & \(/4\)
\\

2
& \(-\xi(\xi-1)(\eta^2-1)\) & \(/2\)
& \(-(2\xi-1)(\eta^2-1)\) & \(/2\)
& \(-\xi\eta(\xi-1)\) & {}
& \(1-\eta^2\) & {}
& \(-\xi(\xi-1)\) & {}
& \(-\eta(2\xi-1)\) & {}
\\

3
& \(\xi\eta(\xi-1)(\eta+1)\) & \(/4\)
& \(\eta(2\xi-1)(\eta+1)\) & \(/4\)
& \(\xi(\xi-1)(2\eta+1)\) & \(/4\)
& \(\eta(\eta+1)\) & \(/2\)
& \(\xi(\xi-1)\) & \(/2\)
& \((2\xi-1)(2\eta+1)\) & \(/4\)
\\

4
& \(-\eta(\xi^2-1)(\eta-1)\) & \(/2\)
& \(-\xi\eta(\eta-1)\) & {}
& \(-(\xi^2-1)(2\eta-1)\) & \(/2\)
& \(-\eta(\eta-1)\) & {}
& \(1-\xi^2\) & {}
& \(-\xi(2\eta-1)\) & {}
\\

5
& \((\xi^2-1)(\eta^2-1)\) & {}
& \(2\xi(\eta^2-1)\) & {}
& \(2\eta(\xi^2-1)\) & {}
& \(2(\eta^2-1)\) & {}
& \(2(\xi^2-1)\) & {}
& \(4\xi\eta\) & {}
\\

6
& \(-\eta(\xi^2-1)(\eta+1)\) & \(/2\)
& \(-\xi\eta(\eta+1)\) & {}
& \(-(\xi^2-1)(2\eta+1)\) & \(/2\)
& \(-\eta(\eta+1)\) & {}
& \(1-\xi^2\) & {}
& \(-\xi(2\eta+1)\) & {}
\\

7
& \(\xi\eta(\xi+1)(\eta-1)\) & \(/4\)
& \(\eta(2\xi+1)(\eta-1)\) & \(/4\)
& \(\xi(\xi+1)(2\eta-1)\) & \(/4\)
& \(\eta(\eta-1)\) & \(/2\)
& \(\xi(\xi+1)\) & \(/2\)
& \((2\xi+1)(2\eta-1)\) & \(/4\)
\\

8
& \(-\xi(\xi+1)(\eta^2-1)\) & \(/2\)
& \(-(2\xi+1)(\eta^2-1)\) & \(/2\)
& \(-\xi\eta(\xi+1)\) & {}
& \(1-\eta^2\) & {}
& \(-\xi(\xi+1)\) & {}
& \(-\eta(2\xi+1)\) & {}
\\

9
& \(\xi\eta(\xi+1)(\eta+1)\) & \(/4\)
& \(\eta(2\xi+1)(\eta+1)\) & \(/4\)
& \(\xi(\xi+1)(2\eta+1)\) & \(/4\)
& \(\eta(\eta+1)\) & \(/2\)
& \(\xi(\xi+1)\) & \(/2\)
& \((2\xi+1)(2\eta+1)\) & \(/4\)
\\

\bottomrule
\end{tabular}%
}
\end{table}

Assume we have a function defined in the natural coordinates $f\left(\xi ,\eta \right)$. The derivatives of the variable $f$ at the isoparametric natural coordinates $\xi$ and $\eta$ can be computed as:
\begin{align}
    \label{eq:natural_derivative_first}
    \frac{\partial f\left(\xi ,\eta \right)}{\partial \xi }=\sum_{n=1}^{9} \frac{\partial N_n\left(\xi ,\eta \right)}{\partial \xi }\,f_{n}
    \\
    \label{eq:natural_derivative_second}
    \frac{{\partial }^{2}f\left(\xi ,\eta \right)}{\partial {\xi }\partial {\eta }}=\sum_{n=1}^{9} \frac{{\partial }^{2}N_n\left(\xi ,\eta \right)}{\partial {\xi }\partial {\eta }}\,f_{n}
\end{align}
\noindent
where $N_n$ and $f_n$ are the shape function and the variable $f$ value at the isoparametric node $n$, respectively. ${\partial f\left(\xi ,\eta \right)}/{\partial \eta }$ can be computed following the same approach in Eq.~\eqref{eq:natural_derivative_first}. Similarly, ${\partial^2 f\left(\xi ,\eta \right)}/{\partial \xi^2 }$ and ${\partial^2 f\left(\xi ,\eta \right)}/{\partial \eta^2 }$ can be computed following the same approach in Eq.~\eqref{eq:natural_derivative_second}.

Now, we derive the derivatives in the physical coordinates $x$ and $y$.  Applying the chain rule to ${\partial f\left(\xi ,\eta \right)}/{\partial \xi }$ and ${\partial f\left(\xi ,\eta \right)}/{\partial \eta }$, we can write the following expressions:
\begin{align}
    \label{eq:df_dxi_chain}
    \frac{\partial f}{\partial \xi}
    =
    \frac{\partial f}{\partial x}
    \frac{\partial x}{\partial \xi}
    +
    \frac{\partial f}{\partial y}
    \frac{\partial y}{\partial \xi}
    \\
    \label{eq:df_deta_chain}
    \frac{\partial f}{\partial \eta}
    =
    \frac{\partial f}{\partial x}
    \frac{\partial x}{\partial \eta}
    +
    \frac{\partial f}{\partial y}
    \frac{\partial y}{\partial \eta}
\end{align}

These two expressions (Eq.~\eqref{eq:df_dxi_chain} and Eq.~\eqref{eq:df_deta_chain}) can be combined into a matrix form as follows:
\begin{align}
    \label{eq:df_dx_matrix}
    \begin{Bmatrix}
    \frac{\partial f}{\partial \xi } 
    \\[8pt] %
    \frac{\partial f}{\partial \eta }
    \end{Bmatrix}
    &=
    \begin{bmatrix}
    \frac{\partial x}{\partial \xi } & \frac{\partial y}{\partial \xi } 
    \\[8pt] %
    \frac{\partial x}{\partial \eta } & \frac{\partial y}{\partial \eta }
    \end{bmatrix}
    \begin{Bmatrix}
    \frac{\partial f}{\partial x} 
    \\[8pt] %
    \frac{\partial f}{\partial y}
    \end{Bmatrix}
\end{align}

For second-order derivatives, in the physical coordinates $x$ and $y$, we differentiate Eq.~\eqref{eq:df_dxi_chain} with respect to $\xi$, as follows: 
\begin{align}
    \label{eq:d2f_dxi2_chain}
    \frac{\partial^2 f}{\partial \xi^2}
    &=
    \frac{\partial}{\partial \xi}
    \left(
        \frac{\partial f}{\partial \xi}
    \right)
    =
    \frac{\partial}{\partial \xi}
    \left(
        \frac{\partial f}{\partial x}
        \frac{\partial x}{\partial \xi}
        +
        \frac{\partial f}{\partial y}
        \frac{\partial y}{\partial \xi}
    \right)
    \notag\\
    &=
    \frac{\partial}{\partial \xi}
    \left(\frac{\partial f}{\partial x}\right)
    \frac{\partial x}{\partial \xi}
    +
    \frac{\partial f}{\partial x}
    \frac{\partial^2 x}{\partial \xi^2}
    +
    \frac{\partial}{\partial \xi}
    \left(\frac{\partial f}{\partial y}\right)
    \frac{\partial y}{\partial \xi}
    +
    \frac{\partial f}{\partial y}
    \frac{\partial^2 y}{\partial \xi^2}
\end{align}

The two derivatives $\frac{\partial}{\partial \xi} \left(\frac{\partial f}{\partial x}\right)$ and $ \frac{\partial}{\partial \xi} \left(\frac{\partial f}{\partial y}\right)$ in Eq.~\eqref{eq:d2f_dxi2_chain} can be computed, using the chain rule as follows:
\begin{align}
    \label{eq:df_dx_dxi}
    \frac{\partial}{\partial \xi}
    \left(\frac{\partial f}{\partial x}\right)
    &=
    \frac{\partial^2 f}{\partial x^2}
    \frac{\partial x}{\partial \xi}
    +
    \frac{\partial^2 f}{\partial x\partial y}
    \frac{\partial y}{\partial \xi},
    \\
    \label{eq:df_dy_dxi}
    \frac{\partial}{\partial \xi}
    \left(\frac{\partial f}{\partial y}\right)
    &=
    \frac{\partial^2 f}{\partial x\partial y}
    \frac{\partial x}{\partial \xi}
    +
    \frac{\partial^2 f}{\partial y^2}
    \frac{\partial y}{\partial \xi}
\end{align}

Substituting Eq.~\eqref{eq:df_dx_dxi} and Eq.~\eqref{eq:df_dy_dxi} into Eq.~\eqref{eq:d2f_dxi2_chain}, the second-order derivative term can be expressed as:

\begin{align}
    \label{eq:d2f_dxi2_chain2}
    \frac{\partial^2 f}{\partial \xi^2}
    &=
    \frac{\partial^2 f}{\partial x^2}
    \left(\frac{\partial x}{\partial \xi}\right)^2
    +
    \frac{\partial^2 f}{\partial y^2}
    \left(\frac{\partial y}{\partial \xi}\right)^2
    \notag\\
    &\quad+
    2\frac{\partial^2 f}{\partial x\partial y}
    \frac{\partial x}{\partial \xi}
    \frac{\partial y}{\partial \xi}
    +
    \frac{\partial f}{\partial x}
    \frac{\partial^2x}{\partial \xi^2}
    +
    \frac{\partial f}{\partial y}
    \frac{\partial^2y}{\partial \xi^2}.
\end{align}

Isolating the second order derivatives of $f$ with respect to physical coordinates ($x$ and $y$), we get:

\begin{align}
    \label{eq:d2f_dxi2_isolate}
    \frac{\partial^2 f}{\partial x^2}
    \left(\frac{\partial x}{\partial \xi}\right)^2
    +
    \frac{\partial^2 f}{\partial y^2}
    \left(\frac{\partial y}{\partial \xi}\right)^2
    +
    \frac{\partial^2 f}{\partial x\partial y}
    \left(2\frac{\partial x}{\partial \xi}
    \frac{\partial y}{\partial \xi}\right)
    = 
    \frac{\partial^2 f}{\partial \xi^2}
    -
    \frac{\partial f}{\partial x}
    \frac{\partial^2x}{\partial \xi^2}
    -
    \frac{\partial f}{\partial y}
    \frac{\partial^2y}{\partial \xi^2}
\end{align}

Following the same approach for $\frac{\partial^2 f}{\partial \eta^2}$ and $\frac{\partial^2 f}{\partial \xi \partial \eta}$, the isolated  second order terms with respect to physical coordinates can be expressed as follows 
\begin{align}
    \label{eq:d2f_deta2_isolate}
    \frac{\partial^2 f}{\partial x^2}
    \left(\frac{\partial x}{\partial \eta}\right)^2
    +
    \frac{\partial^2 f}{\partial y^2}
    \left(\frac{\partial y}{\partial \eta}\right)^2
    +
    \frac{\partial^2 f}{\partial x\partial y}
    \left(2\frac{\partial x}{\partial \eta}
    \frac{\partial y}{\partial \eta}\right)
    &= 
    \frac{\partial^2 f}{\partial \eta^2}
    -
    \frac{\partial f}{\partial x}
    \frac{\partial^2x}{\partial \eta^2}
    -
    \frac{\partial f}{\partial y}
    \frac{\partial^2y}{\partial \eta^2}
    \\[8pt] %
    \label{eq:d2f_dxideta_isolate}
    \frac{\partial^2 f}{\partial x^2}
    \frac{\partial x}{\partial\xi}
    \frac{\partial x}{\partial\eta}
    +
    \frac{\partial^2 f}{\partial y^2}
    \frac{\partial y}{\partial\xi}
    \frac{\partial y}{\partial\eta}
    +
    \frac{\partial^2 f}{\partial x\partial y}
    \left(
        \frac{\partial x}{\partial\xi}
        \frac{\partial y}{\partial\eta}
        +
        \frac{\partial y}{\partial\xi}
        \frac{\partial x}{\partial\eta}
    \right)
    &=
    \frac{\partial^2 f}{\partial\xi\partial\eta}
    -
    \frac{\partial f}{\partial x}
    \frac{\partial^2x}{\partial\xi\partial\eta}
    -
    \frac{\partial f}{\partial y}
    \frac{\partial^2y}{\partial\xi\partial\eta}
\end{align}

The three equations Eq.~\eqref{eq:d2f_dxi2_isolate}, Eq.~\eqref{eq:d2f_deta2_isolate}, and Eq.~\eqref{eq:d2f_dxideta_isolate} can be assembled in a matrix form as follows:

\begin{align}
\label{eq:d2f_dx2_matrix}
\begin{bmatrix}{
\left(\frac{\partial x}{\partial \xi }\right)}^{2} & {\left(\frac{\partial y}{\partial \xi }\right)}^{2} & 2\frac{\partial x}{\partial \xi }\frac{\partial y}{\partial \xi } 
\\[8pt] %
{\left(\frac{\partial x}{\partial \eta }\right)}^{2} & {\left(\frac{\partial y}{\partial \eta }\right)}^{2} & 2\frac{\partial x}{\partial \eta }\frac{\partial y}{\partial \eta } 
\\[8pt] %
\frac{\partial x}{\partial \xi }\frac{\partial x}{\partial \eta } & \frac{\partial y}{\partial \xi }\frac{\partial y}{\partial \eta } & \frac{\partial x}{\partial \xi }\frac{\partial y}{\partial \eta }+\frac{\partial x}{\partial \eta }\frac{\partial y}{\partial \xi }
\end{bmatrix}
\begin{Bmatrix}
\frac{{\partial }^{2}f}{\partial x^{2}} 
\\[8pt] %
\frac{{\partial }^{2}f}{\partial y^{2}} 
\\[8pt] %
\frac{{\partial }^{2}f}{\partial x\partial y}
\end{Bmatrix}
=
\begin{Bmatrix}
\frac{{\partial }^{2}f}{\partial {\xi }^{2}}-\frac{\partial f}{\partial x}\frac{{\partial }^{2}x}{\partial {\xi }^{2}}-\frac{\partial f}{\partial y}\frac{{\partial }^{2}y}{\partial {\xi }^{2}} 
\\[8pt] %
\frac{{\partial }^{2}f}{\partial {\eta }^{2}}-\frac{\partial f}{\partial x}\frac{{\partial }^{2}x}{\partial {\eta }^{2}}-\frac{\partial f}{\partial y}\frac{{\partial }^{2}y}{\partial {\eta }^{2}} 
\\[8pt] %
\frac{{\partial }^{2}f}{\partial \xi \partial \eta }-\frac{\partial f}{\partial x}\frac{{\partial }^{2}x}{\partial \xi \partial \eta }-\frac{\partial f}{\partial y}\frac{{\partial }^{2}y}{\partial \xi \partial \eta }
\end{Bmatrix}
\end{align}

For more efficient calculations, the first-order derivatives $\frac{\partial f}{\partial x}$ and $\frac{\partial f}{\partial y}$ on the right hand side of Eq.~\eqref{eq:d2f_dx2_matrix} can be isolated as follows:

\begin{align}
\label{eq:d2f_dx2_matrix_2}
\begin{bmatrix}{
\left(\frac{\partial x}{\partial \xi }\right)}^{2} & {\left(\frac{\partial y}{\partial \xi }\right)}^{2} & 2\frac{\partial x}{\partial \xi }\frac{\partial y}{\partial \xi } 
\\[8pt] %
{\left(\frac{\partial x}{\partial \eta }\right)}^{2} & {\left(\frac{\partial y}{\partial \eta }\right)}^{2} & 2\frac{\partial x}{\partial \eta }\frac{\partial y}{\partial \eta } 
\\[8pt] %
\frac{\partial x}{\partial \xi }\frac{\partial x}{\partial \eta } & \frac{\partial y}{\partial \xi }\frac{\partial y}{\partial \eta } & \frac{\partial x}{\partial \xi }\frac{\partial y}{\partial \eta }+\frac{\partial x}{\partial \eta }\frac{\partial y}{\partial \xi }
\end{bmatrix}
\begin{Bmatrix}
\frac{{\partial }^{2}f}{\partial x^{2}} 
\\[8pt] %
\frac{{\partial }^{2}f}{\partial y^{2}} 
\\[8pt] %
\frac{{\partial }^{2}f}{\partial x\partial y}
\end{Bmatrix}
=
\begin{Bmatrix}
\frac{{\partial }^{2}f}{\partial {\xi }^{2}} 
\\[8pt] %
\frac{{\partial }^{2}f}{\partial {\eta }^{2}} 
\\[8pt] %
\frac{{\partial }^{2}f}{\partial \xi \partial \eta }
\end{Bmatrix}
-
\begin{bmatrix}
\frac{{\partial }^{2}x}{\partial {\xi }^{2}} & \frac{{\partial }^{2}y}{\partial {\xi }^{2}} 
\\[8pt] %
\frac{{\partial }^{2}x}{\partial {\eta }^{2}} & \frac{{\partial }^{2}y}{\partial {\eta }^{2}} 
\\[8pt] %
\frac{{\partial }^{2}x}{\partial \xi \partial \eta } & \frac{{\partial }^{2}y}{\partial \xi \partial \eta }
\end{bmatrix}
\begin{Bmatrix}
\frac{\partial f}{\partial x} 
\\[8pt] %
\frac{\partial f}{\partial y}
\end{Bmatrix}
\end{align}

Given the previous definitions, a straightforward implementation would compute $f_n$ at the element nodes, then evaluate the function derivatives at natural coordinates using the definitions in Table~\ref{tab:q9_sf_derivatives}, Eq.~\eqref{eq:natural_derivative_first}, and Eq.~\eqref{eq:natural_derivative_second}. After evaluating the derivatives with respect to the natural coordinates, the derivatives with respect to the physical coordinates can be computed utilizing the inverse of the system matrices in Eq.~\eqref{eq:df_dx_matrix} and Eq.~\eqref{eq:d2f_dx2_matrix_2}, as shown in Eq.~\eqref{eq:physical_derivative_first} and Eq.~\eqref{eq:physical_derivative_second}, respectively.

\begin{align}
    \label{eq:physical_derivative_first}
    \begin{Bmatrix}
    \frac{\partial f}{\partial x} 
    \\[8pt] %
    \frac{\partial f}{\partial y}
    \end{Bmatrix}
    &=
    \begin{bmatrix}
    \frac{\partial x}{\partial \xi } & \frac{\partial y}{\partial \xi } 
    \\[8pt] %
    \frac{\partial x}{\partial \eta } & \frac{\partial y}{\partial \eta }
    \end{bmatrix}^{-1}
    \begin{Bmatrix}
    \frac{\partial f}{\partial \xi } 
    \\[8pt] %
    \frac{\partial f}{\partial \eta }
    \end{Bmatrix}
    \\
    \label{eq:physical_derivative_second}
    \begin{Bmatrix}
    \frac{{\partial }^{2}f}{\partial x^{2}} 
    \\[8pt] %
    \frac{{\partial }^{2}f}{\partial y^{2}} 
    \\[8pt] %
    \frac{{\partial }^{2}f}{\partial x\partial y}
    \end{Bmatrix}
    &=
    \begin{bmatrix}{
    \left(\frac{\partial x}{\partial \xi }\right)}^{2} & {\left(\frac{\partial y}{\partial \xi }\right)}^{2} & 2\frac{\partial x}{\partial \xi }\frac{\partial y}{\partial \xi } 
    \\[8pt] %
    {\left(\frac{\partial x}{\partial \eta }\right)}^{2} & {\left(\frac{\partial y}{\partial \eta }\right)}^{2} & 2\frac{\partial x}{\partial \eta }\frac{\partial y}{\partial \eta } 
    \\[8pt] %
    \frac{\partial x}{\partial \xi }\frac{\partial x}{\partial \eta } & \frac{\partial y}{\partial \xi }\frac{\partial y}{\partial \eta } & \frac{\partial x}{\partial \xi }\frac{\partial y}{\partial \eta }+\frac{\partial x}{\partial \eta }\frac{\partial y}{\partial \xi }
    \end{bmatrix}^{-1}
    \begin{Bmatrix}
    \begin{Bmatrix}
    \frac{{\partial }^{2}f}{\partial {\xi }^{2}} 
    \\[8pt] %
    \frac{{\partial }^{2}f}{\partial {\eta }^{2}} 
    \\[8pt] %
    \frac{{\partial }^{2}f}{\partial \xi \partial \eta }
    \end{Bmatrix}
    -
    \begin{bmatrix}
    \frac{{\partial }^{2}x}{\partial {\xi }^{2}} & \frac{{\partial }^{2}y}{\partial {\xi }^{2}} 
    \\[8pt] %
    \frac{{\partial }^{2}x}{\partial {\eta }^{2}} & \frac{{\partial }^{2}y}{\partial {\eta }^{2}} 
    \\[8pt] %
    \frac{{\partial }^{2}x}{\partial \xi \partial \eta } & \frac{{\partial }^{2}y}{\partial \xi \partial \eta }
    \end{bmatrix}
    \begin{Bmatrix}
    \frac{\partial f}{\partial x} 
    \\[8pt] %
    \frac{\partial f}{\partial y}
    \end{Bmatrix}
    \end{Bmatrix}
\end{align}

\section{Additional results and parametric studies}
\label{section:numerical_examples_appendix}

\subsection{2D injection example with constant \texorpdfstring{permeability $K_p$ and time increment $\Delta t$}{permeability and time increment}}
\label{section:numerical_examples_appendix:inj_flux_2d}

\subsubsection{Impact of using adaptive weights and dimensionless forms and other hyperparameters}
\label{section:numerical_examples_appendix:inj_flux_2d:hyperparameters}

We start our parametric study by listing the different techniques explored in this study and their options, as shown in Table~\ref{tab:inj_param_abb}. For example, if a method is referred to as ``SF-GN-DS-L2", this means it is using shape-function derivatives for spatial gradients, GradNorm for adaptive weights, dimensionless scaling, and L2 norm for the residual terms. Different tuning-parameter variations for GradNorm and ReLoBRaLo are tested. For this test, only 3 sampled increments are used for training, and the $L_2^{all}$ error metric for these increments is computed after training, as listed in Table~\ref{tab:inj_3inc_GN_param} and Table~\ref{tab:inj_3inc_RB_param} for ``SF-GN-DS-L2" and ``SF-RB-DS-L2", respectively. It should be noted that the tested values for the tuning parameters ($\alpha_g$, $\beta_g$, $\alpha_r$, $\theta_r$, and $\rho_r$) are based on recommended values in the literature~\cite{chen_gradnorm_2017, bischof_multi-objective_2025}, and some exploration trials. 

\begin{table}[hbt!]
\renewcommand{\arraystretch}{1.2}
\centering
\caption{Different techniques explored in PI ML and their options and abbreviations}

\begin{threeparttable}
\begin{tabular}{ccc}
\hline
ML Technique & Options & Abbreviations  \\ 
\hline
Spatial gradients calculation &&\\ 
& Automatic differentiation  & AD \\
& Shape-function derivatives & SF \\
Updating weights for loss terms&&\\ 
& GradNorm                  & GN \\ 
& ReLoBRaLo                 & RB \\
& Uniform weights \tnote{a} & UW \\
Scaling loss residuals&&\\ 
& Dimensionless scaling      & DS \\ 
& Custom scaling \tnote{b}   & CS \\ 
& Uniform scaling \tnote{a}  & US \\ 
Residual norm computation &&\\ 
& L2 Norm                   & L2 \\ 
& Mean squared value Norm   & MS \\ 
\hline
\end{tabular}

\begin{tablenotes}
\footnotesize
\item[a] Constant values of unity.
\item[b] Scales are computed to make the initial loss terms almost unity, regardless of their physical units.
\end{tablenotes}
\vspace{-7pt}
\rule{\linewidth}{0.1pt}
\end{threeparttable}
\label{tab:inj_param_abb}
\end{table}

\begin{table}[hbt!]
\renewcommand{\arraystretch}{1.2}
\centering
\caption{The $L_2^{all}$ error metric for different GradNorm settings, used to train 3 sampled increments, using SF-GN-DS-L2}
\begin{tabular}{cccc}
\hline
Trial name & $\alpha_g$ & $\beta_g$  & $L_2^{all}$   \\ 
\hline
GN1 & $1.0$ & $0.01$ & $6.1 \times 10^{-3}$ \\
GN2 & $0.1$ & $0.01$ & $4.1 \times 10^{-3}$ \\
GN3 & $1.5$ & $0.01$ & $7.8 \times 10^{-3}$ \\
GN4 & $1.0$ & $0.10$ & $2.3 \times 10^{-2}$ \\
GN5 & $1.0$ & $0.05$ & $2.1 \times 10^{-2}$ \\
\hline
\end{tabular}
\label{tab:inj_3inc_GN_param}
\end{table}

\begin{table}[hbt!]
\renewcommand{\arraystretch}{1.2}
\centering
\caption{The $L_2^{all}$ error metric for different ReLoBRaLo settings, used to train 3 sampled increments, using SF-RB-DS-L2. ``$-$" means the method produced ``NaN" values and could not complete the training.}
\begin{tabular}{ccccc}
\hline
Trial name & $\alpha_r$ & $\theta_r$ & $E[\rho_r]$ & $L_2^{all}$   \\ 
\hline
RB1 & $0.99$  & $1.0 \times 10^{-5}$ & 0.99   & $-$ \\
RB2 & $0.99$  & $5.0 \times 10^{-1}$ & 0.9    & $7.3 \times 10^{-3}$ \\
RB3 & $0.999$ & $1.0 \times 10^{-1}$ & 0.9999 & $1.0$ \\
RB4 & $0.999$ & $1.0 \times 10^{-2}$ & 0.9999 & $-$ \\
RB5 & $0.99$  & $1.0 \times 10^{-2}$ & 0.99   & $-$ \\
RB6 & $0.999$ & $1.0$                & 0.99   & $1.0$ \\
\hline
\end{tabular}
\label{tab:inj_3inc_RB_param}
\end{table}

\begin{figure}[hbt!]
\centering
\includegraphics[width=1.0\textwidth]{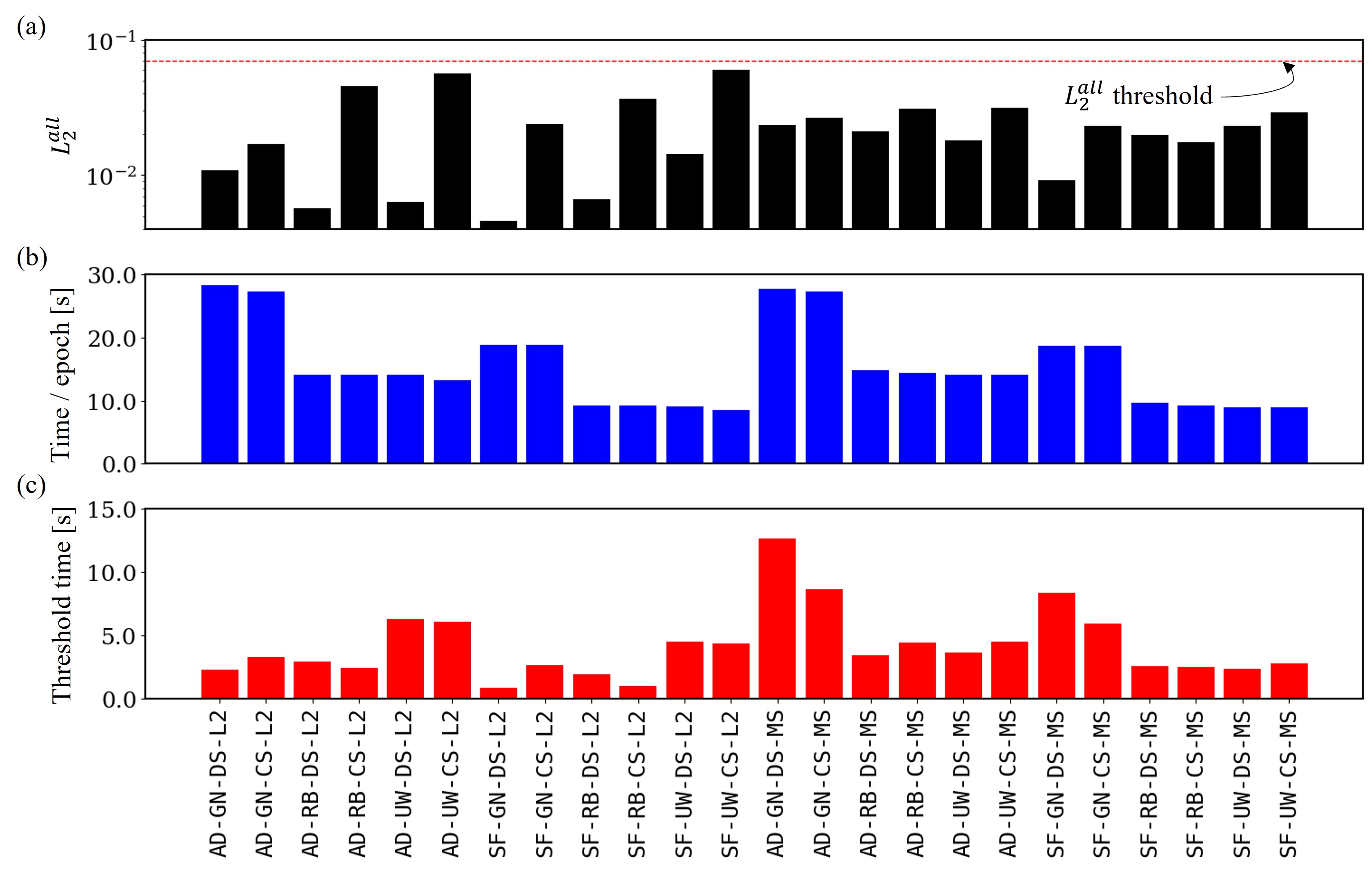}
\caption{Parametric study summary, considering multiple training variations: a) The final $L_2^{all}$ after training for $40000$ epochs, b) The average computational time per epoch, c) The computational time required to reach a threshold accuracy of $7\times 10^{-2}$ }
\label{fig:inj_param_study}
\end{figure}

Given the results in Table~\ref{tab:inj_3inc_GN_param} and Table~\ref{tab:inj_3inc_RB_param}, the tuning parameters for trials GN1 and RB2 are chosen for a more detailed parametric study comparing multiple variations of the methods listed in Table~\ref{tab:inj_param_abb}. GN2 has relatively lower error than GN1; however, it showed more training instability and was therefore not chosen for further analysis. Fig.~\ref{fig:inj_param_study} provides some insights on the parametric study results. Notably, all combinations that used a uniform scale of unity (US) either produced ``nan" values during training or had very high error values after training; therefore, all combinations using uniform scaling are excluded from the plot. 

In terms of accuracy after a given number of epochs, AD-RB-DS-L2, AD-UW-DS-L2, SF-GN-DS-L2, and SF-RB-DS-L2 showed the lowest error value, as shown in Fig.~\ref{fig:inj_param_study}a. In addition, the results (in Fig.~\ref{fig:inj_param_study}b) show that the combination between automatic differentiation (AD) and GradNorm (GN) is the most expensive one in terms of time per epoch. On the other hand, shape-function derivatives (SF) and ReLoBRaLo (RB) provide the least expensive combination. 

Some combinations are relatively expensive; however, they are efficient, reaching low error values in fewer epochs. This efficiency can be inferred by comparing the time required to reach a specific accuracy for all combinations, as shown in Fig.~\ref{fig:inj_param_study}c. The threshold value was reasonably set to be a value that is already satisfied by all the compared combinations. The combinations SF-GN-DS-L2, SF-RB-DS-L2, and SF-RB-CS-L2 provide the highest efficiency by reaching the threshold error in the least amount of time. Given the final accuracy and efficiency of the different combinations, we select SF-GN-DS-L2 and SF-RB-DS-L2 for a more detailed training/validation process for the injection example, as explained in section~\ref{section:numerical_examples:inj_flux_2d:training}.

\subsubsection{Additional plots on training using SF-GN-DS-L2}
\label{section:numerical_examples_appendix:inj_flux_2d:SF-GN-DS-L2}
\begin{figure}[hbt!]
\centering
\includegraphics[width=1.0\textwidth]{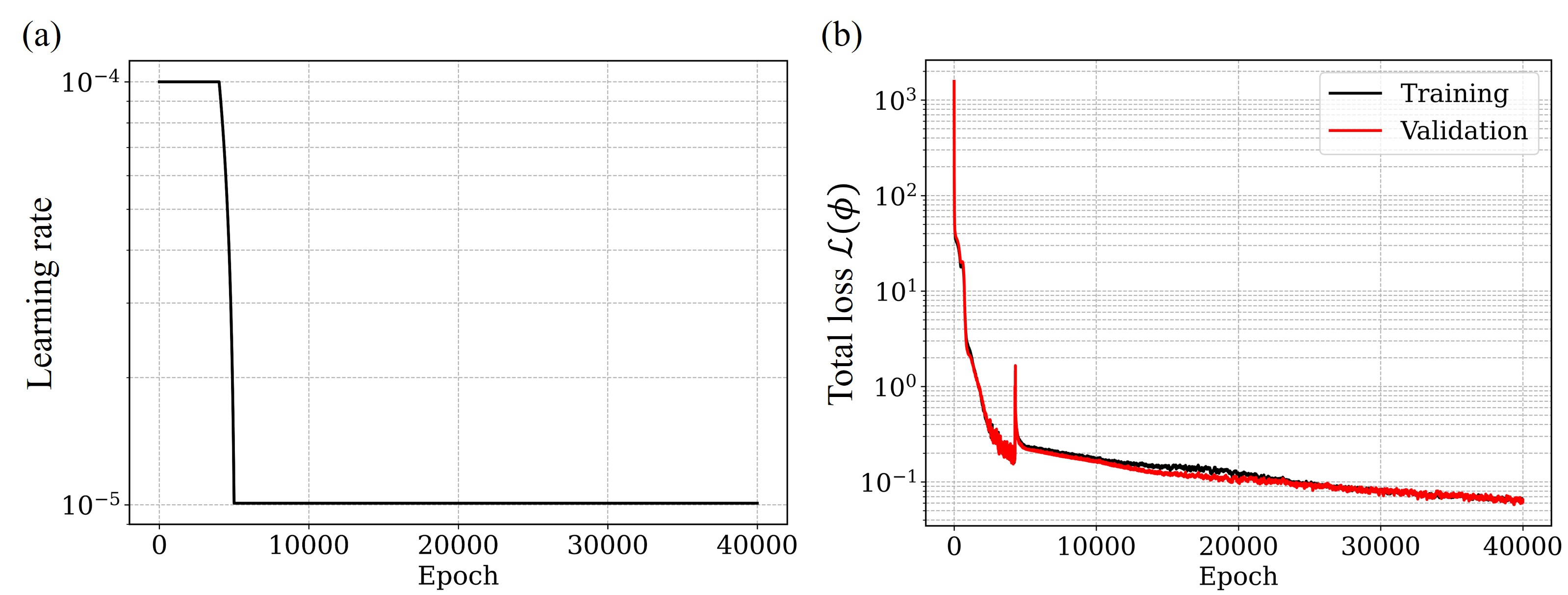}
\caption{Training metrics for the 2D injection example, using SF-GN-DS-L2: (a) Custom learning rate,  (b) Training and validation losses $\mathcal{L}(\phi)$.}
\label{fig:inj_2d_training}
\end{figure}

The adopted custom learning rate is provided in Fig.~\ref{fig:inj_2d_training}a. The training and validation loss ($\mathcal{L}(\phi)$) for this combination are depicted in Fig.~\ref{fig:inj_2d_training}b, showing no signs of overfitting with a common error-descending pattern for both curves.

\subsubsection{Training results using SF-RB-DS-L2}
\label{section:numerical_examples_appendix:inj_flux_2d:SF-RB-DS-L2}
\begin{figure}[hbt!]
\centering
\includegraphics[width=1.0\textwidth]{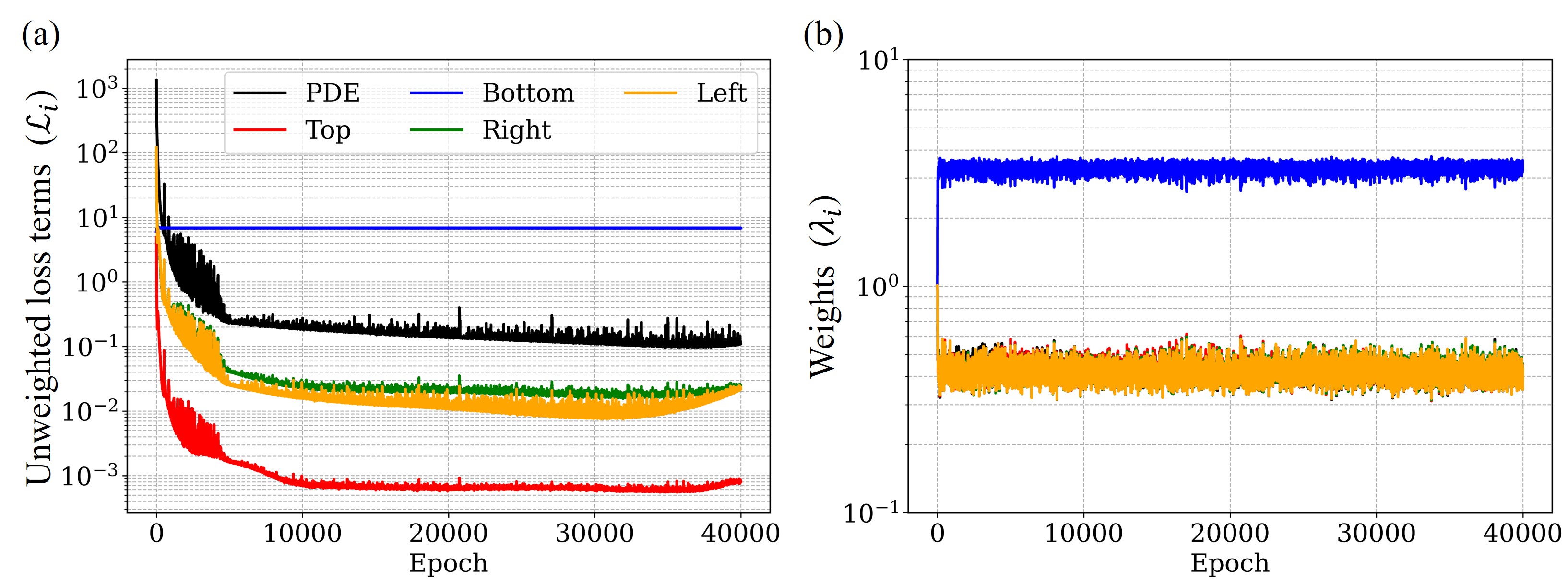}
\caption{Training metrics for the 2D injection example, using SF-RB-DS-L2: (a) unweighted loss terms $\mathcal{L}_i$, (b) the corresponding adaptive weights $\lambda_i$. Only PDE residual and boundary condition losses are included; no data loss is used.}
\label{fig:inj_2d_terms_RB}
\end{figure}

Fig.~\ref{fig:inj_2d_terms_RB} depicts the loss terms and their adaptive weights using SF-RB-DS-L2 for the 2D injection problem discussed in section~\ref{section:numerical_examples:inj_flux_2d:setup} and section~\ref{section:numerical_examples:inj_flux_2d:training}. All error terms were successfully reduced during training, except for the loss term $\mathcal{L}_3^{bottom}$, which corresponds to the injection-flux boundary condition. The adaptive weights algorithm tried to overcome this by increasing the weight assigned to this term; however, this did not affect the loss term. A general conclusion can be drawn from these results, along with the results in Appendix~\ref{section:numerical_examples_appendix:inj_flux_2d:hyperparameters}, that the ReLoBRaLo algorithm can be very efficient; however, it requires fine-tuning whenever changes are introduced to the training process.

\subsubsection{Additional IFENN-deal.II and IFENN-GEOS results.}
\label{section:numerical_examples_appendix:inj_flux_2d:ifenn_dealii_geos_2d}

\begin{figure}[hbt!]
\centering
\includegraphics[width=1.0\textwidth]{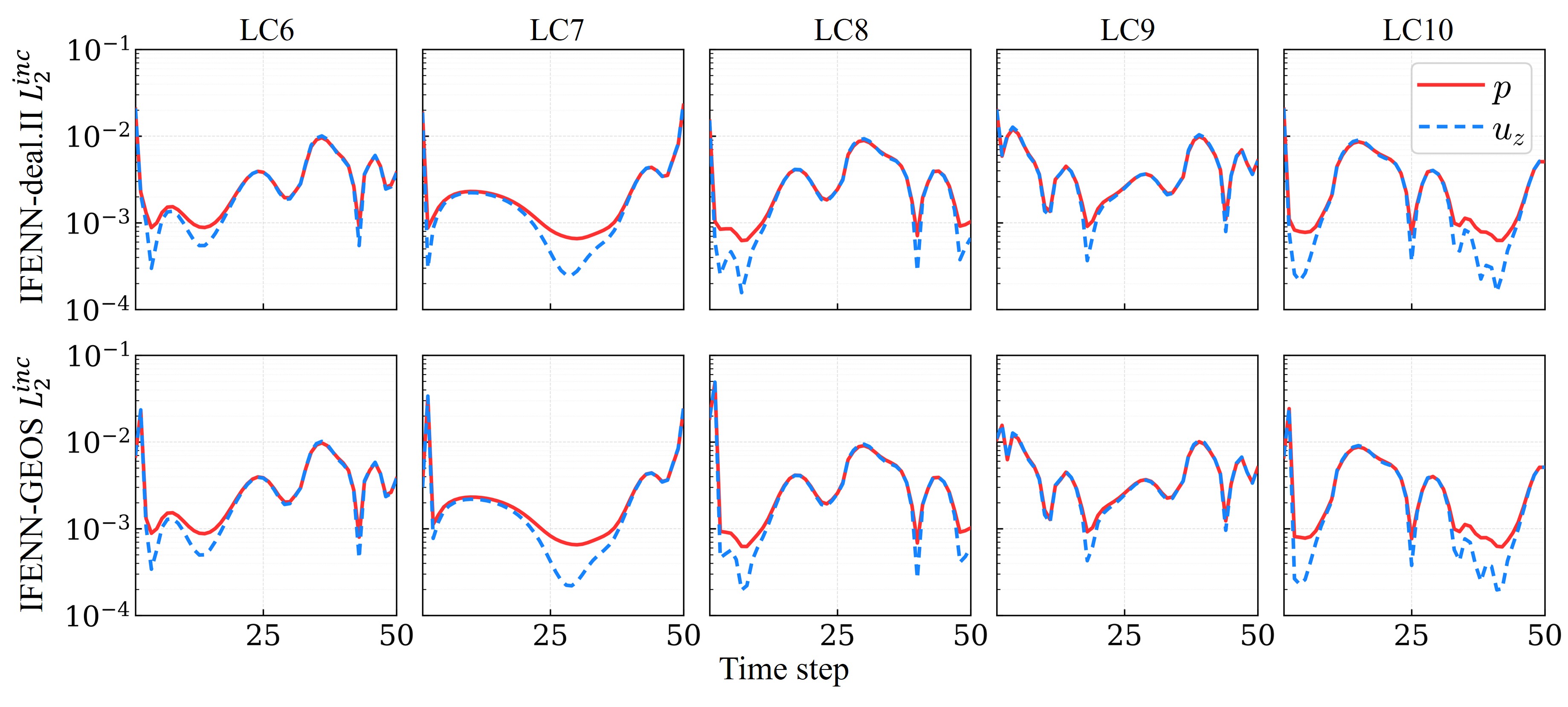}
\caption{Testing error $L_2^{inc}$ of pressure $(p)$ and displacement in the z-direction $(u_z)$ for testing load cases LC6-LC10 of the 2D injection problem, using IFENN-deal.II and IFENN-GEOS.}
\label{fig:inj_2D_ifenn_dealii_vs_geos_6_10}
\end{figure}

\begin{figure}[hbt!]
\centering
\includegraphics[width=1.0\textwidth]{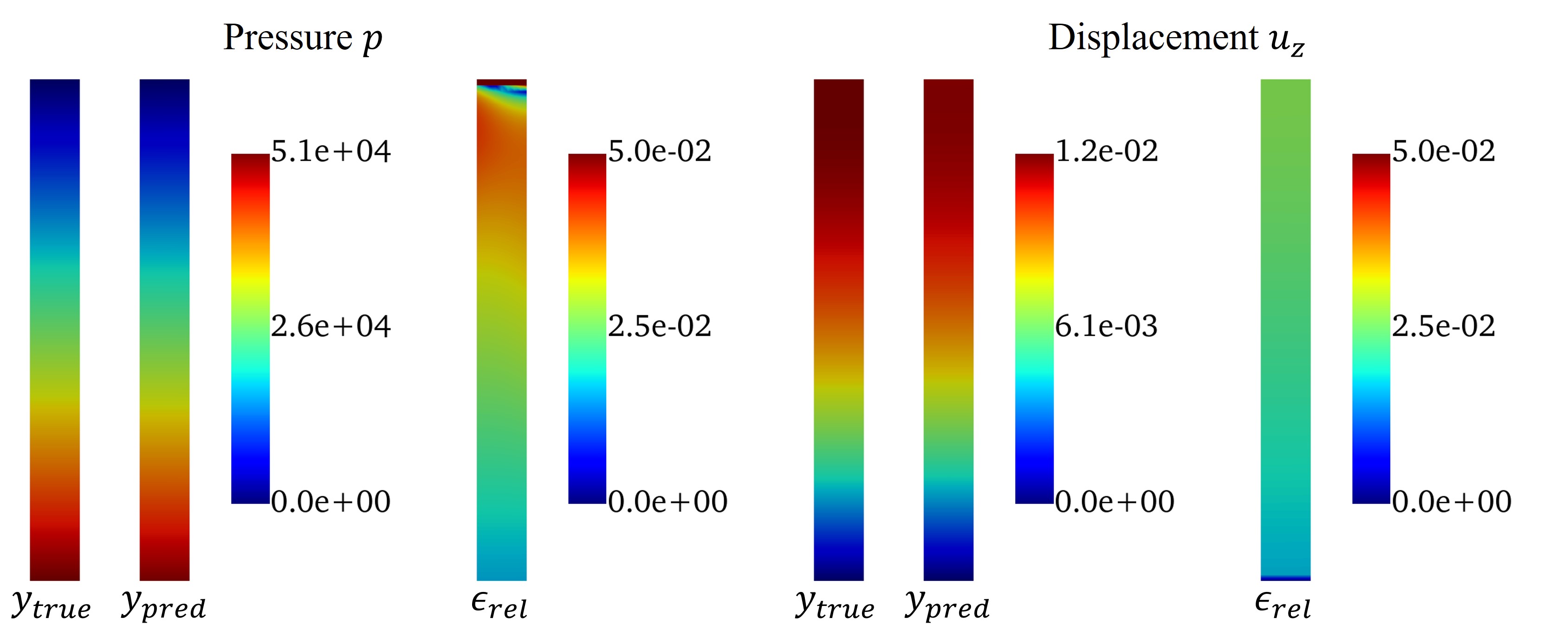}
\caption{Solution obtained using the fully coupled FEM solver ($y_{true}$) and IFENN-deal.II ($y_{pred}$), with the corresponding relative error ($\epsilon_{rel}$), for the 7th testing load case for the 2D injection example, at the 50th time step.}
\label{inj_ifenn_dealii_2D_LC7-T50}
\end{figure}

\begin{figure}[hbt!]
\centering
\includegraphics[width=1.0\textwidth]{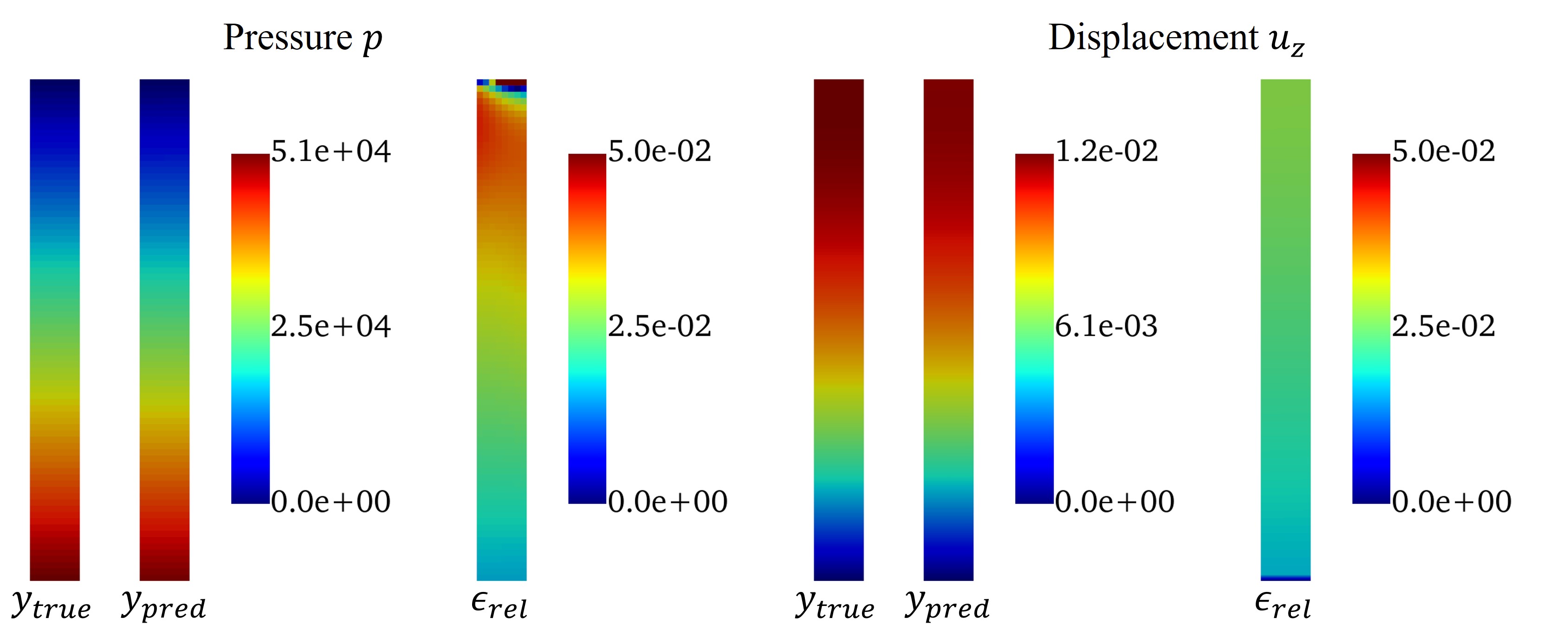}
\caption{Solution obtained using the fully coupled FEM solver ($y_{true}$) and IFENN-GEOS ($y_{pred}$), with the corresponding relative error ($\epsilon_{rel}$), for the 7th testing load case at the 50th time step for the 2D injection example.}
\label{fig:inj_ifenn_geos_2D_LC7-T50}
\end{figure}

To further investigate the results, we select a sample increment from Fig.~\ref{fig:inj_ifenn_dealii_2D_Norms} to plot the relative error metric $\epsilon_{rel}$, defined in Eq.~\eqref{eq:rel_error}. The seventh testing load case (LC7) had one of the highest $L_2^{inc}$ error values at the last time increment and, hence, is selected for plotting the point-wise relative error ($\epsilon_{rel}$) at this increment. Fig.~\ref{inj_ifenn_dealii_2D_LC7-T50} and Fig.~\ref{fig:inj_ifenn_geos_2D_LC7-T50} depict the response ($y_{pred}$) for IFENN-deal.II and IFENN-GEOS, respectively, for LC7 at the 50th time increment. The IFENN results show strong agreement with the coupled FEM results, maintaining a relative error of less than $5\%$ across most of the domain. The relative error $\epsilon_{rel}$ is computed by dividing the error by the true value $y_{true}$. Thus, when the true value is close to zero, the relative error values are magnified.

\subsection{2D injection example with variable \texorpdfstring{permeability $K_p$ and time increment $\Delta t$}{permeability and time increment}}
\label{section:numerical_examples_appendix:inj_flux_2d_k_dt}

\begin{figure}[hbt!]
\centering
\includegraphics[width=1.0\textwidth]{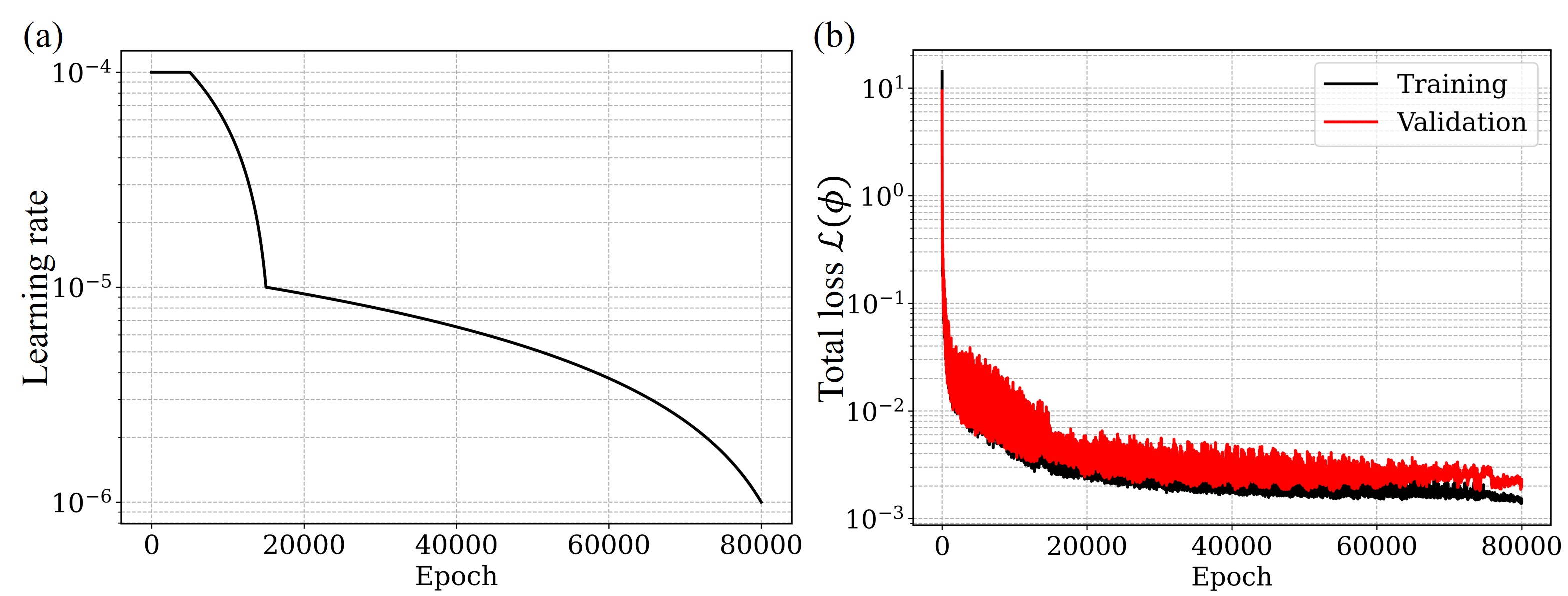}
\caption{Training metrics for the 2D injection example considering variable permeability $K_p$ and time increment $\Delta t$: (a) Custom learning rate,  (b) Training and validation losses $\mathcal{L}(\phi)$.}
\label{fig:inj_gen_training}
\end{figure}

\begin{figure}[hbt!]
\centering
\includegraphics[width=1.0\textwidth]{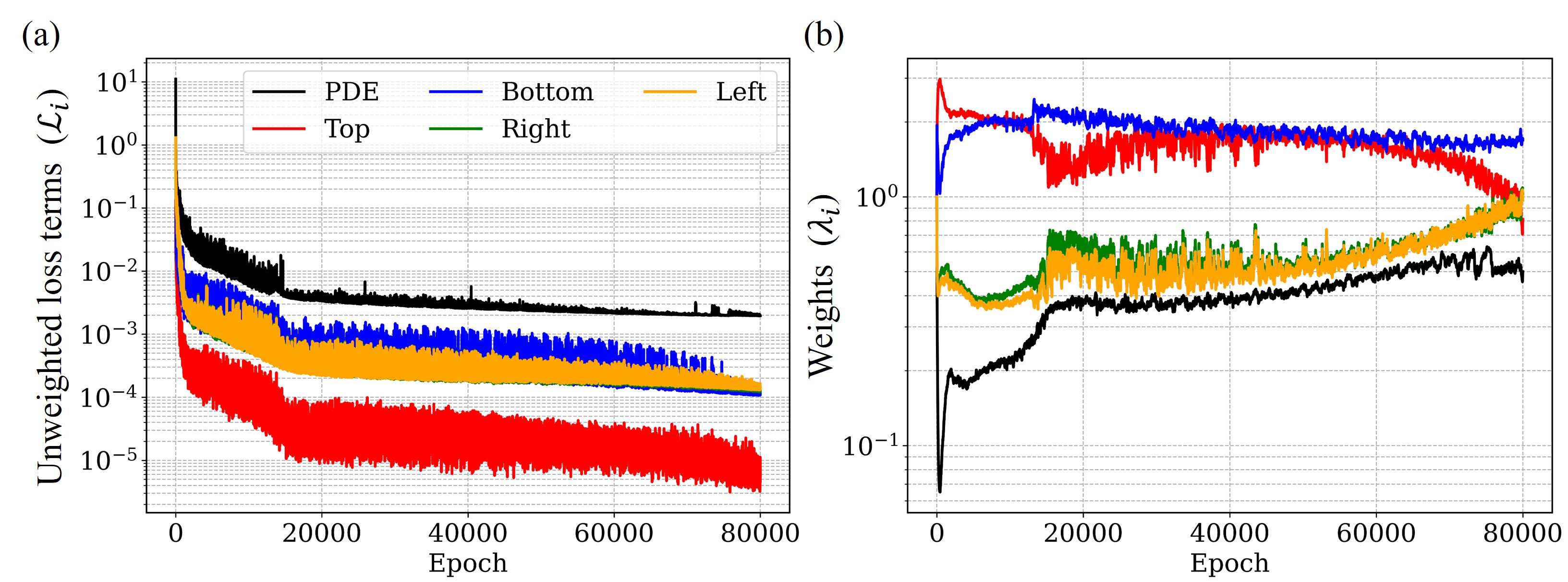}
\caption{Training metrics for the 2D injection example considering variable permeability $K_p$ and time increment $\Delta t$: (a) unweighted loss terms $\mathcal{L}_i$, (b) the corresponding adaptive weights $\lambda_i$. Only PDE residual and boundary condition losses are included; no data loss is used.}
\label{fig:inj_gen_terms}
\end{figure}

The model parameters are listed in Table~\ref{tab:inj_pifmionet_arch_gen}. The $K_p$-$\Delta t$ generalized PIF-MIONet is trained for 80,000 epochs using the Adam optimizer with a custom learning rate, as shown in Fig.~\ref{fig:inj_gen_training}a. Training and validation loss curves in Fig.~\ref{fig:inj_gen_training}b show a consistent downward trend, with fewer oscillations toward the end of training, consistent with the decreasing learning rate. The unweighted loss terms and their adaptive weights are provided in Fig.~\ref{fig:inj_gen_terms}.

\begin{table}[hbt!]
\renewcommand{\arraystretch}{1.2}
\centering
\caption{Hyperparameters of the generalized PIF-MIONet model trained for variable $K_p$ and $\Delta t$}

\begin{threeparttable}
\begin{tabular}{ccccc}
\hline
Network     & Input size & Sizes of hidden layers\tnote{a} &  $N_{ch}$\tnote{b}  & Output size \\
\hline
Branch \# 1 &$N_q=$1               & [8,16]   & -  & 16 \\
Branch \# 2 &$N_{K}=$1             & [8,8]    & -  & 8 \\
Branch \# 3 &$N_{t}=$1             & [8,8]    & -  & 8 \\
Branch \# 4 &$N_{\varepsilon}=$205 & [32]     & 16 & 8 \\
Branch \# 5 &$N_p=$205             & [32]     & 16 & 8 \\
Fusion      &$N_{fi}=$48           & [48,48]  & -  & 48 \\
Trunk       &$N_c=$2               & [48,48]  & -  & 48 \\
\hline
\end{tabular}

\begin{tablenotes}
\footnotesize
\item[a] The length of the sizes array is the number of hidden layers.
\item[b] Group normalization using $N_{ch}$ channels is implemented on the output of the input layer. ``-" means no group normalization is considered. 
\end{tablenotes}
\vspace{-7pt}
\rule{\linewidth}{0.1pt}
\end{threeparttable}

\label{tab:inj_pifmionet_arch_gen}
\end{table}

\begin{figure}[hbt!]
\centering
\includegraphics[width=1.0\textwidth]{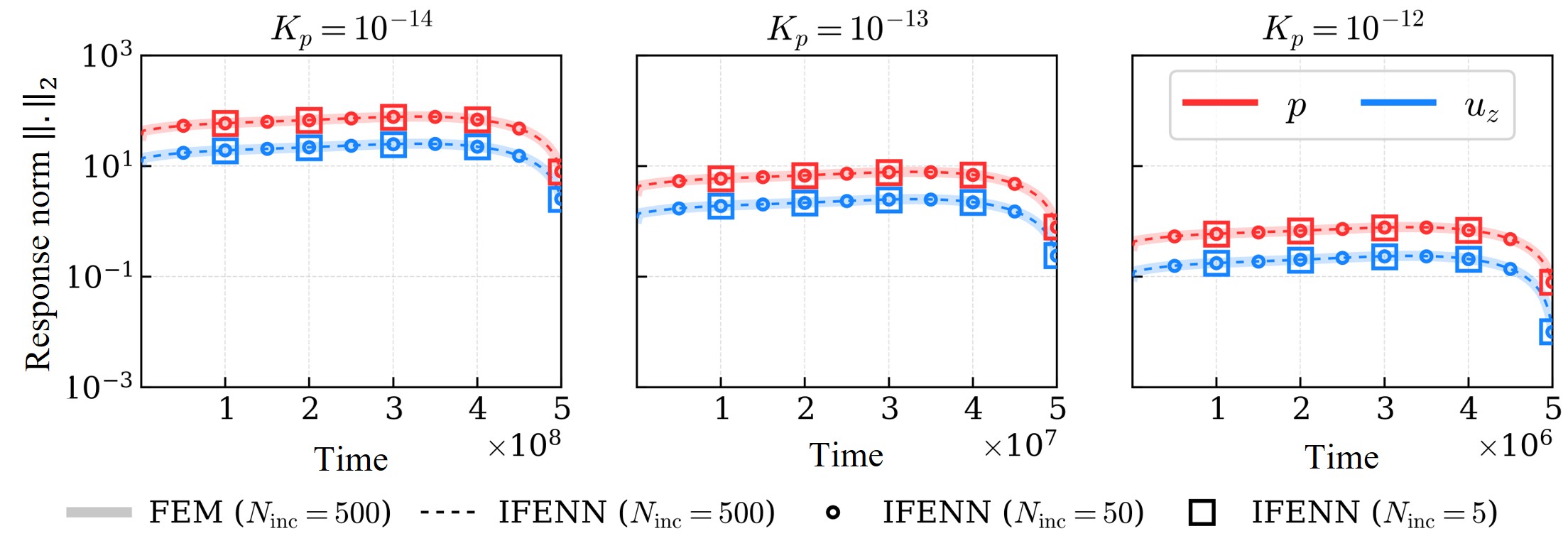}
\caption{Pressure norm ${\left\| p \right\|_2}$ and displacement norm ${\left\| u_z \right\|_2}$ obtained using the fully coupled FEM solver and IFENN-deal.II for 9 different setups (3 permeability values $\times$ 3 time increment values) in the 2D injection problem, subjected to the 7th testing load case. Response norm units are MPa and m for pressure and displacement, respectively.}
\label{fig:inj_gen_3x1_LC7}
\end{figure}

Fig.~\ref{fig:inj_gen_3x1_LC7} provides additional testing results for the 7th testing load case, showing consistent agreement between IFENN-deal.II and the fully coupled FEM solver across the different permeability and time-discretization configurations, further supporting the observations presented in Fig.~\ref{fig:inj_gen_3x1_LC9}.

\subsection{3D injection example}
\label{section:numerical_examples_appendix:inj_flux_3d}
\subsubsection{Training}
\label{section:numerical_examples_appendix:inj_flux_3d:training}

\begin{figure}[hbt!]
\centering
\includegraphics[width=1.0\textwidth]{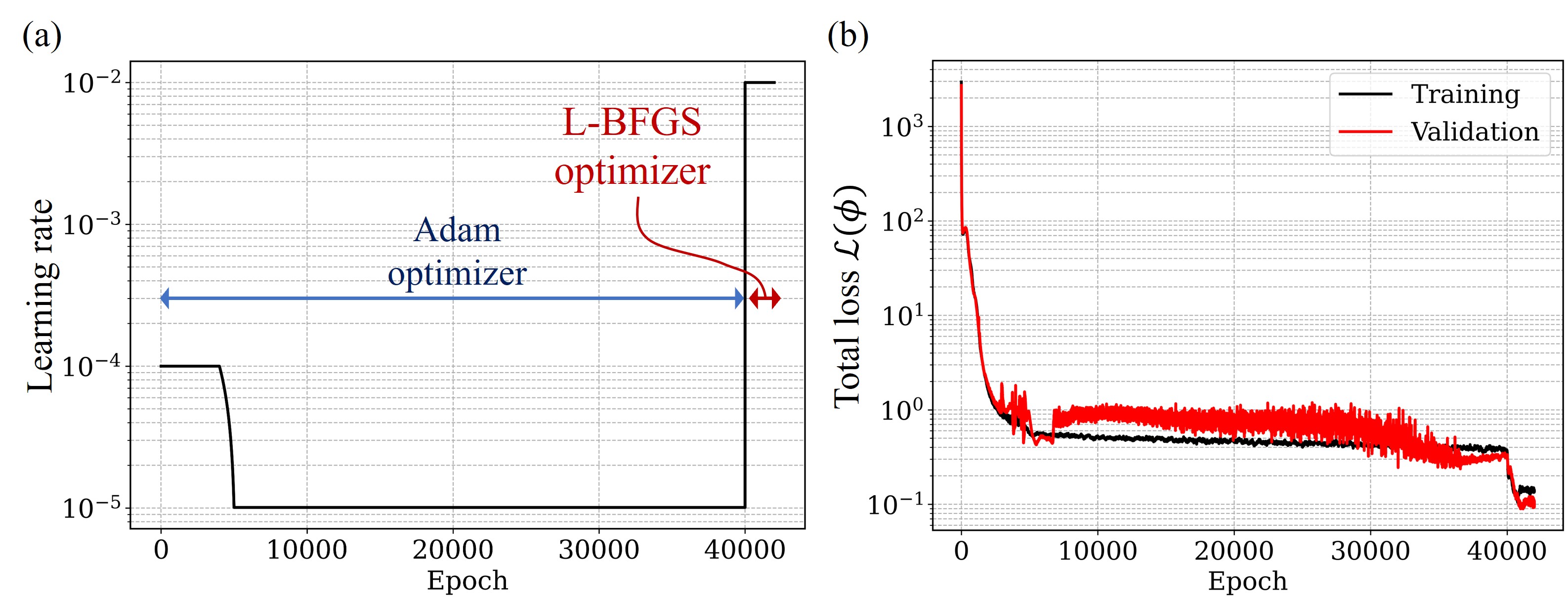}
\caption{Training metrics for the 3D injection example, using AD-GN-DS-L2: (a) Custom learning rate,  (b) Training and validation losses $\mathcal{L}(\phi)$}
\label{fig:inj_3d_training}
\end{figure}

The same optimizer setup used for the 2D case is used, followed by $2000$ epochs with L-BFGS and a learning rate of $1.0\times10^{-2}$, as shown in Fig.~\ref{fig:inj_3d_training}a. The training and validation histories in Fig.~\ref{fig:inj_3d_training}b show a consistent decrease in the loss after the Adam and L-BFGS stages.

\subsubsection{Additional results}
\label{section:numerical_examples_appendix:inj_flux_3d:results}

\begin{figure}[hbt!]
\centering
\includegraphics[width=1.0\textwidth]{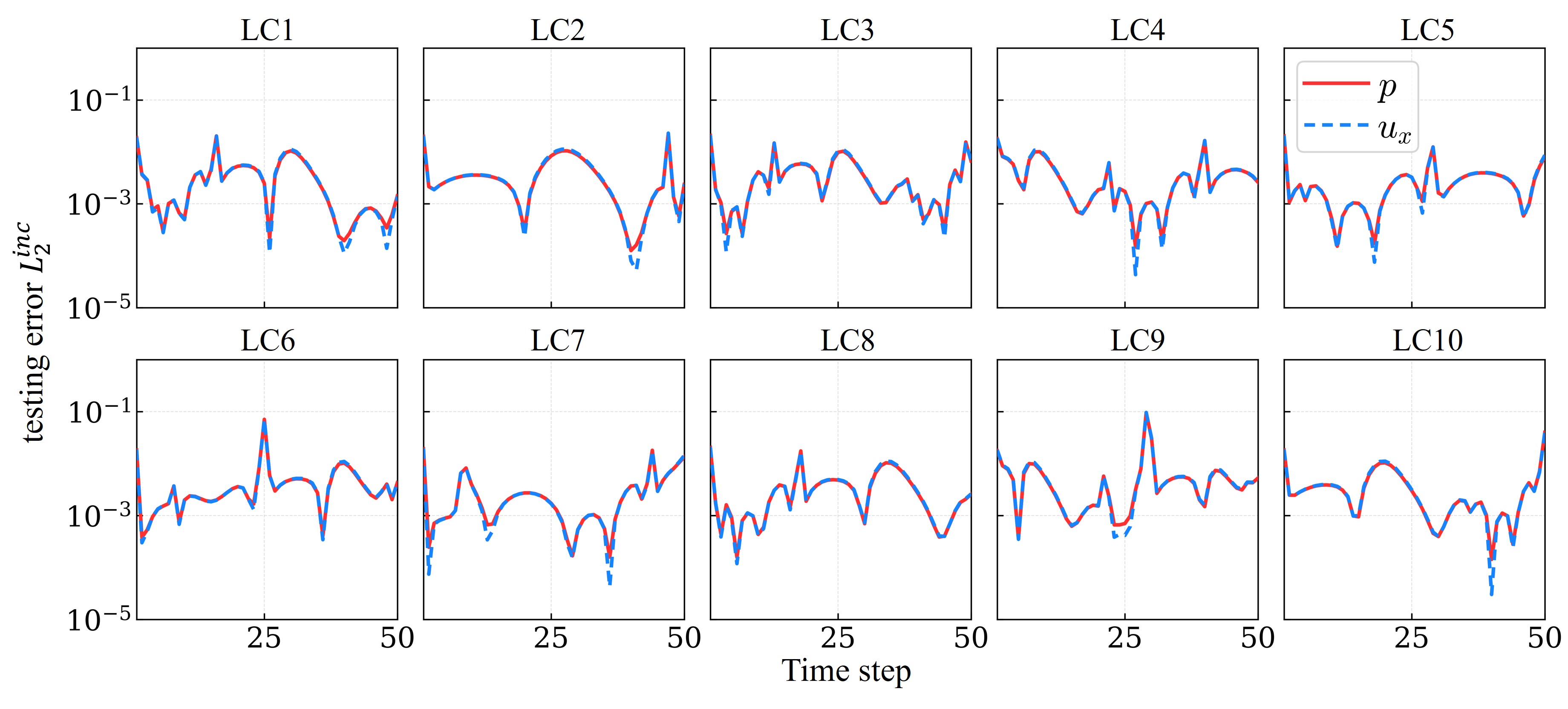}
\caption{Normalized testing error $L_2^{inc}$ for pressure $(p)$ and displacement in the z-direction $(u_z)$ of all testing load cases (LCs) using IFENN-GEOS for the 3D injection problem.}
\label{fig:inj_ifenn_geos_3D}
\end{figure}

\begin{figure}[hbt!]
\centering
\includegraphics[width=1.0\textwidth]{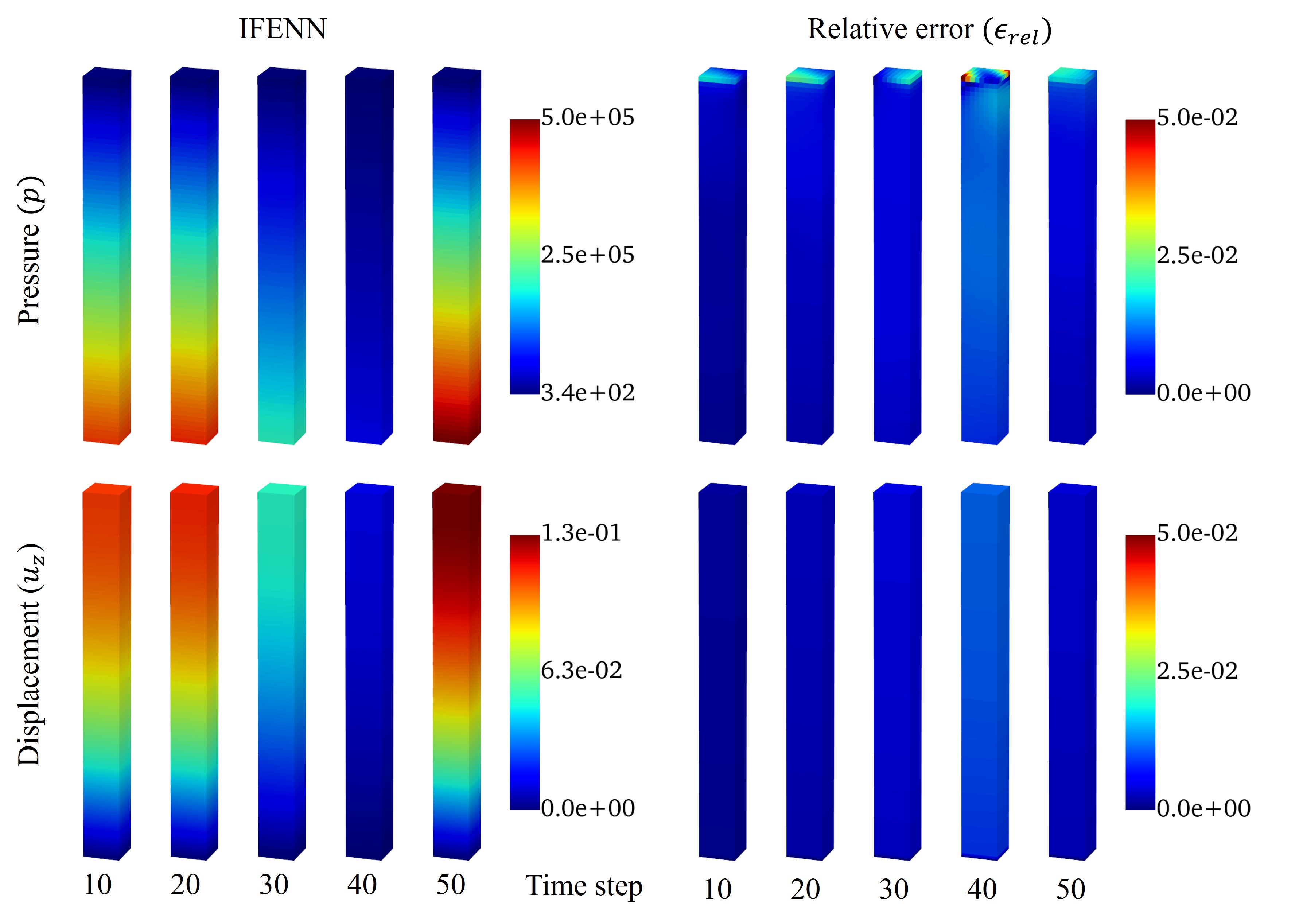}
\caption{Pressure $p$ and displacement $u_z$ obtained using IFENN-GEOS and the corresponding relative error ($\epsilon_{rel}$) for the 3D injection example at different time steps of the 7th testing load case.}
\label{fig:inj_ifenn_geos_3D_LC6}
\end{figure}

Additional results for the 3D injection example using IFENN-GEOS are presented in Fig.~\ref{fig:inj_ifenn_geos_3D} and Fig.~\ref{fig:inj_ifenn_geos_3D_LC6}. The errors in Fig.~\ref{fig:inj_ifenn_geos_3D} remain generally low across the test load cases, while the predictions in Fig.~\ref{fig:inj_ifenn_geos_3D_LC6} show that the largest pointwise relative errors occur in regions where response magnitudes are close to zero.

\subsection{Excavation problem}
\label{section:numerical_examples_appendix:excav}

\subsubsection{Additional discussion on training}
\label{section:numerical_examples_appendix:excav:param}

\begin{table}[hbt!]
\renewcommand{\arraystretch}{1.2}
\centering
\caption{Hyperparameters of the single-trunk PIF-MIONet model trained for the 2D excavation example}

\begin{threeparttable}
\begin{tabular}{ccccc}
\hline
Network     & Input size & Sizes of hidden layers\tnote{a} &  $N_{ch}$\tnote{b}  & Output size \\
\hline
Branch \# 1 &$N_q=$1               & [64,64]        & - & 64 \\
Branch \# 2 &$N_{\varepsilon}=$499 & [64,64]        & - & 64 \\
Branch \# 3 &$N_p=$499             & [64,64]        & - & 64 \\
Fusion      &$N_{fi}=$192          & [192]          & - & 256 \\
Trunk  \#1  &$N_c=$2               & [256,256,256]  & - & 256 \\
\hline
\end{tabular}

\begin{tablenotes}
\footnotesize
\item[a] The length of the sizes array is the number of hidden layers.
\item[b] Group normalization using $N_{ch}$ channels is implemented on the output of the input layer. ``-" means no group normalization is considered. 
\end{tablenotes}
\vspace{-7pt}
\rule{\linewidth}{0.1pt}
\end{threeparttable}

\label{tab:excav_pifmionet_arch_1_trunk}
\end{table}

Table~\ref{tab:excav_training_summary_bcs} lists a summary of the different training trials and their validation loss metric ($L_2^{all}$), computed as provided in Eq.~\eqref{eq:all_err}. The results show that the three-trunk model achieves a higher accuracy than the single-trunk model, despite having almost the same number of trainable parameters. In addition, including the data loss term has a positive impact on all cases. Finally, using AD to compute spatial derivatives in complex geometries results in unstable convergence and training failure.

\begin{table}[hbt!]
\renewcommand{\arraystretch}{1.2}
\centering
\caption{Different trained models for the excavation problem and their final validation loss values ($L_2^{all}$).}

\begin{threeparttable}
\begin{tabular}{cccccc}
\hline

Model & Spatial & Number & \multirow{2}{*}{$\mathcal{L}^{data}$} & $\mathcal{L}^{right}$, $\mathcal{L}^{wall1}$, & Validation loss  \\ 

code & gradients* & of trunks &  & and $\mathcal{L}^{wall2}$ & ($L_2^{all}$)  \\ 

\hline
M1    & SF  & 3 & $\checkmark$ & $\times$ & $3.7\times 10^{-3}$ \\
M2    & SF  & 3 & $\checkmark$ & $\checkmark$ & $4.5\times 10^{-3}$ \\
M3    & SF  & 3 & $\times$     & $\times$ & $1.0\times 10^{-1}$ \\
M4    & SF  & 3 & $\times$     & $\checkmark$ & $1.2\times 10^{-1}$ \\
M5    & SF  & 1 & $\checkmark$ & $\times$ & $1.2\times 10^{-2}$ \\
M6    & SF  & 1 & $\checkmark$ & $\checkmark$ & $1.8\times 10^{-2}$ \\
M7    & SF  & 1 & $\times$     & $\times$ & $1.3\times 10^{-1}$ \\
M8    & SF  & 1 & $\times$     & $\checkmark$ & $9.6\times 10^{-2}$ \\
M1-AD & AD  & 3 & $\checkmark$ & $\times$ & $4.6\times 10^{1}$  \\
\hline
\end{tabular}

\begin{tablenotes}
\footnotesize
\item * Spatial gradient calculation methods: AD for automatic differentiation and SF for shape-function derivatives. 
\item $\checkmark$ for included terms.
\item $\times$ for excluded terms.
\end{tablenotes}
\vspace{-7pt}
\rule{\linewidth}{0.1pt}
\end{threeparttable}

\label{tab:excav_training_summary_bcs}
\end{table}

\begin{figure}[hbt!]
\centering
\includegraphics[width=1.0\textwidth]{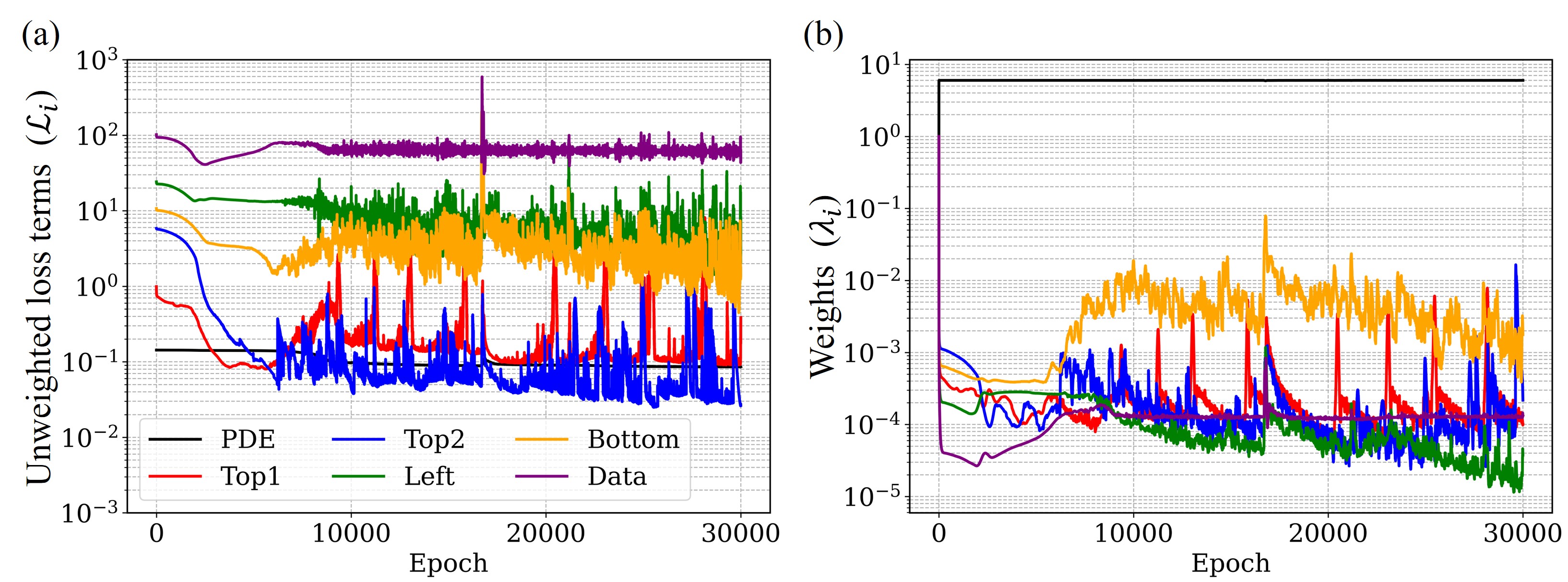}
\caption{Training metrics for the excavation example, using automatic differentiation (AD): (a) unweighted loss terms $\mathcal{L}_i$, (b) the corresponding adaptive weights $\lambda_i$.}
\label{fig:excav_terms_AD}
\end{figure}

The training behavior using automatic differentiation is further illustrated in Fig.~\ref{fig:excav_terms_AD}, where the loss terms exhibit persistent oscillations and unstable convergence. One possible reason is the singular behavior near the impermeable wall tip, as shown in Fig.~\ref{fig:excav_fem_strain_LC9_T30} and discussed in Section~\ref{section:numerical_examples:excav:training}. With AD, the PDE residual is evaluated directly from the network output, producing large residuals at collocation points near the wall-tip corners. In contrast, SF derivatives use 9-node macro-element interpolation of nodal predictions (Section~\ref{section:pif_mionet_design:sf_derivatives}), providing mesh-based regularization similar to FEM discretization. This approach remains stable for singular solutions. This likely explains the divergence of model M1-AD ($L_2^{all}=4.6\times10^{1}$) compared to the otherwise identical model M1 using SF derivatives ($L_2^{all}=3.7\times10^{-3}$). AD-based training performed well for the injection examples (Appendix~\ref{section:numerical_examples_appendix:inj_flux_2d:hyperparameters} and Section~\ref{section:numerical_examples:inj_flux_3D}), where the solution is smooth, supporting this interpretation. This observation also aligns with methods that avoid pointwise second derivatives by using finite element interpolation, as in~\cite{grekas_pinn-dg_2025}.

\begin{figure}[hbt!]
\centering
\includegraphics[width=1.0\textwidth]{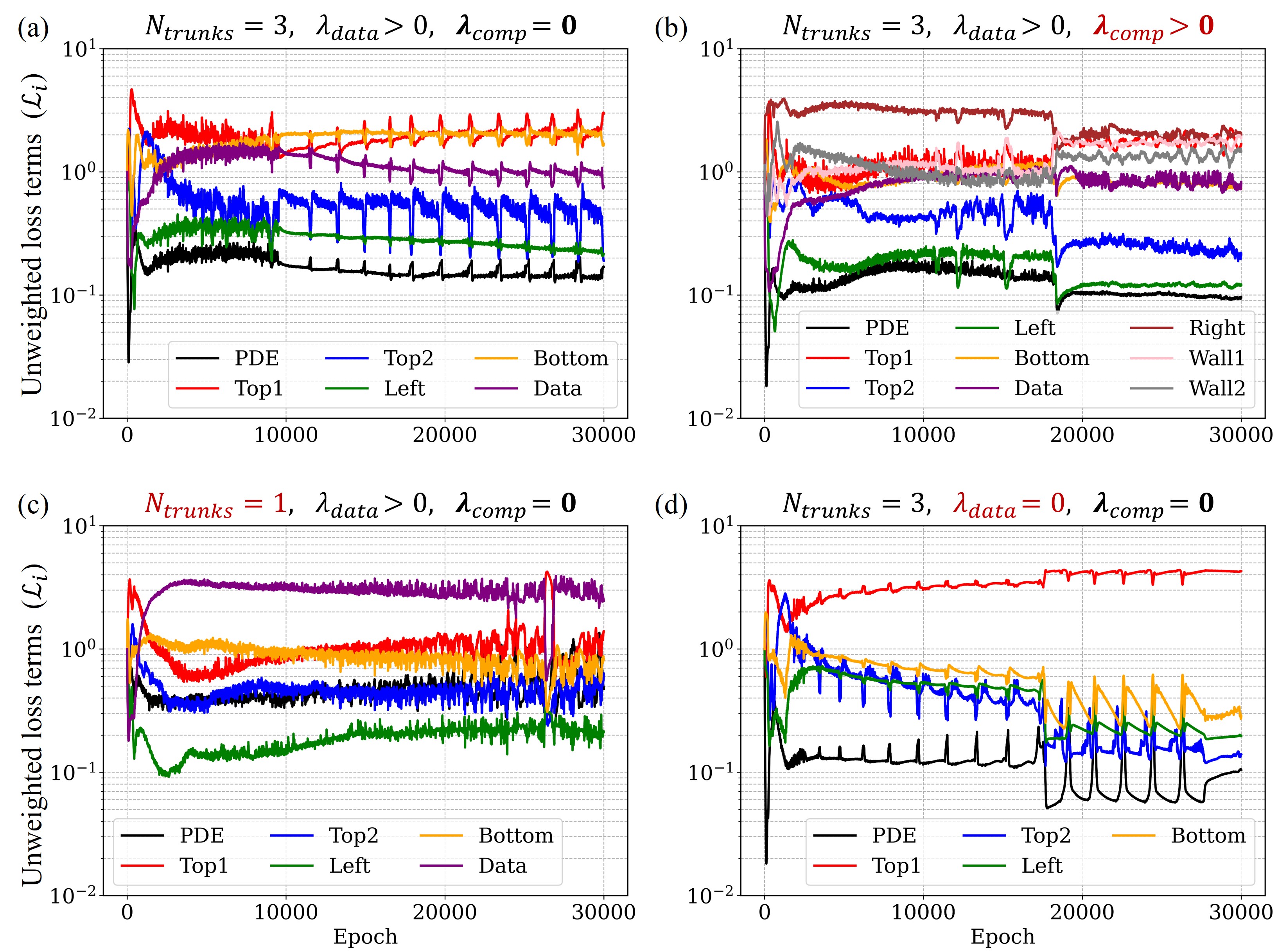}
\caption{Adaptive training weights $\lambda_i$ of the excavation problem under four training configurations: 
(a) A reference model of three trunks with
$\mathcal{L}^{data}$ ($\lambda_{data}>0$)
and without $\mathcal{L}^{comp}$
($\boldsymbol{\lambda}_{comp}=\boldsymbol{0}$),
(b) three trunks with both $\mathcal{L}^{data}$
($\lambda_{data}>0$) and $\mathcal{L}^{comp}$
($\boldsymbol{\lambda}_{comp}>\boldsymbol{0}$),
(c) one trunk with $\mathcal{L}^{data}$
($\lambda_{data}>0$) and without
$\mathcal{L}^{comp}$
($\boldsymbol{\lambda}_{comp}=\boldsymbol{0}$),
and (d) three trunks without either term
($\lambda_{data}=0$ and
$\boldsymbol{\lambda}_{comp}=\boldsymbol{0}$).}
\label{fig:excav_weights_4_models}
\end{figure}

Fig.~\ref{fig:excav_weights_4_models} compares the adaptive weights for the different training configurations, highlighting how changes in the loss terms and network architecture influence the weighting process. These observations complement the accuracy comparisons in Table~\ref{tab:excav_training_summary_bcs} and support the selection of the reference training configuration.

\subsubsection{Testing load cases and additional IFENN results}
\label{section:numerical_examples_appendix:excav:lcs_ifenn}
\begin{figure}[hbt!]
\centering
\includegraphics[width=1.0\textwidth]{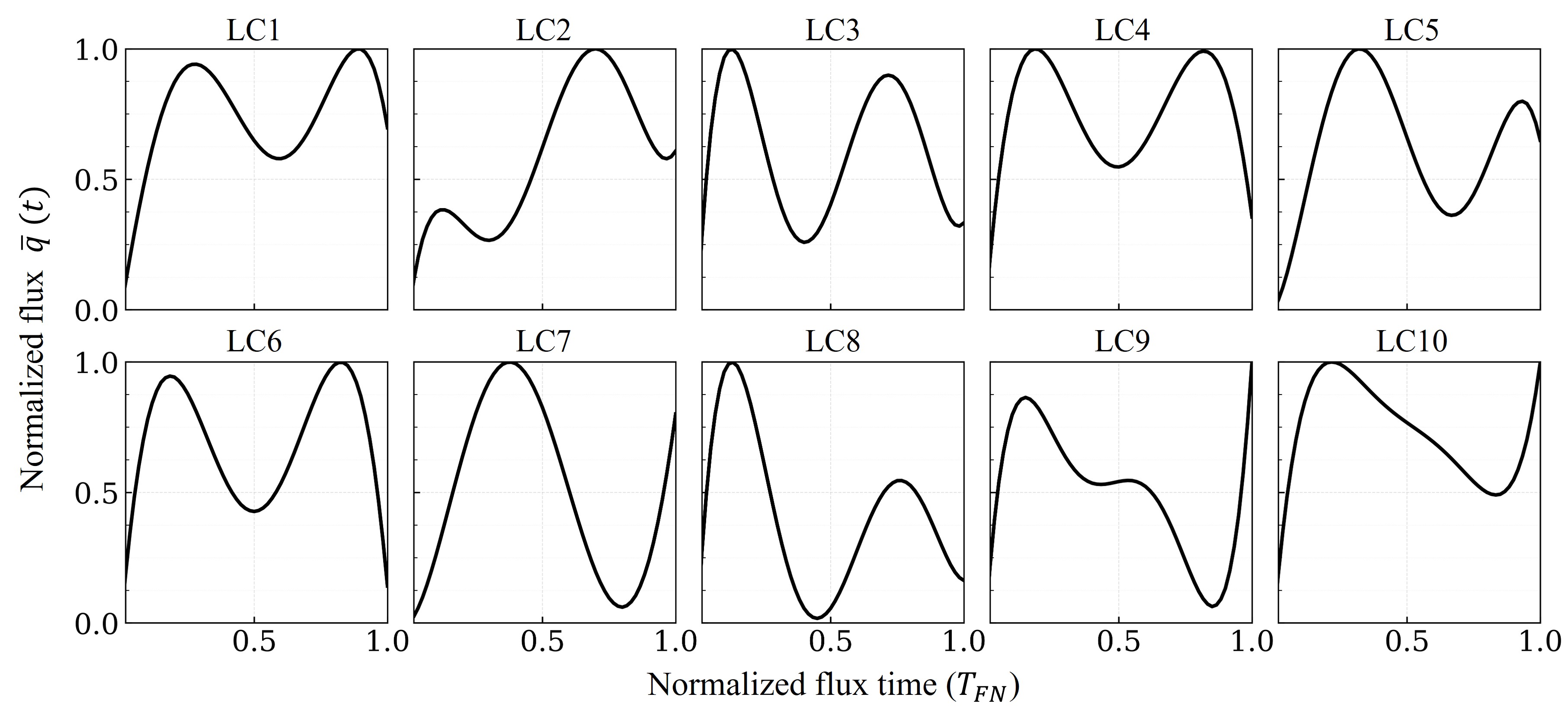}
\caption{Normalized flux amplitude versus normalized flux time ($T_{FN}$) for all testing load cases (LCs) of the excavation problem. Flux and time are normalized by their respective maximum values.}
\label{fig:excav_flux_testing_lcs}
\end{figure}

\begin{figure}[hbt!]
\centering
\includegraphics[width=0.5\textwidth]{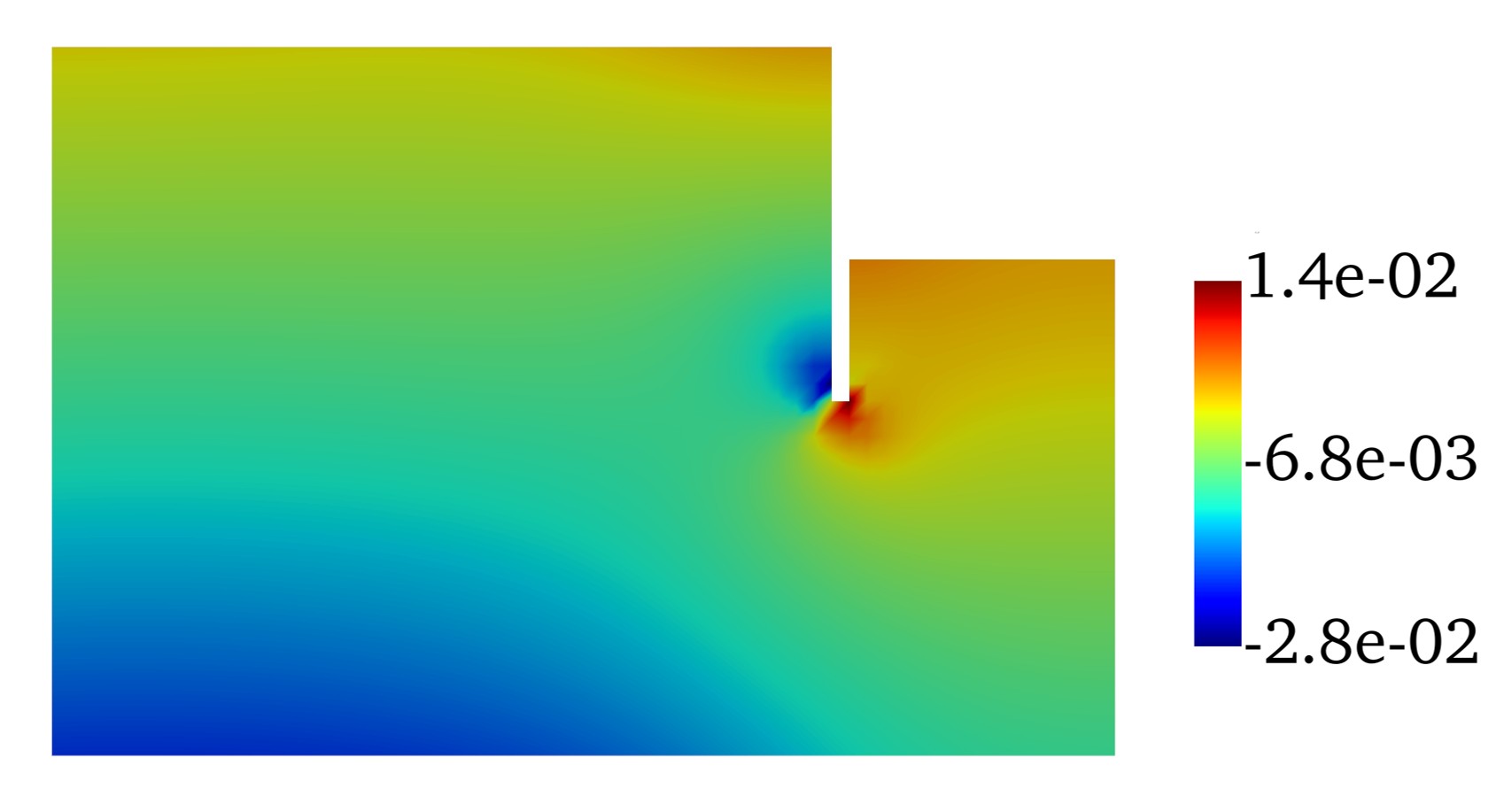}
\caption{Strain trace values $\text{tr}(\boldsymbol{\varepsilon})$ computed using the coupled FEM solver for the 30th increment of the 9th testing load case of the excavation example.}
\label{fig:excav_fem_strain_LC9_T30}
\end{figure}

\begin{figure}[hbt!]
\centering
\includegraphics[width=1.0\textwidth]{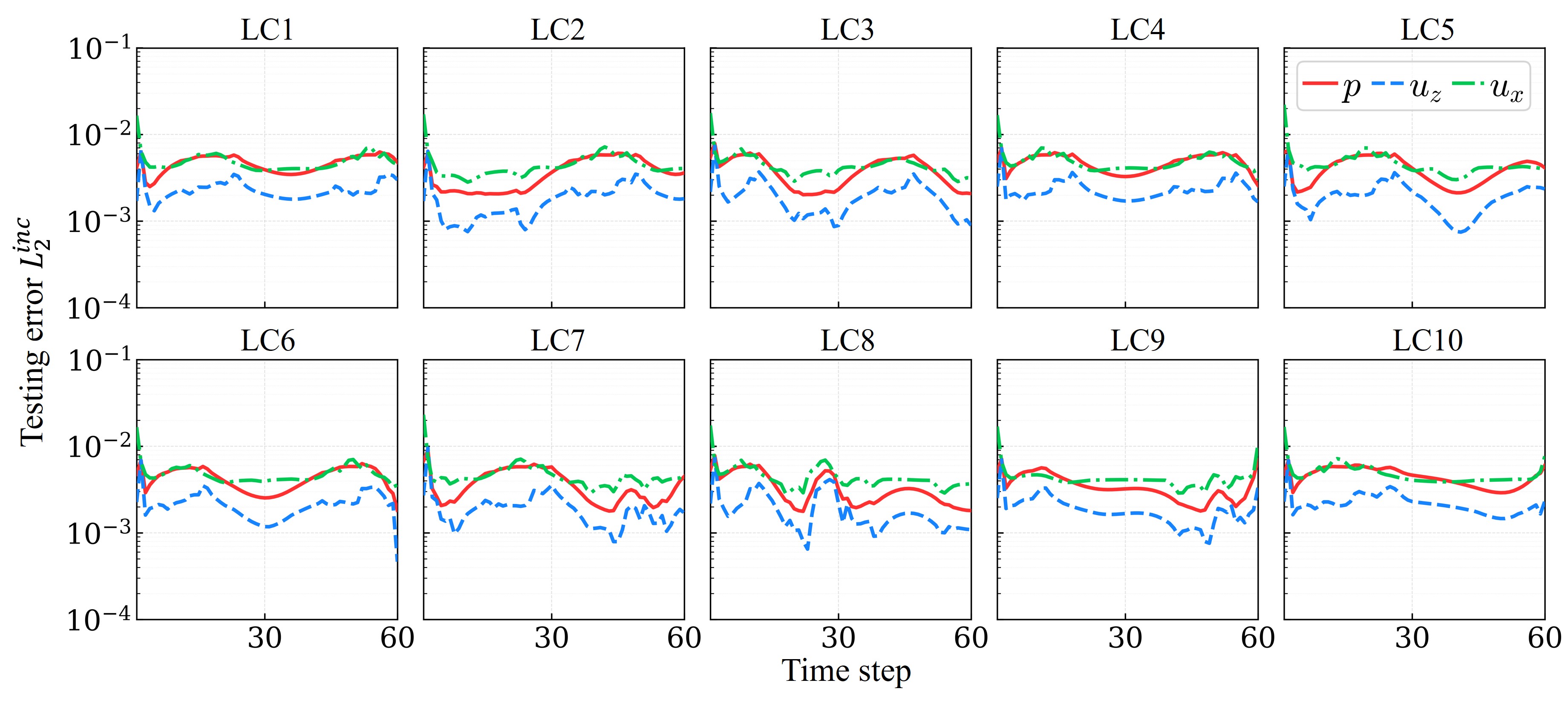}
\caption{Normalized testing error $L_2^{inc}$ for pressure $(p)$ and displacement in the z and x-directions $(u_z, u_x)$ of all testing load cases (LCs) using IFENN-deal.II with four predictor-multi-corrector subcycles ($N_{sub}=4$), for the excavation problem.}
\label{fig:excav_ifenn_4SC}
\end{figure}

\begin{figure}[hbt!]
\centering
\includegraphics[width=1.0\textwidth]{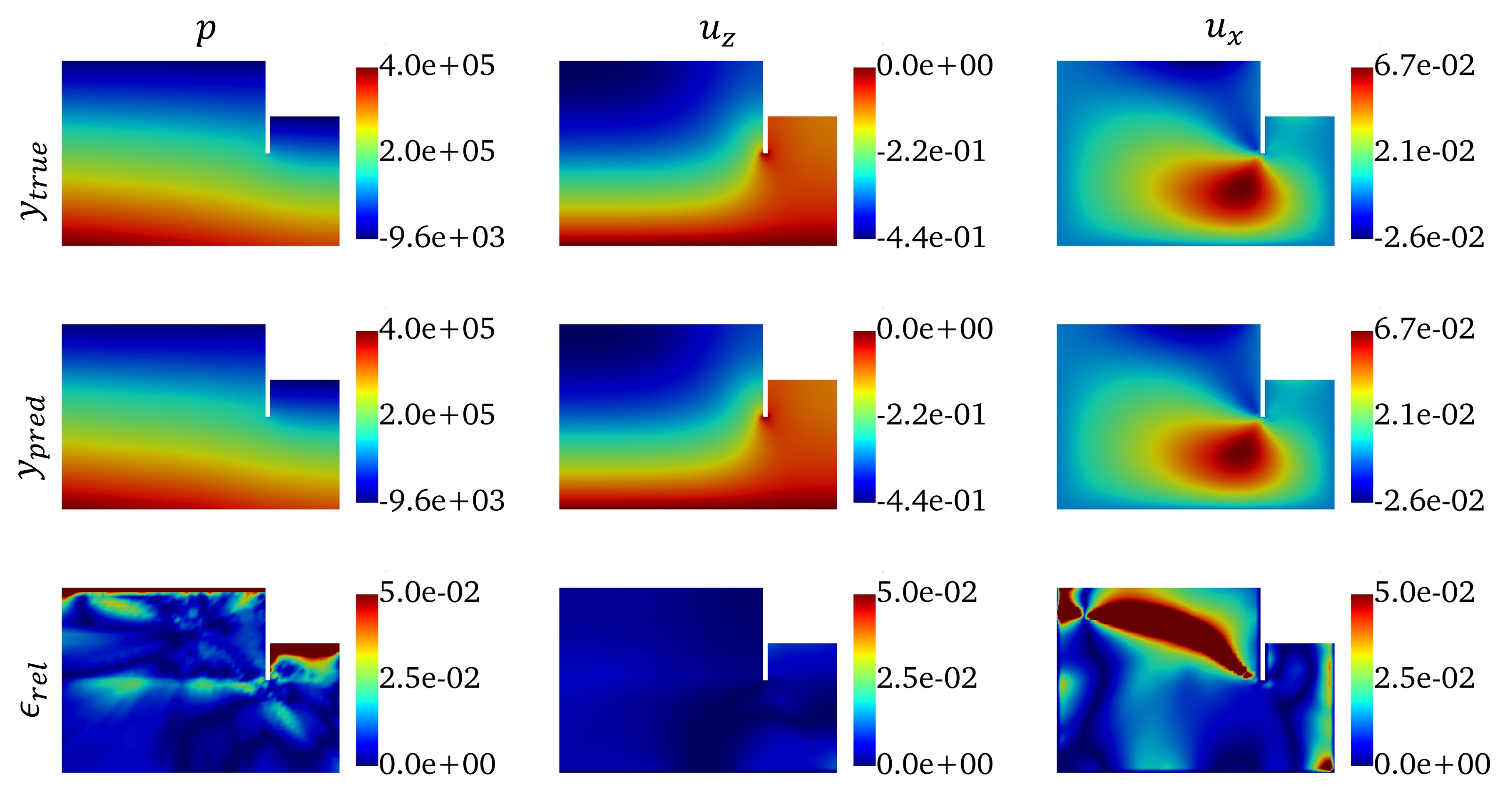}
\caption{Solution obtained using the fully coupled FEM solver ($y_{true}$) and IFENN-deal.II ($y_{pred}$), with the corresponding relative error ($\epsilon_{rel}$), for the 9th testing load case at the 60th time step for the excavation example.}
\label{fig:excav_ifenn_4SC_LC9_T60}
\end{figure}

The testing load cases for the excavation example are presented in Fig.~\ref{fig:excav_flux_testing_lcs}. Fig.~\ref{fig:excav_fem_strain_LC9_T30} illustrates the localized strain concentration near the excavation wall tip, highlighting the problem's geometric complexity. The testing errors for pressure $(p)$ and displacement in the z and x-directions $(u_z, u_x)$ using four subcycles  ($N_{sub}=4$) are shown in Fig.~\ref{fig:excav_ifenn_4SC}, while Fig.~\ref{fig:excav_ifenn_4SC_LC9_T60} provides a spatial field comparison between IFENN-deal.II and the fully coupled FEM solution for the 9th testing load case. These results further demonstrate the accuracy of the proposed framework for the excavation problem. 

\subsubsection{Effect of using multiple subcycles}
\label{section:numerical_examples_appendix:excav:subcycles}

\begin{figure}[hbt!]
\centering
\includegraphics[width=1.0\textwidth]{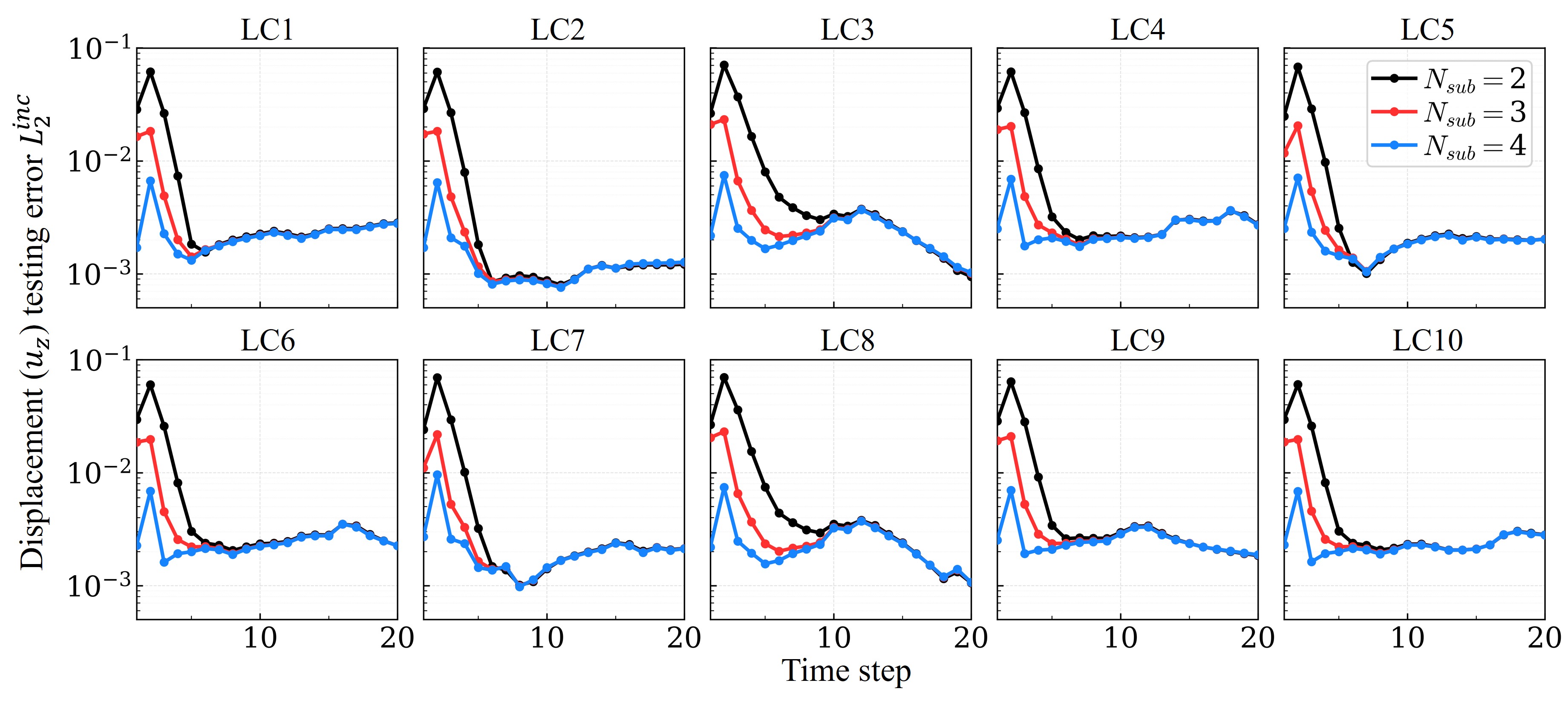}
\caption{Normalized testing error $L_2^{inc}$ for the displacement in z-direction $(u_z)$ of all testing load cases (LCs) using IFENN-deal.II with variations of predictor-multi-corrector subcycles ($N_{sub}=2,3,4$), for the excavation problem.}
\label{fig:excav_2_3_4_subcylces_uz}
\end{figure}

\begin{figure}[hbt!]
\centering
\includegraphics[width=1.0\textwidth]{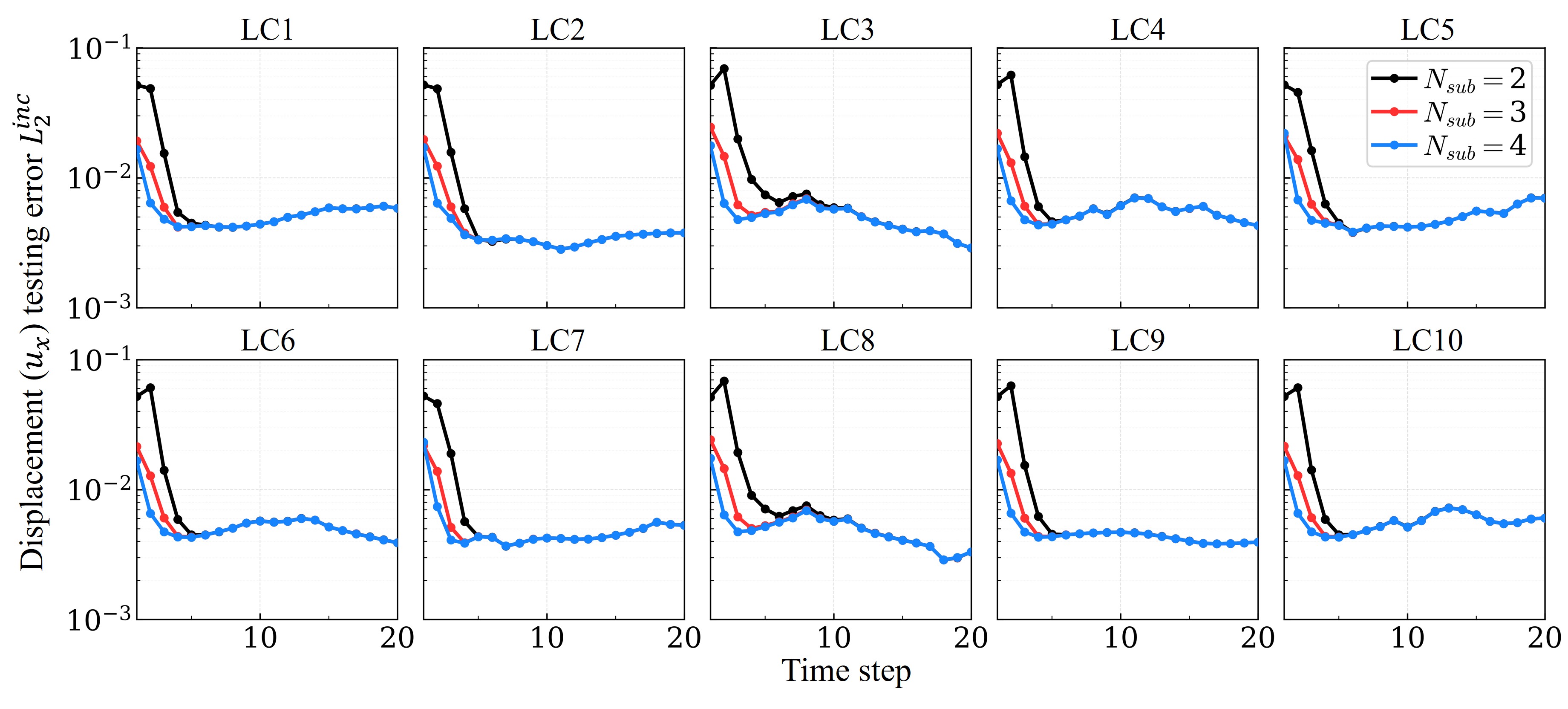}
\caption{Normalized testing error $L_2^{inc}$ for the displacement in x-direction $(u_x)$ of all testing load cases (LCs) using IFENN-deal.II with variations of predictor-multi-corrector subcycles ($N_{sub}=2,3,4$), for the excavation problem.}
\label{fig:excav_2_3_4_subcylces_ux}
\end{figure}

Fig.~\ref{fig:excav_2_3_4_subcylces_uz} and Fig.~\ref{fig:excav_2_3_4_subcylces_ux}  present the effect of the number of predictor-multi-corrector subcycles on the vertical and horizontal displacement errors, respectively. Consistent with the pressure results in Fig.~\ref{fig:excav_2_3_4_subcylces_p}, increasing the number of subcycles primarily improves accuracy during the initial time increments, while the differences become less pronounced at later increments.

\FloatBarrier
\bibliographystyle{style/elsarticle-num-no-url}
\bibliography{references/references}

\end{document}